\documentclass[12pt,preprint]{aastex}

\usepackage{lscape}
\usepackage{indentfirst}
\usepackage{graphicx}
\usepackage{longtable}
\usepackage[varg]{txfonts}
\usepackage[left,pagewise,mathlines]{lineno}

\renewcommand{\baselinestretch}{1.0}
\begin{document}

\begin{center}
\textbf{\large A multilayer model for thermal infrared emission of Saturn's rings. III: 
 Thermal inertia inferred from Cassini CIRS} 

 Ryuji Morishima$^{1,2,3}$,   Linda Spilker$^{2}$,  Keiji Ohtsuki$^{3,4}$
  
Ryuji.Morishima@jpl.nasa.gov

1: University of California, Los Angels, Institute of Geophysics and Planetary Physicc, Los Angeles, CA  90095, USA

2: Jet Propulsion Laboratory/California Institute of Technology,
Pasadena, CA 91109, USA

3: Laboratory for Atmospheric and Space Physics, University of Colorado, Boulder, CO 80309, USA

4: Department of Earth and Planetary Sciences and Center for Planetary Science, Kobe University, 657-8501, Japan

\end{center}
\vspace{5em}

Icarus 215, 107--127
\vspace{3em}

Manuscript total pages: 47

Table: 7

Figures: 13

\clearpage

\hspace{-2em}\textbf{Proposed Running Head:}
 
\textbf{Thermal inertia of Saturn's ring particles} 

\vspace{2em}

\noindent \textbf{Editorial correspondence to:}

\noindent Ryuji Morishima

\noindent Jet Propulsion Laboratory  

\noindent M/S 230-205, 4800 Oak Grove Drive 

\noindent Pasadena, CA 91109Duane Physics - 392 UCB

\noindent Phone:+1 818 393 1014; Fax:+1 818 393 4495 

\noindent e-mail: Ryuji.Morishima@jpl.nasa.gov
\clearpage

\begin{abstract}
{The thermal inertia values of Saturn's main rings (the A, B, and C rings and 
the Cassini division) are derived by applying our thermal model
to azimuthally scanned spectra taken by the Cassini Composite Infrared Spectrometer (CIRS).
Model fits show the thermal inertia of ring particles to be 
16, 13, 20, and 11 Jm$^{-2}$K$^{-1}$s$^{-1/2}$ for the A, B, and C rings, and the Cassini division, respectively.
However, there are systematic deviations between modeled and observed temperatures in Saturn's shadow depending 
on solar phase angle, and these deviations indicate that the apparent thermal inertia increases 
with solar phase angle. This dependence is likely to be explained if large slowly spinning particles 
have lower thermal inertia values than those for small fast spinning particles because 
the thermal emission of slow rotators is relatively stronger than that of fast rotators at low phase and vise versa. 
Additional parameter fits, which assume that slow and fast rotators have different thermal inertia values, 
show the derived thermal inertia values of slow (fast) rotators to be 8 (77), 8 (27), 9 (34), 5 (55) Jm$^{-2}$K$^{-1}$s$^{-1/2}$ 
for the A, B, and C rings, and the Cassini division, respectively. 
The values for fast rotators are still much smaller 
than those for solid ice with no porosity. Thus, 
fast rotators are likely to have surface regolith layers, but these may not be as fluffy as 
those for slow rotators, probably because the capability of holding regolith particles is limited for fast rotators
due to the strong centrifugal force on surfaces of fast rotators. 
Other additional parameter fits, in which radii of fast rotators are varied, indicate 
that particles less than  $\sim$ 1 cm should not occupy more than a half of the cross section
for the A, B, and C rings.  
 
}

\end{abstract}

\section{Introduction}
This is our third paper analyzing thermal infrared data of Saturn's rings.
In Morishima et al. (2009; hereafter Paper I), we introduced our new multilayer model, and applied it to ground-based data obtained before the Cassini mission.
In Morishima et al. (2010; hereafter Paper II), we improved our code, and estimated the values of 
the albedo and the ratio of fast and slow rotators using thermal data radially scanned by the Cassini Composite Infrared Spectrometer (CIRS).
In the present paper, we estimate the thermal inertia values of Saturn's rings using 
thermal data azimuthally scanned by Cassini CIRS including those in Saturn's shadow.

Saturn's main rings (the A, B, and C rings and the Cassini division) consist of a numerous number of icy particles. 
The range of the particle size deduced from radio and stellar light occultations is roughly 1 cm to 10 m 
(Marouf et al., 1983; Zebker et al., 1985; French and Nicholson, 2000; Cuzzi et al., 2009).
The ring spectra in ultra-violet (Bradley et al., 2010), near-infrared (Poulet et al., 2003; Nicholson et al., 2008), and far-infrared (Spilker et al., 2005) light 
indicate that surfaces of ring particles are covered by tiny regolith particles ranging from micron to cm sizes.\footnote{The estimated size of regolith particles increases with wavelength. 
It may indicate the regolith particle size increases with depth in a single ring particle, as longer waves reach to deeper in a ring particle. On the other hand, 
the estimated size may be biased by observed wavelengths even though regolith particles with different sizes are well mixed regardless of depth.} 
This means that the largest regolith particles are comparable to the smallest free floating ring particles. 
The abundance of free floating micron-size grains is likely to be negligible (Dones et al., 1993), except when
spokes form (e.g., D'Aversa et al., 2010). 
Most likely, most of small grains ($<$ cm) stick to larger particles due to the surface energy (e.g., Choksi et al., 1993).
The composition of ring (or regolith) particles is mostly crystalline water ice and the mass fraction of contaminants
(Tholins, PAHs, or nanohematite) is less than a percent (Poulet et al., 2003; Cuzzi et al., 2009). 

Besides the spectroscopy, thermal infrared observations can provide unique information about the surface layers of icy particles and 
about the particle size. 
The ring temperature drops during particles pass through Saturn's shadow and increases after they exit from the shadow to the sunshine.
The degree of the temperature drop increases with decreasing thermal inertia of the particle, if the particle size is larger than the thermal skin depth for eclipse.
The thermal inertia, which is given by the square root of the product of  the physical density, the specific heat, and the thermal conductivity, represents 
how much the particle suppresses its temperature change against illumination flux change.
While the uncertainties in the density and the specific heat may be only by a factor of 2-3, 
the thermal conductivity varies over several orders of magnitude depending on structures of surface layers.
A numerical model of Shoshany et al. (2002) showed that the thermal conductivity decreases by a factor of $10^{4}$  
with increasing porosity from 0 to $\sim 80 \%$. There has been no direct estimation of the porosity of ring particles. 
However, the internal density of ring particles of $\sim 450$ kg m$^{-3}$ or less is favorable in 
dynamical simulations for formation of wakes in the A and B rings (Salo et al., 2001; Stuart et al., 2010), 
because signatures of wakes become too strong in simulations using the density
of the solid ice with no porosity ($\sim 900$ kg m$^{-3}$). This indicates that the porosity of ring particles is likely to be 
larger than $50\%$ and  the surface layers' porosity may be even higher. 
An importance of a network of microcracks was also pointed out by Kouchi et al. (1992), who found very low values of 
the thermal conductivity of amorphous and crystalline ices 
(thermal inertia values less than 10 Jm$^{-2}$K$^{-1}$s$^{-1/2}$) in their laboratory experiments even though the porosity was low. 
Overall, the thermal inertia value is a good indication for the structure of regolith layers represented by poorly known quantities such as the porosity.
If the particle size is smaller than the thermal skin depth, the interior of the particle is roughly entirely isothermal. In this case, the temperature drop 
in the shadow increases with decreasing  particle size and is rather independent of the thermal inertia. If the particle size is comparable to the thermal
skin depth, the temperature drop depends on both thermal inertia and particle size.

Prior to the Cassini mission, the thermal inertia values were estimated using thermal data mostly from ground-based observations. 
Morrison (1974) and Froidevaux et al. (1981) 
estimated the thermal inertia of the B ring particles to be $>$ 40 Jm$^{-2}$K$^{-1}$s$^{-1/2}$ 
and $\sim$ 13 Jm$^{-2}$K$^{-1}$s$^{-1/2}$, respectively, using their ground-based data and applying a simple cooling and heating model developed by Aumann and Kieffer (1973). 
They also found that particles less than about a centimeter should not occupy a significant fraction of the ring cross section.
Ferrari et al. (2005) applied their thermal model to their ground based data and estimated the thermal inertias of the B and C rings 
to be 5$^{+18}_{-2}$ Jm$^{-2}$K$^{-1}$s$^{-1/2}$ and 6$^{+12}_{-4}$ Jm$^{-2}$K$^{-1}$s$^{-1/2}$, respectively.
Leyrat et al. (2008a) analyzed both their ground-based observation data and Voyager IRIS data for the C ring, 
using their new monolayer model (Ferrari and Leyrat, 2006), and confirmed the very low thermal inertia estimated 
by Ferrari et al. (2005).  The thermal inertia values for the A ring and the Cassini division have not been estimated 
so far using ring temperatures in Saturn's shadow.

Since the Saturn orbit insertion of the Cassini spacecraft in July 2004, Cassini CIRS
has obtained more than a million spectra (7 $\mu$m -- 1mm) of Saturn's rings (Flasar et al., 2005; Spilker et al., 2005, 2006; 
Altobelli et al., 2007, 2008, 2009; Leyrat et al., 2008b; Flandes et al., 2010). 
In Paper II, we estimated the albedo and the fraction of fast rotators with a good radial resolution,
applying our multilayer model developed in Paper I to the radially scanned CIRS data selected in Spilker et al. (2006). 
Although the thermal inertia values were also estimated in Paper II,  the data sets were not very suitable for accurate measurements of the thermal inertia.
For optically thick rings, the thermal inertia was constrained from the temperature difference between the lit and unlit faces,
but there was degeneracy between the thermal inertia and the fraction of particles bouncing at the midplane. 
For optically thin rings,  the thermal inertia was constrained from the temperature difference between the morning and evening temperatures, 
but the accuracy of the estimated thermal inertia was not good because the azimuthal locations of the data points were far from Saturn's shadow.  
In the present paper,  we analyze azimuthally scanned CIRS data including those in Saturn's shadow and precisely estimate the thermal inertia values 
for Saturn's main rings (the A, B, and C rings and the Cassini division), extensively using our multilayer model.  
The azimuthally scanned CIRS data have advantages in their spatial and spectral resolutions, 
azimuthal coverages both inside and outside Saturn's shadow, and solar phase coverages as compared with previous ground-based and 
spacecraft (Pioneer and Voyager) observations. Some of azimuthally scanned CIRS data obtained in the early Cassini mission 
have been already published by Leyrat et al. (2008b).  In addition to these data, we use some more data obtained later. 

In Sec.~2,  we discuss how the thermal relaxation time is related to 
the thermal inertia and the particle size.
In Sec.~3,  we explain data selections and fitting procedures and show fitted parameter values.
In Sec.~4, we discuss implication from estimated thermal inertia values and future works.
A summary is given in Sec.~5.

\section{Thermal relaxation time and thermal inertia}
The physics of eclipse cooling is summarized as follows.
When the illumination on a ring particle changes  with a typical frequency $\omega$, 
its temperature is modified in a surface layer of characteristic thickness (i.e., thermal skin depth) given as 
\begin{equation}
\ell_{\rm s} = \sqrt{\frac{K}{\rho C \omega}},
\end{equation}
where $K$, $\rho$, and $C$ are the thermal conductivity, the physical density, and the specific heat of the particle, respectively.
For the case of eclipse cooling of Saturn's rings, the time scale of the cycle, $2\pi/\omega$, may be the
twice of the eclipse time $2t_{\rm eclip}$, which is typically $\sim$ 3 hr. 
The timescale to release the heat in the skin depth when the particle is in Saturn's shadow (i,e., thermal relaxation time for eclipse) is given by 
\begin{equation}
t_{\rm rel}  = \frac{4\pi r^2 \ell_{\rm s} \rho C (T_{\rm p}-T_{\rm p,0}) }{4 \pi r^2 \sigma_{\rm SB} \epsilon (T_{\rm p}^4-T_{\rm p,0}^4)}
= \frac{\Gamma  (T_{\rm p}-T_{\rm p,0})}{\epsilon \sigma_{\rm SB} (T_{\rm p}^4-T_{\rm p,0}^4) \sqrt{\omega}} \hspace{0.3em} (r \gg \ell_{\rm s}),  \label{eq:trel}
\end{equation}
where $r$ is the radius of the particle, $T_{\rm p}$ is the particle temperature, $T_{\rm p,0}$ is the particle temperature calculated only
using the Saturn's thermal flux,  
$\sigma_{\rm SB}$ is the Stefan-Boltzmann constant, $\epsilon$ is the thermal emissivity,
and $\Gamma = \sqrt{\rho C K}$ is the thermal inertia.
The temperature $T_{\rm p, 0}$ is expected to be close to the temperature at
equinox (when the solar elevation is zero), which occurred in August, 2009 (Spilker et al., 2009; Flandes et al., 2010).
If $t_{\rm rel} \ll t_{\rm eclip} $, $T_{\rm p}$ 
quickly drops to $T_{\rm p,0}$ after particles enter Saturn's shadow.
On the other hand, if $t_{\rm rel} \gg t_{\rm eclip}$, the temperature drop is negligible.
For the typical thermal inertia values of Saturn's ring particles reported in previous works 
($\sim 10$ Jm$^{-2}$K$^{-1}$s$^{-1/2}$; Ferrari et al., 2005), $t_{\rm rel} \sim t_{\rm eclip} $, so the degree of 
the temperature drop sensitively depends on $\Gamma$.

Eq.~(\ref{eq:trel}) assumes that the particle size is sufficiently larger than the skin depth ($r \gg \ell_{\rm s}$). 
If $r \ll \ell_{\rm s}$,  the thermal relaxation time is given by the larger one of the thermal diffusion time
\begin{equation}
t_{\rm diff} = \frac{r^2}{K/ \rho C}
\end{equation}
or the time to release the internal energy of whole the hemisphere by radiation 
\begin{equation}
t_{\rm rad}  = \frac{\frac{4}{3} \pi r^3 \rho C  (T_{\rm p}-T_{\rm p,0})}{4 \pi r^2 \sigma_{\rm SB} \epsilon (T_{\rm p}^4-T_{\rm p,0}^4)}
= \frac{1}{3}\frac{\rho C r (T_{\rm p}-T_{\rm p,0})}{\epsilon \sigma_{\rm SB} (T_{\rm p}^4 -T_{\rm p,0}^4)}  \hspace{0.3em}.  \label{eq:trel2}
\end{equation}
 The ratio of the two time scale is roughly given as $t_{\rm diff}/t_{\rm rad} \sim \sigma_{\rm SB}T_{\rm p}^3r/K$. 
 Substituting typical values for Saturn's rings and considering a particle comparable to the skin depth 
 ($K = 10^{-3}$ W m$^{-2}$ K, $T_{\rm p}$ = 90 K, $r  = \ell_{\rm s} \sim$ 1 mm), 
one can find that  $t_{\rm diff} \ll t_{\rm rad}$.  This means that 
 the interior of a small particle quickly reaches to isothermal due to heat conduction before the particle cools down by radiation.  
Thus, the temperature drop of small particles in the shadow depends on the product  $\rho C r$ instead of $\Gamma$ in Eq.~(\ref{eq:trel}).
When  $r \sim \ell_{\rm s}$,  $t_{\rm rel}$ depends on both $\rho C r$ and $\Gamma$.
In the present paper, we first consider particles sizes sufficiently larger than the skin depth and estimate the thermal inertia values.
Then, we make additional fits where size of small particles is varied. Comparison between Eqs.~(\ref{eq:trel}) and (\ref{eq:trel2})
suggests that decreasing $r$ has an effect similar to decreasing $\Gamma$ (i.e., particles tend to react to the input illumination instantaneously in both cases). 
We will discuss this relation in Sec.~3.3.3 after we estimate $\Gamma$ and $r$.

Our thermal model assumes a bimodal spin distribution consisting of slow and fast rotators, while an actual spin distribution 
of a ring may be continuous. 
A discussion for the criterion for fast and slow rotators is similar to the above (Farinella et al., 1998).
The time scale of the illumination change for a particle spin is the spin period $t_{\rm spin}$ and 
the corresponding frequency is given by $\omega$ = $2\pi/t_{\rm spin}$.
The particle is regarded as a fast rotator if $t_{\rm spin} \ll t_{\rm rel}$, and as a slow rotator if $t_{\rm spin} \gg t_{\rm rel}$.
Dynamical simulations for Saturn's rings (Ohtsuki, 2005; Morishima and Salo, 2006) 
showed that, for Saturn's rings, the spin period of the largest particles is comparable to the orbital period $t_{\rm orb}$ 
and is usually longer than the relaxation time 
($t_{\rm spin} \sim t_{\rm orb }> t_{\rm rel}$), while the spin period of small particles is much shorter,  
as the spin period is roughly proportional to their sizes due to mutual collisions. 
The fraction of fast rotators $f_{\rm fast}$ (the normal optical depth of fast rotators relative to the total normal optical depth, $\tau$), 
which is one of the parameters in our model,
may be approximately given by the total cross section of particles with $t_{\rm spin} <  t_{\rm rel}$ relative to the total cross section of all particles.
Since $t_{\rm rel}$ depends on $\Gamma$,   $f_{\rm fast}$ depends on $\Gamma$, if 
a spin (and size) distribution is given.
In this paper, we first estimate $f_{\rm fast}$ and $\Gamma$ as independent parameters, as we did in Papers I and II. 
Then, we compare estimated values of  $f_{\rm fast}$ and $\Gamma$ with the relation predicted from the above theory
using the size distribution estimated in previous observations (French and Nicholson, 2000) (see Sec.~4.3).

\section{Parameter fits}

\subsection{Data selection}
The azimuthally scanned CIRS data for the main rings are summarized in Tables~1-4 (observational geometries) 
and Figs.~1-4 (observed temperatures). 
\marginpar{\textbf{[Figs.~1-4]}}\marginpar{\textbf{[Tables~1-4]}}
Only the scans selected for parameter fits (see below) are shown in these tables and figures, 
although there are more azimuthal scans.
All azimuthally scanned data up to Feb.18 (DOY 49), 2007, including those in our Tables~1, 2, and 4, are shown in Leyrat et al. (2008b).
We use some additional data obtained later:  the C10 scan for the C ring (Table~1) and all the data for the Cassini division (Table~3).
The ring spectra from the CIRS far-infrared channel, named Focal Plane 1 (FP1) are used. 
As done in all previous works using FP1 spectra (e.g., Spilker et al., 2005, 2006), 
the ring temperature is obtained by fitting a black body spectrum multiplied by a scale factor to the observed 
spectrum between 100-400 cm$^{-1}$ (100-25 $\mu$m). 
The scale factor includes the geometric filling factor, the thermal emissivity, and a scalar factor that arises when observing a system comprised of 
particles at more than one temperatures (Altobelli et al., 2007, Paper II).
The spectral resolution for the data ranges from 0.5 to 15.5 cm$^{-1}$ (Flasar et al., 2004).

We select the data for parameter fits taking into account the following points.
\begin{enumerate}
\item Data at high solar elevation are favorable (if they exist) since a large temperature drop in Saturn's shadow allows us to 
estimate the thermal inertia accurately. 
For the A ring and the Cassini division, 
the eclipse time is too short if solar elevation is large and  
the data at intermediately high ($\sim 15^{\circ}$) or lower solar elevations are used.
\item In order to resolve the degeneracy between the albedo and the fraction of fast rotators, it is necessary to make fits to data at low and 
high solar phases at the same radial location simultaneously.  It is desirable that spatial resolutions and solar elevations of these scans are similar.
\item The data for the lit face only are used for the A and B rings. The temperature of the unlit face of an optically thick ring 
is determined by how efficiently the heat is transported from the lit face to the unlit face and is controlled 
not only by the thermal inertia but also by the fraction of particles which are hindered from their vertical excursions through the midplane 
by mutual collisions (Paper II). Therefore, there will be a strong degeneracy between  the thermal inertia and the fraction of bouncing particles if we include
the unlit face data. We have a more detailed discussion in Sec.~4.2.   
\item For the A ring scans, there are azimuthal modulations not only in the absolute thermal brightness but also in the physical temperature 
due to wakes (Leyrat et al., 2008b; Ferrari et al., 2009). 
The exact reason why wakes cause temperature modulations is not well known, 
but this effect is most pronounced when the elevation angle of an observer, $|B|$, is $\sim 10^{\circ}$ as well as the brightness modulation.   
Therefore, we need to use data with $|B| \ll 10^{\circ}$ or $|B| \gg 10^{\circ}$. 
For the B ring, the temperature modulation is much smaller even at 
$|B| \sim 10^{\circ}$, so we do not remove data by $B$ values.   
\end{enumerate}

We find that there are not many data sets available with the above restrictions. 
The data sets in four radial locations will be used (one location for each ring): 83,000 km, 105,000 km, 120,000 km, and 129,000 km for the 
C ring, the B ring, the Cassini division, and the A ring, respectively.    
Since the scans CD1 and CD2 were obtained in the same day and these data have little overlaps in local hour angles, 
we merge these two scans to a single scan in parameter fits.  
The azimuthal coverages in the scans CD4, CD5, and CD6 were originally slightly larger than $2\pi$.  We removed some of footprints 
in these scans in order to avoid azimuthal overlaps in a single scan (see Fig.~3). This is just for convenience for our fitting code, and 
we believe that the impact of removing a small fraction of data on estimated values of parameters is very small.

The optical depth, $\tau$, of each footprint is obtained by averaging a high resolution $\tau$ profile over the CIRS FP1 footprint 
(see Appendix~A of Paper II).\footnote{In Paper II, the observational geometries used in $\tau$ smoothing are those for occultations by PPS and UVIS. Therefore, the smoothed $\tau$ is 
uniquely determined with a given saturnocentric radius, $r_{\rm p}$. 
On the other hand, in this work,  the geometries of CIRS observations are used. Thus, the smoothed $\tau$ is not always the same
even if $r_{\rm p}$'s of the footprint centers are the same. The use of CIRS geometries improves fits particularly for the Cassini division.}
The high resolution $\tau$ profile from Voyager PPS (Esposito et al., 1983) is used for all rings and 
another profile from Cassini UVIS (Colwell et al., 2006, 2007) is also used for the A and B rings for comparison.
The mean value of $\tau$ averaged over all footprints
for each scan is shown in Tables~1 to 4 as $\tau_{\rm PPS}$ and $\tau_{\rm UVIS}$. 
These optical depths are used in simulations.
The larger value of $\tau_{\rm UVIS}$ than $\tau_{\rm PPS}$ is due to wakes. 
Although our model does not take into account spatial non-uniformity due to wakes,  we check how estimated parameters vary with $\tau$ for the A and B rings.

\subsection{Fitting procedures}
We use our multilayer model developed in Paper I and improved in Paper II.
Our model solves the equation of classical radiative transfer both in thermal and visible light, 
taking into account all the heat sources for ring particles including the thermal and reflected visible fluxes from Saturn and the inter-particle heating and scattering.
A plane-parallel approximation is adopted, so spatial non-uniformity due to wakes is ignored.   
Our model adopts a bimodal size distribution consisting of small fast rotators which isotropically emit thermal emissions
and large slow rotators represented by non-spinning smooth Lambertian particles. 
Fast rotators are assumed to have a larger vertical extension than that for slow rotators. 
Note that the zero-volume filling factor of a ring assumed in classical radiative transfer, a plane-parallel approximation,
and  smooth spherical particles are probably inappropriate for Saturn's rings, 
and fitted parameters may be potentially modified in more advanced models (see also Paper II). 
The thermal emission from a ring is calculated by integrating thermal emissions from a large number of individual fast and slow rotators. 
About 3000 particles are used in a single simulation, which is roughly three times larger than those used in Paper II. 
The most important parameters are the bolometric Bond albedo, $A_{\rm V}$, the fraction of fast rotators, $f_{\rm fast}$, and 
the thermal inertia, $\Gamma$. These three parameters are simultaneously estimated in model fits.
Since $A_{\rm V}$ and $f_{\rm fast}$ were well estimated in Paper II and ring temperatures vary smoothly and 
monotonically with these two parameters,  coarse grids (the grid sizes are typically 0.1 for $A_{\rm V}$ and 0.2 for $f_{\rm fast}$) 
with a few points around the values estimated in Paper II are used for them.  
On the other hand, fine grids are used for $\Gamma$.
Ring temperatures in coarse parameter grids are interpolated to those for finer grids for the chi-square fits shown below.

There are some other important input parameters: the size of fast rotators $r_{\rm fast}$,  
the optical depth $\tau$, the ratio of the vertical scale height of fast rotators to that of slow rotators $h_{\rm r}$,  and the type of the vertical motion
(whether particles rebound at the midplane or not). As did in Paper II, we vary these parameters and check how sensitively estimated values of 
$A_{\rm V}$, $f_{\rm fast}$, and $\Gamma$ are affected. 
Although there are some uncertainties in these other parameters, the base-line cases, for which the estimated values of parameters are 
marked boldly in Table~5,  give the most reasonable results in Paper II and this work.\footnote{
In the base-line cases, we assume that $h_{\rm r} = 3$, except  $h_{\rm r} = 1$ for the B ring,
based on dynamical studies (Salo et al., 2001; Morishima and Salo, 2006), and $r_{\rm fast}$ = 10 cm.
This size of fast rotators is sufficiently larger than the skin depth, $\ell_{\rm s}$.
The size of slow rotators is fixed to be 5 m in all cases.}
\marginpar{\textbf{[Table~5]}}
The error size of simulated temperatures is estimated to be less than 0.2 K  (while 0.5 K for Paper II).
The physical density of 450 kg m$^{-3}$ and the specific heat of 760 J kg$^{-1}$ K$^{-1}$ are used as in Papers I and II.
The total time for simulations, by which we estimated parameters shown in Table~5, was about two cpu years.

For simulations of a single scan,  we use single values of $r_{\rm p}$ and $\tau$, both averaged over all footprints.
This is valid as the dispersions in $r_{\rm p}$ and $\tau$ are usually small enough. 
Exceptions are scans for the Cassini division, where small changes in radii of footprint centers or spacecraft ranges result in
large changes in smoothed $\tau$'s because the $\tau$ profiles, obtained in high radial resolution occultations, show large radial variations
in this region. 
Therefore, we take into account the $\tau$ variation in different footprints for scans of the Cassini division 
by conducting two simulations with low and high $\tau$'s and linearly interpolating calculated spectra to the one for a footprint. 
One clear effect of the $\tau$ variation is seen in the CD6 scan (Fig.~3), 
in which the temperature drops near the noon.  This is caused by the mutual shadowing effect,
as $\tau$ near the noon is largest. Our model succeeds in reproducing this temperature drop (fitting curves are explained in Sec.~3.3).

We make about 100 azimuthal bins and average the elevation $B$ and longitude $\phi_{\rm Cas}$ of the Cassini spacecraft
over the bins. These averaged values are used in simulations.
Fine-size bins are used near the shadow lines, as azimuthal temperature variations are large around there.
We first calculate thermal spectra in the range of 100-400 cm$^{-1}$, which is the same rage used in deriving observed temperatures,  
with the spectral resolution of 6 cm$^{-1}$.
Then, these spectra are smoothed over the CIRS FP1 footprints (elongated ellipses on a ring plane) 
with a gaussian weight.\footnote{The full-width half-maximum (FWHM) of CIRS FP1 is 2.51 mrad with a Gaussian fit (Flasar et al. 2004) 
and the diameter we assume is 5.2 mrad, which covers more than 99$\%$ of the signal.  
The azimuthal extension of the diameter projected on a ring is typically 2-3 degs but exceeds 10 degs for some data with low spatial resolutions.
The smoothing procedure is particularly necessary for numerical data whose corresponding observational data have such large azimuthal extensions.} 
This smoothing needs to be done not only in the azimuthal direction but also in the radial direction. 
Since we use a single averaged $r_{\rm p}$ value for each scan, 
we assume that ring spectra are uniform in the radial direction and  vary only in the azimuthal direction in a footprint.
This allows us to avoid conducting different simulations at slightly different radii and save computational time. 
Applying a black body fit to the smoothed spectrum,
we obtain the ring effective (or physical) temperature. 
In Papers I and II, we did not make smoothing of spectra obtained by simulations because we analyzed thermal data which were 
azimuthally far from Saturn's shadow and azimuthal temperature variations in footprints were small there. 

Using the effective temperatures from observations and simulations,
the reduced chi-square $\chi_0^2$ in the three dimensional 
parameter space ($A_{\rm V}$, $f_{\rm fast}$, $\Gamma$) is evaluated. 
We first calculate the reduced chi-squares for low phase $\chi_{\rm low}^2$ and high phase $\chi_{\rm high}^2$ separately, 
and take the average of them:
\begin{equation}
\chi_{\rm low}^2 (A_{\rm V}, f_{\rm fast},  \Gamma) = \frac{1}{N_{\rm low}-M}\sum_{n=1}^{N_{\rm low}} 
\left(\frac{T_{{\rm obs},n} - T_{{\rm sim},n}(A_{\rm V}, f_{\rm fast},  \Gamma)}{\sigma_{{\rm obs},n}}\right)^2 \hspace{0.3em} ({\rm for} \hspace{0.3em}  \alpha \le \alpha_{\rm crit}), \label{eq:cl}
\end{equation}
\begin{equation}
\chi_{\rm high}^2 (A_{\rm V}, f_{\rm fast},  \Gamma) = \frac{1}{N_{\rm high}-M}\sum_{n=1}^{N_{\rm high}} 
\left(\frac{T_{{\rm obs},n} - T_{{\rm sim},n}(A_{\rm V}, f_{\rm fast},  \Gamma)}{\sigma_{{\rm obs},n}}\right)^2 \hspace{0.3em} ({\rm for} \hspace{0.3em}  
\alpha > \alpha_{\rm crit}), , \label{eq:ch}
\end{equation} 
\begin{equation}
\chi^2_0 (A_{\rm V}, f_{\rm fast},  \Gamma) = \frac{1}{2} \left(\chi_{\rm low}^2 (A_{\rm V}, f_{\rm fast},  \Gamma)  + \chi_{\rm high}^2 (A_{\rm V}, f_{\rm fast},  \Gamma)\right),
\end{equation}
where $N_{\rm low}$ and $N_{\rm high}$ are the total number of footprints from multiple azimuthal scans at low and high phases, $M (= 3)$ is the number of parameters,
$n$ is the index for a footprint, $T_{\rm obs,n}$ and $T_{\rm sim,n}$ are the observed and simulated temperatures for the $n$-th footprint, 
$\sigma_{\rm obs,n}$ is the standard error size of the observed temperature (see below),
$\alpha$ is the solar phase angle, and $\alpha_{\rm crit}$ is the angle which discriminates low and high phases.
We set the error size $\sigma_{\rm obs,n}$ to be larger one of the one obtained from black body fits or the standard deviation of 
temperatures in azimuthal bins; in most cases, the latter is several times larger than the former (0.2-2.0K).
We choose $\alpha_{\rm crit} = 83.71^{\circ}$ because a slow rotator illuminated by a single heat source 
has stronger thermal emission than a fast rotator when $\alpha \le \alpha_{\rm crit}$ and vise versa. 
Values of fitted parameters do not sensitively depend on choice of $\alpha_{\rm crit}$.
The combination of best-fit values of $(A_{\rm V}, f_{\rm fast}, \Gamma)$ is derived to minimize $\chi_0^2$. 
The error bars are estimated after Press et al. (1986) as follows.
The chi-square subtracted by its minimum value at the best-fit parameters is defined as $\Delta{\chi}_0^2$. 
The lower and upper limits of a parameter is given by the maximum and minimum values of the parameter 
on the surface of $\Delta{\chi}_0^2=1$  (68$\%$ confidence level of a parameter without regard to other parameters). 
When the upper or lower limits of  parameters  are out of coverages in simulations, we extrapolate $\chi_0^2$ to some degree.  
The B ring data are split into two sets, one at high solar elevations (the scans B1-B4) and another at low elevations (the scans B5- B8),
and $\chi_0^2$'s are calculated separately and the average $\chi_0^2$ is taken. 
The B ring parameters are also estimated for high and low solar elevation cases separately to see whether our model can reproduce the dependence on
solar elevation. 

We first assume that all particles have the same value of $\Gamma$. Later, we consider a case with different values of $\Gamma$ for slow and fast rotators 
($\Gamma_{\rm slow}$ and $\Gamma_{\rm fast}$) and parameter fits will be done similar to the above but in the four dimensional parameter space ($M = 4$).
The reduced chi-square for this case is defined as $\chi_1^2(A_{\rm V}, f_{\rm fast},  \Gamma_{\rm slow},\Gamma_{\rm fast}) $.

In order to see the goodness of the estimated parameters,  the reduced chi-square for an individual scan is also calculated as 
\begin{equation}
\chi^2 (A_{\rm V}, f_{\rm fast},  \Gamma) = \frac{1}{N_{\rm foot}-M}\sum_{n=1}^{N_{\rm foot}} 
\left(\frac{T_{{\rm obs},n} - T_{{\rm sim},n}(A_{\rm V}, f_{\rm fast},  \Gamma)}{\sigma_{{\rm obs},n}}\right)^2 
\hspace{0.3em} ({\rm for}  \hspace{0.2em} {\rm each} \hspace{0.3em} {\rm scan}), \label{eq:ind}
\end{equation}
where $N_{\rm foot}$ is the number of footprints in each scan shown in Tables~1-4.

\subsection{Results}
\subsubsection{A case with a same thermal inertia for all particles}
We first assume that all particles have the same $\Gamma$.
The coverage of $\Gamma$ in simulations is 2-79 Jm$^{-2}$K$^{-1}$s$^{-1/2}$ with twelve grid points 
($\Gamma_i = i(i+1)/2+1$ Jm$^{-2}$K$^{-1}$s$^{-1/2}$ for $i$ = 1 to 12).
The estimated  values of  $A_{\rm V}$, $f_{\rm fast}$, and $\Gamma$ for all cases are listed in Table~5 as 
$A_{\rm V,0}, f_{\rm fast,0}$,  and $\Gamma_0$, respectively, and those parameters for the base-line cases are shown in Fig.~5.
\marginpar{\textbf{[Fig.~5]}}
The best fit temperature curves for the base-line cases are given by black solid curves in Figs.~1-4. 
The values of $\Gamma_0$ in units of Jm$^{-2}$K$^{-1}$s$^{-1/2}$
for the base-line cases are: 
20.0 for the C ring, 13.0 for the B ring , 11.0 for the Cassini division, and 16.2 for the A ring.
The values of $A_{\rm V,0}$ and $f_{\rm fast,0}$ are consistent with those estimated in Paper II (see Sec.~4.2 for a more detailed discussion).
For the case of the B ring, $A_{\rm V,0}$ estimated with the high $|B'|$ data is lower than that with the low $|B'|$ data. This means that 
our model underestimates the $|B'|$ dependence of the ring temperature as compared with the observations. 
Nevertheless,  the differences of the estimated parameter values for the high and low $|B'|$ cases remain reasonably small 
and the parameters estimated using all data simply take intermediate values between those for the high and low $|B'|$ cases.
In the followings, we will mainly show the results for the high and low $|B'|$ cases separately, as the reduced chi-square values are much smaller 
than those for the case using all data. 

The contour of the reduced chi-square, $\Delta{\chi}_0^2$, sliced on the $A_{\rm V}$ vs. $\Gamma$ plane with 
$f_{\rm fast}=f_{\rm fast,0}$ is shown in Fig.~6. \marginpar{\textbf{[Fig.~6]}}
If the dispersion of observed temperatures at a given range of azimuthal angle is large, $\Delta{\chi}_0^2$ becomes flat around the best fit
parameters and it makes error sizes larger. The Cassini division is a such case; the large temperature dispersion is probably due to 
complicated radial structures whose scales are smaller than the CIRS footprint size.
The dispersion of observed temperatures is usually small for optically thick rings
due to good signal to noise ratio, so that the B ring parameters have small error bars.   However, 
the actual uncertainties of parameters for the B ring are probably 
larger than at least the difference between the values for the high and low $|B'|$ cases which 
are larger than the error bars (particularly for $A_{\rm V}$).
 
\subsubsubsection{3.3.1.1. Dependence on phase angle} 
Comparisons of the observed temperatures with the modeled temperatures (black curves) in Figs.~1-4 
show that the ring temperatures outside the shadow at both low and high phases 
are well reproduced by the model with appropriate combinations of $A_{\rm V,0}$ and $f_{\rm fast,0}$ whereas
there are systematic deviations between the modeled and observed temperatures in the shadow depending on $\alpha$. 
At low $\alpha$, the shadow temperature is overestimated in the model, 
or $\Gamma_0$ is overestimated (see e.g., the C1 scan). On the other hand, the shadow temperature is underestimated in the model at high phase (e.g.,  the C3 scan). 
These facts seem to indicate that $\Gamma$ increases with $\alpha$. 
In order to confirm this trend, we make test fits using fixed $A_{\rm V}$ and $f_{\rm fast}$ for all scans
and varying only $\Gamma$ for different scans.
Note that  the temperature difference between inside and outside the shadow is controlled 
almost solely by $\Gamma$ whereas a change in $A_{\rm V}$ or $f_{\rm fast}$ causes nearly a constant 
temperature change at all azimuthal locations. 

The estimated values of $\Gamma$ of individual scans for the base-line cases are shown in Fig.~7 and listed in Table~6 as $\Gamma_{\rm ind}$.  
\marginpar{\textbf{[Fig.~7]}}\marginpar{\textbf{[Table~6]}}
The corresponding temperature curves are given by  blue dashed curves in Figs.~1-4.
As we expected, $\Gamma_{\rm ind}$ increases with $\alpha$ for all rings.
These trends are verified as the reduced chi-square, $\chi^2$ (shown in Figs.~1-4), 
for each scan is reduced with $\alpha$-dependent $\Gamma$. 
In Fig.~7, we also plot  the $\Gamma$ values for the C and B rings estimated by Ferrari et al. (2005). 
The range of $\alpha$ in their ground based observations is 4.7 - 6.2$^{\circ}$, and their very low $\Gamma$ values
are consistent with our fitting lines for $\alpha$ vs. $\Gamma$. 
For the B ring, if we use  $A_{\rm V}$ and $f_{\rm fast}$ estimated using all data, 
the low $|B'|$ case shows lower $\Gamma_{\rm ind}$ than that for the high $|B'|$ case at similar $\alpha$ (shown only in Table~6 not in Fig.~7), 
because the temperature outside the shadow is underestimated for the high $|B|'$ case and 
overestimated for the low $|B'|$ case. On the other hand, 
the use of different combinations of  $A_{\rm V}$ and $f_{\rm fast}$ gives values of $\Gamma_{\rm ind}$ consistent each other for both cases. 

Although  $\Gamma_{\rm ind}$ depends on $\alpha$ as seen in Fig.~7, it is unlikely that individual particles have
$\alpha$-dependent $\Gamma$ and their low-$\Gamma$ faces align toward the Sun 
because particles are spinning and spin axis orientations change by collisions.
A plausible interpretation for the $\alpha$-dependence of $\Gamma_{\rm ind}$ is that large slow rotators
have lower $\Gamma$ values than those for small fast rotators. 
The thermal emission from slow rotators is relatively stronger than that from fast rotators at low phase and vice versa.
Therefore, particle groups which are dominant in the thermal emission may vary with phase angle and measured thermal inertias 
may vary as well. This possibility will be examined in Sec.~3.3.2.

\subsubsubsection{3.3.1.2. Dependence on optical depth and particle bouncing}
For the A and B rings, $\Gamma_0$ is almost independent of the scale height ratio, $h_{\rm r}$, 
but decreases with increasing optical depth, $\tau$, and takes larger values for the case without bouncing than with bouncing. 
The latter two dependences indicate that the thermal relaxation time for an optically thick ring 
depends not only on $\Gamma$ of individual particles, but also on $\tau$ and type of vertical motion,
whereas the discussion for the thermal relaxation time 
in Sec.~2 is for an isolated particle.
In order to help understand the latter two dependences, we examine how the ring physical temperature decreases in Saturn'
shadow at different vertical locations,  $-3< z/h <3$, where $z$ is the vertical position and is positive on the lit face and $h$ is the vertical scale height  (Fig.~8). 
\marginpar{\textbf{[Fig.~8]}}
Here, the ring physical temperature is given by the averaged temperature over different particles
at a certain $z$ (Eq.~(27) of Paper I), and all particles are assumed to be fast rotators ($f_{\rm fast} = 1$) 
so that the temperature is independent of the direction of emission.
For cases with bouncing of particles at the midplane (the top and middle panels of Fig.~8), 
the energy is transported through the midplane by radiation only, so 
the unlit face temperature ($z<0$) 
is very low and its azimuthal variation is also very small.
On the lit face, the ring temperature increases with increasing $z$ when illuminated by the Sun while 
it decreases with increasing $z$ in Saturn's shadow.  
These trends are explained by the inter-particle shadowing and heating effects, respectively.
The inter-particle heating cases a sort of the green house effect, by which 
an optically thicker ring can retain the heat relatively longer in the planetary shadow 
as compared with a thinner ring (compare the top and middle panels of Fig.~8).
Since the effective thermal relaxation time of a ring increases with increasing $\tau$ for a fixed $\Gamma$,
an estimated $\Gamma$ needs to decrease with increasing $\tau$ in order to reproduce the observed temperature drop in the shadow.  
When particles have sinusoidal vertical motion without bouncing,
the vertical temperature gradient is small both for the lit and unlit faces (the bottom panel of Fig.~8). 
In this case, the heat is efficiently removed from both faces, meaning a short thermal relaxation time, as compared with the case with bouncing.  
Thus, the estimated  $\Gamma$ needs to be relatively larger.

\subsubsection{A case with different thermal inertias for slow and fast rotators}
In the last section, the thermal inertia value for all particles was assumed to be the same.
Here, we make another fit assuming different values of the thermal inertias for slow and fast rotators, $\Gamma_{\rm slow}$ and $\Gamma_{\rm fast}$. 
This means that the number of the free parameters is now four instead of three. 
The grids of $\Gamma_{\rm slow}$ and  $\Gamma_{\rm fast}$ used in simulations are 2, 5, 9,  and 15 Jm$^{-2}$K$^{-1}$s$^{-1/2}$
and 10, 20, 35, 55, 80, and 120 Jm$^{-2}$K$^{-1}$s$^{-1/2}$, respectively. 
The estimated values of $A_{\rm V}$ $f_{\rm fast}$, $\Gamma_{\rm slow}$, and $\Gamma_{\rm fast}$ are shown in Fig.~5 and listed in Table~5.
The corresponding temperature curves are given by red solid curves in Figs.~1-4. 
The estimated values of  $A_{\rm V}$ and $f_{\rm fast}$ are almost the same with those for the case with a single thermal inertia, $\Gamma_0$. 
The estimated values of $\Gamma_{\rm slow}$ and $\Gamma_{\rm fast}$ for the base-line cases are in the range of  5-9 Jm$^{-2}$K$^{-1}$s$^{-1/2}$ and 
19-77 Jm$^{-2}$K$^{-1}$s$^{-1/2}$, respectively. 
Although the values of $\Gamma_{\rm fast}$ are large,  they are still much lower than that for solid ice with no porosity 
($\sim$ 2600 Jm$^{-2}$K$^{-1}$s$^{-1/2}$). 
Therefore, fast rotators are likely to have surface regolith layers, but
they may be not as fluffy as those for slow rotators.
The most likely reason for larger $\Gamma_{\rm fast}$ than $\Gamma_{\rm slow}$ is 
that the centrifugal force for fast rotators is so strong that fluffy regolith particles
cannot stay on fast rotators.

Unfortunately, the error bars for $\Gamma_{\rm slow}$ and $\Gamma_{\rm fast}$ are larger than those for $\Gamma_0$,
because of the increased freedom in adjusting temperature drops in the planetary shadow:
although slow rotators are dominant emitters at low phase, the contribution from fast rotators is not unimportant and 
similarly the slow rotators' contribution is not unimportant at high phase.
Figure~9 shows contours of $\Delta \chi_1^2$ on the $\Gamma_{\rm slow}$ vs. $\Gamma_{\rm fast}$ plane 
with the best-fit values of $A_{\rm V}$ and $f_{\rm fast}$. \marginpar{\textbf{[Fig.~9]}}
One can find that $\Delta \chi_1^2$ is very flat along the $\Gamma_{\rm fast}$ axis and the error bars 
for $\Gamma_{\rm fast}$ are not well determined for some cases. 
This happens particularly if $f_{\rm fast}$ is small (B ring, low $|B'|$) or  the scatter in the observed temperature 
is large (the Cassini division).
However, it is obvious that $\Gamma_{\rm fast} > \Gamma_{\rm slow}$ for all rings.

Parameter fits are clearly improved by using different values of $\Gamma_{\rm slow}$ and $\Gamma_{\rm fast}$, as 
 the lowest $\chi_1^2$ (with the best-fit parameters) is lower than the lowest $\chi_0^2$ for all rings. 
 Indeed, the chi-square for an individual scan, $\chi^2$,   
is also reduced for almost all scans for all rings. 
Although the $\alpha$-dependent systematic deviations
between the modeled and observed temperatures in the shadow are much reduced, 
they are not completely removed, as $\chi^2$ is still larger than that for the $\alpha$-dependent $\Gamma$ case. 
This is probably because our model assumes a bimodal size distribution
whereas an actual ring has a continuous size distribution.
The distribution of the thermal inertia is likely to be continuous as well. 
Therefore, $\Gamma$ for the largest particles may be even smaller than $\Gamma_{\rm slow}$, which is the thermal inertia averaged over 
a certain size range of large particles. Similarly,  $\Gamma$ for the smallest particles may be larger than $\Gamma_{\rm fast}$. 
This effect needs to be investigated in future works.
 
\subsubsection{Lower limit of the size of fast rotators}
As discussed in Sec.~2, fitting results shown in Secs.~3.3.1 and 3.3.2 are not altered even if we use different particle sizes as long as 
the particle size is sufficiently larger than the thermal skin depth. If we use particle sizes as small as or even smaller than the skin depth,
reducing particle size causes an effect similar to reducing thermal inertia. 
Eqs.~(\ref{eq:trel}) and (\ref{eq:trel2}) suggest the following relation:
\begin{equation}
r \simeq \frac{3\Gamma}{\rho C \sqrt{\omega}}. \label{eq:rg} \label{eq:r2gm}
\end{equation} 
Since the use of a too small $r$ makes fits worse, the lower limit of the size of particles can be constrained. 
Here we make additional parameter fits varying size of fast rotators, $r_{\rm fast}$. 
The number of parameters to be fitted is now five
instead of four in Sec.~3.3.2.
We define the lower limit size of fast rotators,  $r_{\rm fast, min}$, as the lowest $r_{\rm fast}$ giving 
$\Delta{\chi}_1^2 = 1$ in the five dimensional parameter space.\footnote{Correction factors $(N_{\rm low}-4)/(N_{\rm low}-5)$ and $(N_{\rm high}-4)/(N_{\rm high}-5)$
are ignored, as  $N_{\rm low}$ and $N_{\rm high}$ are sufficiently large.}
This is similar to the definitions of the lower limits of other parameters estimated in previous sections. 
Since a comprehensive coverage of the five dimensional parameter 
space is computationally so intense that the following effective method to estimate $r_{\rm fast, min}$ is adopted.
We first fix all four parameters ($A_{\rm V}$, $f_{\rm fast}$, $\Gamma_{\rm slow}$, and $\Gamma_{\rm fast}$) estimated in Sec.~3.3.2 and 
search for a size of fast rotators, $r_2$, which gives  $\Delta{\chi}_1^2 \sim 1$. Then, we make parameter fits varying four other parameters 
at $r_{\rm fast} = r_2$ and 0.5$r_2$. When other parameters vary,  $\Delta{\chi}_1^2$ is usually smaller than $1$ for $r_{\rm fast} = r_2$. This means 
that $r_{\rm fast, min}$ should be at least smaller than $r_2$.
Linearly interpolating $\Delta{\chi}_1^2$ for $r_{\rm fast} = r_2$ and  $0.5r_2$,  we estimate  $r_{\rm fast, min}$, where $\Delta{\chi}_1^2 = 1$.

The values of $\chi_1^2$ for $r_{\rm fast} \sim r_2$ and other estimated parameters are listed in Table~5. 
We find that $r_{\rm fast, min} $ ranges from 1-10 mm. 
The temperature curves for  $r_{\rm fast} \sim r_{\rm fast, min}$ are shown as purple dotted curves in Figs.~1-4.
Figure~10 compares $r_{\rm fast, min}$ with the lower limit of $\Gamma_{\rm fast}$, $\Gamma_{\rm fast, min}$, for the base-line cases 
($r_{\rm fast} =$ 10 cm)
(note if $\Gamma_{\rm fast} = 33.5^{+71.6}_{-15.5}$ Jm$^{-2}$K$^{-1}$s$^{-1/2}$, 
$\Gamma_{\rm fast, min}$ is 18.0 Jm$^{-2}$K$^{-1}$s$^{-1/2}$). 
\marginpar{\textbf{[Fig.~10]}}
For the B ring at low $|B'|$ and the Cassini division, we were not able to estimate $\Gamma_{\rm fast, min}$. 
For other three rings, for which the values of $\Gamma_{\rm fast, min} $ are well estimated,  
the relations between $r_{\rm fast, min}$ and $\Gamma_{\rm fast, min}$ are well represented by Eq.~(\ref{eq:rg})
with $r = r_{\rm fast, min}$ and $\Gamma = \Gamma_{\rm fast, min}$.
There,  the cycle of the illumination change, $2\pi/\omega$, needs to be specified to determine the thermal skin depth. 
We find that a cycle which gives a reasonable fit is somewhere between the twice of the eclipse time 
 $2t_{\rm eclip}$ (see Tables~1-4 for $t_{\rm eclip}$) and the orbital period 
 $t_{\rm orb}$.\footnote{The orbital periods are 6.78 hr for the C ring (83,000 km), 9.65 hr for the B ring (105,000 km), 
 11.79 hr for the Cassini division (120,000 km), and 13.14 hr for the A ring  (129,000 km).} 
We think that this is reasonable, because temperatures of ring particles recover almost immediately after they exit Saturn's shadow, but not completely, 
and their temperatures gradually increase until they enter Saturn's shadow again.    

\section{Discussion}

\subsection{Interpretations}
The estimation of  $r_{\rm fast, min} $ in Sec.~3.3.3 means that 
particles with sizes equal to or less than $r_{\rm fast, min}$ should not occupy more than $f_{\rm fast}$ of the total cross section.
This condition seems to be satisfied for all rings, as 
the lower cutoff $r_{\rm min}$ for the continuous size distribution estimated in French and Nicholson (2000) (Table~7) is comparable to or larger than $r_{\rm fast, min}$.
\marginpar{\textbf{[Table.~7]}}
Therefore, the quantity constrained by the eclipse cooling is the thermal inertia, not the size of ring particles.

In Sec.~3.3.2, we found $\Gamma_{\rm fast} > \Gamma_{\rm slow}$ for all the rings.
The most likely reason for this is that small fast rotators cannot hold very fluffy regolith layers because of their fast spins.
Dynamical studies (Salo, 1987; Ohtsuki, 2005; Morishima and Salo, 2006) show that 
the spin period is about the orbital period for the largest particles and is roughly proportional to 
the particle size for an extended size distribution. The centrifugal force is comparable to the self-gravity 
on a particle with a synchronous rotation near the Roche limit, around which Saturn's main rings exist.
Therefore, the centrifugal force well exceeds the self-gravity on surfaces of small particles.
Since $\Gamma_{\rm fast}$ is likely to be still much smaller than the thermal inertia of the solid ice with no porosity,  
some regolith particles probably stick to small fast rotators due to the surface energy (Choksi et al., 1993, Albers et al., 2006). 
The regolith layers of fast rotators may also be compacted by mutual impacts whose velocities 
are larger than the velocity dispersion of large particles  (Salo, 1992; Morishima and Salo, 2006).
Alternatively, small particles might be regolith free.
In this case, as suggested by Kouchi et al. (1992), microcracks would be responsible for $\Gamma_{\rm fast}$.
In any case, it is expected that the thermal inertia increases with decreasing size of a ring particle.
A similar size dependence of the thermal inertia is also found in asteroids (Delbo and Tanga, 2009). 
However,  the mechanism causing the size dependence is probably different from that for Saturn's ring particles;
their interpretation is that large asteroids  ($> 100$ km) are primordial and collisional debris build up thick regolith layers on them
whereas small bodies are fragments in catastrophic impacts between large bodies. 

In the interpretation of the variations of $\Gamma_{\rm slow}$ and $\Gamma_{\rm fast}$ in different rings,
we need to consider (1) the distribution of $\Gamma (r)$ as a function of particle size $r$ and (2) 
the ratio of the smallest particle size $r_{\rm min}$ to the largest particle size $r_{\rm max}$ for actual rings with continuous size distribution.
As we discussed in Sec.~3.3.2,  the thermal inertia of the largest particles, $\Gamma (r = r_{\rm max})$, may be close to $\Gamma_{\rm slow}$ or slightly less.
Therefore,  $\Gamma (r = r_{\rm max})$ seems to be similar for all the rings, except  the Cassini division has a lower $\Gamma (r = r_{\rm max})$. 
Because $\Gamma$ is likely to increase with decreasing $r$, as discussed above,  $\Gamma_{\rm fast}$ is expected 
to increase with decreasing $r_{\rm min}/r_{\rm max}$.
On the other hand, $\Gamma_{\rm fast}$ for the A ring is larger than that for the C and B rings, even though 
$r_{\rm min}/r_{\rm max}$ (A ring) =  $r_{\rm min}/r_{\rm max}$ (B ring) $> r_{\rm min}/r_{\rm max}$ (C ring) (Table~7).
This means that with a same $r/r_{\rm max}$ ($< 1$), an A ring particle has the largest $\Gamma$.
Thus, the size dependence of $\Gamma$ for the A ring is steeper than those for the B and C rings, whereas
$\Gamma_{\rm fast}$ for the C ring larger than that for the B ring may be explained by the smaller $r_{\rm min}/r_{\rm max}$ for the C ring.
The large $\Gamma_{\rm fast}$ for the A ring is probably due to wakes (Salo, 1995; Colwell et al., 2006; Hedman et al., 2007).
Dynamical simulations (Morishima and Salo, 2006) including the self-gravity 
show that the spin frequency increases with increasing $r_{\rm p}$ for a fixed dynamical optical depth. 
The self-gravity also steepens size dependence of the spin frequency. The collision velocity between large particles
are usually slow as they are in the same wake, while small particles floating in inter-wake spaces suffer collisions with impact velocities 
as fast as the velocity dispersion of wakes, which is much larger than the escape velocity of individual particles. 
Therefore, for small particles with a certain $r/r_{\rm max}$ ($\ll 1$), 
both the spin frequency and the collision velocity are largest in the A ring.
These facts may explain large $\Gamma_{\rm fast}$ for the A ring, while
 $\Gamma_{\rm slow}$ for the A ring is similar to those for the C and B rings. 
 It should be noted again that  the spatial non-uniformity due to wakes is not directly taken into account in our model.
The discussion here was based on the estimated values of $\Gamma_{\rm fast}$ for the standard cases and 
readers are cautioned that the uncertainty in $\Gamma_{\rm fast}$ is very large.

The Cassini division has the smallest $\Gamma_{\rm slow}$ among  the Saturn's main rings. 
This smallest value seems to be real, as  $\Gamma_{\rm ind}$ in Fig.~7 at a given $\alpha$ always shows the smallest value,
except at very large $\alpha$ (i.e., the CD3 scan).
The very small value of  $\Gamma_{\rm slow}$ may indicate that there are some dust sources around this region and ring particles are fluffily coated by these grains. 
There are at least two dusty ringlets in the Cassini division (Hor\'{a}nyi et al., 2009):  one in the Huygens gap (117,490 km) and another in
the Laplace gap (119,940 km) called the Charming ringlet (Hedman et al., 2010). 
In particular, the location of the latter one is very close to the centers of the CIRS footprints of azimuthal scans. 
These dusty ringlets themselves are very tenuous and unimportant for the direct thermal emission. 
However, their existence indicates injection of  small grains into this region by some unknown mechanisms, 
 as they are likely to be very young (less than a few thousand years old; Hedman et al., 2010).
The small lower-cutoff size for the Cassini division found in French and Nicholson (2000)
(Table~7) also seems consistent with a relatively large population of dust. 
They also inferred that there is a substantial variation in the
particle size distribution across this region, with relatively more
small particles in the inner than in the outer Cassini division (reported also in Colwell et al. (2009)). 
This probably indicates that the dust sources are limited to the inner Cassini division.
The value of $\Gamma_{\rm ind}$ for the CD3 scan which looks remarkably 
too large as compared with those for other scans seems to be explained by the fact that  the saturnocentric distance of the 
CD3 scan is slightly larger than those for other scans (by $\sim$ 2000 km at shadow lines).
Howett et al. (2010) found that the thermal inertias of saturnian satellites from Rhea inward have relatively lower values 
($\sim$ 10 Jm$^{-2}$K$^{-1}$s$^{-1/2}$)  than those for Iapetus and Phoebe ($\sim$ 20 Jm$^{-2}$K$^{-1}$s$^{-1/2}$), 
which are located farther away from Saturn. They indicated that this is probably related to    
coating by E-ring material ejected from Enceladus' plumes.  Similar things may happen also in Saturn's rings although dust sources are different. 

The large value of  $\chi_0^2$ for the B ring in the case using all data comes from the fact that our model underestimates $|B'|$ dependence 
of the ring temperature, as we stated in Sec.~3.3.1. By separately estimating parameters for the B ring at the high and low $|B'|$ cases,
the values of $\chi_0^2$ are remarkably reduced. However, $\chi_0^2$ for the high $|B'|$ case is still much larger than unity.
The phase angles for the scans B2, B3, and B4 are low. In Fig.~2, with decreasing phase angle from $\sim 40^{\circ}$ (the B4 scan) 
to $\sim 10^{\circ}$ (the B2 scan), the observed temperature outside the shadow increases by $\sim$ 4K, whereas the modeled temperature increases by only $\sim$ 1K.
Therefore, our model underestimates not only $|B'|$ dependence but also $\alpha$ dependence at low phase.
The steep $\alpha$ dependence observed is probably due to the thermal opposition surge 
caused by packing of ring particles into thin layers (Altobelli et al., 2009).
The effect of particle packing is also known to enhance $|B'|$ dependence of photometric brightness for dense rings 
(Salo and Karjalainen, 2003), but it is uncertain if it is also the case for the thermal emission.
For more accurate estimations of particle properties for optically thick rings such as the B and A rings, 
this packing effect and the effect due to wakes need to be taken into account in future models. 
Nevertheless, we believe that parameters estimated in the present work are not so different from actual values,
because the deviations between observed and modeled temperatures remain reasonably small.   

\subsection{Toward estimations of radial variation of thermal inertia}
We have estimated the thermal inertia values for only four radial points in this study (one point for each ring), 
and it is desirable to examine the radial variation of the thermal inertia in each ring.
Unfortunately,  azimuthal scans were taken at only a limited number of radial points by Cassini CIRS and it is probably difficult to 
cover a sufficiently large number of radial points even in the extended mission.  However, there are some quasi-radial scans 
along the shadow lines and in Saturn's shadow.  
Even if we use reasonable sets of shadow scans,  precise estimation of $\Gamma$ 
is not simple in the optically thick rings (the A and B rings) because an estimated 
$\Gamma$ depends on types of vertical motion of ring particles and adopted $\tau$ (Table~5). 

The degeneracy between $\Gamma$ and the type of vertical motion can be resolved 
if we combine $\Gamma$ estimated from the temperature difference between the lit and unlit faces (Paper II)
and $\Gamma$ estimated from the eclipse cooling (this work).
Figure~11 shows comparisons of parameters estimated in Paper II (solid lines) and this work (diamonds) for the case with 
$\tau_{\rm PPS}$. 
\marginpar{\textbf{[Fig.~11]}}
Here we use $\Gamma_0$ from  this study because Paper II did not consider different 
values of $\Gamma$ for slow and fast rotators.
For the B and A rings, estimated parameters for the bouncing model are shown with black lines. 
Very good agreements between both studies are found if we assume the bouncing model for the B ring
and the non-bouncing model for the other rings. In fact, with the non-bouncing model, 
$\Gamma$ and $A_{\rm V}$ for the A ring from this study are slightly higher and lower than those from Paper II.
These deviations are fixed if we assume a small fraction of particles rebound at the midplane, because
both $\Gamma$ from Paper II and $A_{\rm V}$ from this study increase with increasing fraction of bouncing particles. 

There is still a problem in fitted parameters due to their dependences on $\tau$.
The uncertainty in $\tau$ for the A and B rings comes from wakes (Colwell et al., 2006, 2007), 
so ultimately more sophisticated models, which take into account wakes, are necessary.
However, the uncertainty in estimated values of $\Gamma$ due to the uncertainty of $\tau$ seems very small, as we will see below.
Figure~12 shows $\Gamma$ for the mid B ring (105,000 km) estimated from this study (the mean values of $\Gamma_0$ for high $B'$ and low $B'$ cases) 
and from Paper II ($\Gamma_{\rm PII}$) as a function of the fraction of bouncing particles $f_{\rm b}$
for two different cases of optical depths, $\tau_{\rm PPS}$ and $\tau_{\rm UVIS}$.
\marginpar{\textbf{[Fig.~12]}}
Here, $f_{\rm b}$ = 0 and 1 represent the non-bouncing and bouncing models, respectively, 
and a linear dependence of  $\Gamma$ on $f_{\rm b}$  is assumed 
between $f_{\rm b}$ = 0 and 1. In this work, we did not estimate $\Gamma_0$ for the case with $f_{\rm b}$ = 0 and $\tau_{\rm UVIS}$, 
and we assume that the ratio $\Gamma_0 (\tau_{\rm PPS}) / \Gamma_0 (\tau_{\rm UVIS})$ is independent of $f_{\rm b}$ 
as well as the ratio of the error sizes.
The values of $\Gamma$ and $f_{\rm b}$  
on the crossing points of the solid and dashed lines are the predicted values.
For the case with $\tau_{\rm PPS}$,  the crossing point has $\Gamma \simeq$ 10-13 Jm$^{-2}$K$^{-1}$s$^{-1/2}$ and  $f_{\rm b} \simeq1$. 
On the other hand, for the case with $\tau_{\rm UVIS}$, the crossing point has $\Gamma =$ 14 Jm$^{-2}$K$^{-1}$s$^{-1/2}$ and
$f_{\rm b} = 0.27$. Since we used the possible minimum value, 40 Jm$^{-2}$K$^{-1}$s$^{-1/2}$, for  
$\Gamma_{\rm PII}(\tau_{\rm UVIS})$ at $f_{\rm b} = 1$, $\Gamma$ and $f_{\rm b}$ at the cross point 
will be respectively higher and lower than those shown in Fig.~12. 
Nevertheless,  $\Gamma$ at the crossing point should not be larger than  $\Gamma_0 (\tau_{\rm UVIS})$ at $f_{\rm b} = 0$ ($\sim$15 Jm$^{-2}$K$^{-1}$s$^{-1/2}$).
Therefore, even though we cannot resolve the degeneracy between 
$\tau$ and $f_{\rm b}$, similar values of $\Gamma$ seem to be obtained regardless of the choice of $\tau$.
This discussion needs to be extended to the case with different values of $\Gamma_{\rm slow}$ and $\Gamma_{\rm fast}$
in future works.

\subsection{Criteria of slow and fast rotators}

We have treated the thermal inertia $\Gamma$ and the fraction of fast rotators $f_{\rm fast}$
as independent parameters in data fitting. However, they are not independent of each other if the 
spin (and size) distribution is given. 
Here, we discuss the relation between  $\Gamma$ and $f_{\rm fast}$. 
A similar discussion is found in Sec.~5.2 of Paper I.

We adopt the size distribution estimated in  French and Nicholson (2000) (Table~7).
The number density per unit size $n(r)$ is given by a power-law as 
$n(r) \propto r^{-q} (r_{\rm min} \le r \le r_{\rm max})$,
where $r_{\rm min}$ and $r_{\rm max}$ are the minimum and maximum size of particles, respectively.
They estimated the size distribution using diffracted stellar light, and their results are consistent with those 
from radio occultation (Zebker et al., 1985; Cuzzi et al., 2009).
From dynamical studies (Ohtsuki, 2005, 2006; Morishima and Salo, 2006),  
the particle spin period is approximately given as 
\begin{equation}
t_{\rm spin}(r) =  C_1 \frac{r}{r_{\rm max}}t_{\rm orb}, \label{eq:tspin}
\end{equation}
where $C_1$ is the numerical constant and depends on $r_{\rm p}$ and $\tau$;
we take $C_1 = $ 2.0, 1.0, 2.0, and 0.5 for the C ring, the B ring, the Cassini division, and the A ring from 
Fig.~11 of Morishima and Salo (2006).  
The critical particle size, which discriminates slow and fast rotators, is defined as 
\begin{equation}
t_{\rm spin}(r_{\rm crit})= t_{\rm rel}, \label{eq:rcrit0}
\end{equation}
where $t_{\rm rel}$ is the thermal relaxation time given by Eq.~(\ref{eq:trel}) with $\omega$ = $2\pi/t_{\rm spin}$.
The critical size $r_{\rm crit}$ is given as
\begin{equation}
\frac{r_{\rm crit}}{r_{\rm max}} = 
\frac{1}{2 \pi C_1 t_{\rm orb}}
\left(\frac{\Gamma(T_{\rm p}-T_{\rm p,0})}{\epsilon \sigma_{\rm SB}(T_{\rm p}^4-T_{\rm p,0}^4)}\right)^{2}.
\label{eq:rcrit}
\end{equation}
Here, $\Gamma$ is the thermal inertia for particles with $r = r_{\rm crit}$, which should be somewhere between
$\Gamma_{\rm slow}$ and $\Gamma_{\rm fast}$. In a comparison below, 
we assume that $\Gamma =  \Gamma_0$ as a nominal value.
The temperature,  $T_{\rm p,0}$ is supposed to be derived from the flux to the night-side hemisphere of a particle,
and we take its value from the equinox data assuming the night-side hemisphere is illuminated only by Saturn: 65 K for the C ring and 48 K for other rings 
(Spilker et al., 2009). If we simply regard a particle with $r < r_{\rm crit}$
as a fast rotator and a particle with $r \ge r_{\rm crit}$ as a slow rotator,
the fraction of fast rotators in cross section $f_{\rm fast}$ is given as
\begin{equation}
f_{\rm fast} = \frac{(r_{\rm crit}/r_{\rm max})^{3-q} - (r_{\rm min}/r_{\rm max})^{3-q}}
{1 - (r_{\rm min}/r_{\rm max})^{3-q}}. \label{eq:ff2}
\end{equation}
Eqs.(\ref{eq:rcrit}) and (\ref{eq:ff2}) indicate that $f_{\rm fast}$ 
increases with increasing $\Gamma$ and deceasing 
$T_{\rm p}$ and $t_{\rm orb}$, for given $r_{\rm min}/r_{\rm max}$ and $q$. 

Figure~13 shows the relation between $\Gamma$ and $f_{\rm fast}$ derived from 
Eqs.~(\ref{eq:tspin}) - (\ref{eq:ff2}) with $r_{\rm min}/r_{\rm max}$ and $q$ 
from French and Nicholson (2000). 
\marginpar{\textbf{[Fig.~13]}}
The estimated values, $\Gamma_0$ and $f_{\rm fast,0}$, from our parameter fits (Table~5) are also plotted.
We find good agreements between values from our parameter fits and the theoretical prediction for all rings.
In Paper II, we discussed that our model probably underestimates
$f_{\rm fast}$ due to the lack of the effects of packing of ring particles and surface roughness.
The good agreements seen in Fig.~13 may indicate that these effects do not largely change estimated values of parameters such as $f_{\rm fast}$ and $\Gamma$.
For the A and B rings, we may need to use a larger $T_{\rm p,0}$ due to the mutual heating. In this case, $f_{\rm fast}$ predicted from Eq.~(\ref{eq:ff2}) will be lower
than those shown in Fig.~13 and the agreements will be even better unless  $T_{\rm p,0}$ is very close to $T_{\rm p}$.

\section{Conclusions}
In the present paper, we have estimated 
the thermal inertia values of Saturn's main rings by applying our thermal model to azimuthally scanned spectra taken by the Cassini CIRS instrument.
Model fits show the thermal inertia of ring particles to be: 16$^{+10}_{-5} $, 13$^{+4}_{-4}$, 20$^{+10}_{-6}$, and 11$^{+68}_{-6}$ Jm$^{-2}$K$^{-1}$s$^{-1/2}$ 
for the A, B, and C rings and the Cassini division, respectively.
However, there are systematic deviations between modeled and observed temperatures in Saturn's shadow depending 
on solar phase angle. Test fits, which allow individual scans taken at different solar phase angles to have
different thermal inertia values, show that the apparent thermal inertia increases with solar phase angle. 
This dependence is likely to be explained, if large slowly spinning particles 
have lower thermal inertia values than those for small fast spinning particles, because 
the thermal emission of the slow rotators is relatively stronger than that of fast rotators at low phase and vise versa. 
We made additional parameter fits, which assume that slow and fast rotators have different thermal inertia values.
The estimated thermal inertia values of slow rotators are 8$^{+19}_{-4}$, 8$^{+10}_{-3}$, 
9$^{+23}_{-9}$, and 5$^{+37}_{-5}$ Jm$^{-2}$K$^{-1}$s$^{-1/2}$ for the A, B, and C rings and the Cassini division, respectively,
and those of fast rotators are 
77$^{+123}_{-66}$,  27$^{+173}_{-27}$,  34$^{+71}_{-16}$, 55$^{+145}_{-55}$ Jm$^{-2}$K$^{-1}$s$^{-1/2}$.
The values for fast rotators are still much smaller 
than those for the solid ice with no porosity. Thus, 
fast rotators are likely to have surface regolith layers, but these may be not as fluffy as 
those for slow rotators, probably because the capability of holding regolith particles is limited for fast rotators 
due to the centrifugal force. 
The large thermal inertia of fast rotators for the A ring is probably due to wakes, 
which largely enhance spin and collision velocities of small particles.
The low thermal inertia of the Cassini division probably indicates that there are some dust sources around this region and 
ring particles are fluffily coated by dust.
Other additional parameter fits, in which size of fast rotators is varied, indicate 
that particles less than  $\sim$ 1 cm should not occupy more than a half of the cross section
for the A, B, and C rings.  This is consistent with previous estimations of the ring particle size distribution.

\section*{Acknowledgments}
We thank Cedric Leyrat and an anonymous reviewer for their comments, which improved our manuscript.  
We are grateful for the support by the Cassini project and the NASA's OPR and PGG Programs.
K.O. is also grateful for the support by JSPS KAKENHI (22340125). 
We thank Nicholas Altobelli and Stu Pilorz for developing the CIRS database, 
Scott Edington, Shawn Brooks and Mark Showalter for designing the CIRS ring observations.
R.M. thanks the UVIS ring team in LASP for fruitful discussions.
Numerical simulations were carried out  with the supercomputers, Nebula and Galaxy, at JPL.

\begin{center}
\section*{REFERENCES}
\end{center}

\begin{description}
\item
Albers, N., Spahn, F., 2006.
The influence of particle adhesion on the stability of agglomerates in Saturn's rings.
Icarus 181,  292--301.

\item 
Altobelli, N., Spilker, L.J., Pilorz, S., Brooks, S., Edgington, S., Wallis, B., Flasar, M., 2007.
C ring fine structures revealed in the thermal infrared.
Icarus 191, 691--701. 

\item 
Altobelli, N., Spilker, L.J., Leyrat, C., Pilorz, S.,  2008.
Thermal observations of Saturn's main rings by Cassini CIRS: 
Phase, emission and solar elevation dependence.
Planet. Space. Sci. 56, 134--146.

\item 
Altobelli, N., Spilker, L.J., Pilorz, S., Leyrat, C., Edgington, S., Wallis, B., Flandes, A., 2009.
Thermal phase curves observed in Saturn's main rings by Cassini-CIRS: Detection of an opposition effect?
Geophys. Res. Lett. 36, L10105.

\item
Aumann, H.H., Kieffer, H.H., 1973.
Determination of particle sizes in Saturn's rings from their
eclipse cooling and heating curves.
Astrophys. J. 186, 305--311.

\item
Bradley, E.T. Colwell, J.E., Esposito, L.W., Cuzzi, J.N., Tollerud, H., Chambers, L., 2010.
Far ultraviolet spectral properties of Saturn's rings from Cassini UVIS. 
Icarus 206, 458--466.

\item 
Choksi, A.,  Tielens, A.G.G.M., Hollenbach, D., 1993
Dust coagulation.
Astrophys. J. 407, 806--819.

\item 
Colwell, J.E., Esposito, L.W., Srem\v{c}evi\'{c}, M., 2006.
Self-gravity wakes in Saturn's A ring measured by stellar 
occultations from Cassini.
Geophys. Res. Lett. 33, L07201.

\item 
Colwell, J.E., Esposito, L.W., Srem\v{c}evi\'{c}, M., Stewart, G.R., McClintock, W.E.,  2007.
Self-gravity wakes and radial structure of Saturn's B ring.
Icarus, 190, 127--144.

\item 
Colwell, J.E., Cooney, J.H., Esposito, L.W., Srem\v{c}evi\'{c}, M., 2009.
Density waves in Cassini UVIS stellar occultations: 1. The Cassini Division.
Icarus 200, 574--580.

\item
Cuzzi, J.N., Clark, K., Filacchione, G., French, R., Jhonson, R., Marouf, E., Spilker, L., 2009.
Ring particle composition and size distribution.
In: Dougherty, M.K., Esposito, L.W.,  Krimigis, S.M. (Eds.), Saturn from Cassini-Huygens.
Springer, Berlin, pp. 459--509.

\item 
D'Aversa, E., Bellucci, G., Nicholson, P.D., Hedman, M.M., Brown, R.H., 
Showalter, M.R., Altieri, F., Carrozzo, F.G., Filacchione, G., Tosi, F., 2010. 
The spectrum of a Saturn ring spoke from Cassini/VIMS.
Geophys. Res. L. 37, L01203.

\item
Delbo, M., Tanga, P., 2009.
Thermal inertia of main belt asteroids smaller than 100 km from IRAS data.
Planetary and Space Sci. 57, 259-265.  

\item 
Dones, L., Cuzzi, J.N., Showalter, M.R., 1993.
Voyager photometry of Saturn's A ring.
Icarus 105, 184--215.

\item 
Esposito, L.W., O'Callaghan, M., Simmons, K.E.,
Hord, C.W., West, R.A., Lane, A.L., Pomphrey, R.B., 
Coffeen, D.L., Sato, M., 1983. 
Voyager photopolarimeter steller occultation of Saturn's rings.
J. Geophys. Res. 88, 8643--8649.   

\item 
Farinella, P., Vokorouhlick\'{y}, D., Hartmann, W.K., 1998.
Meteorite delivery via Yarkovsky orbital drift.
Icarus 132, 378--387.

\item 
Ferrari, C., Leyrat, C., 2006.
Thermal emission of spherical spinning ring particles: The standard model.
Astron. Astrophys. 447, 745--760.

\item   
Ferrari, C., Galdemard, P., Lagage, P.O., Pantin, E., Quoirin, C., 2005.
Imaging Saturn's rings with CAMIRAS: thermal inertia of B and C rings.
Astron. Astrophys. 441, 379--389.

\item 
Ferrari, C., Brooks, S., Edgington, S., Leyrat, C., Pilorz, S., Spilker, L., 2009.
Structure of self-gravity wakes in Saturn's A ring as measured by Cassini CIRS.
Icarus 199, 145--153.

\item
Flandes, A., Spilker, L.J., Morishima, R., Pilorz, S., Leyrat, C., Altobelli, N., Brooks, S.M., Edgington, S.G., 2010.
Brightness of Saturn's rings with decreasing solar elevation.
Planet and Space Science 58, 1758--1765. 

\item 
Flasar, F.M., and 44 colleagues, 2004.
Exploring the Saturn system in the thermal infrared: The composite infrared spectrometer.
Space Science Rev. 115, 169--297.

\item 
Flasar, F.M., and 45 colleagues, 2005.
Temperatures, winds, and composition in the Saturnian system.
Science 307, 1247--1251.

\item 
French, R.G., Nicholson, P.D., 2000.
Saturn's rings II.
Particle sizes inferred from stellar occultation data.
Icarus 145, 502--523.

\item 
Froidevaux, L., 1981.
Saturn's rings: Infrared brightness variation with solar elevation.
Icarus 46, 4--17.

\item 
Froidevaux, L., Matthews, K., Neugebauer, G., 1981.
Thermal response of Saturn's ring particles during and after eclipse.
Icarus 46, 18--26.

\item 
Hedman, M.M., Nicholson, P.D., Salo, H., Wallis, B.D. Buratti, B.J.,
Baines, K.H., Brown, R.H., Clark, R.N.,  2007.
Self-gravity wake  structures in Saturn's A ring revealed by Cassini VIMS.
 Astron. J. 133, 2624--2629.
 
 \item 
Hedman, M.M., Burt, J.A., Burns, J.A., Tiscareno, M.S.,  2010.
The shape and dynamics of a heliotropic dusty ringlet in the Cassini division.
Icarus 210, 284--297.

 \item
Hor\'{a}nyi, M., Burns, J.A., Hedman, M.M., Jones, G.H., Kempf, S., 2009.
Diffuse rings.
In: Dougherty, M.K., Esposito, L.W.,  Krimigis, S.M. (Eds.), Saturn from Cassini-Huygens.
Springer, Berlin, pp. 511--536.
 
 \item Howett, C.J.A., Spencer, J.R., Peral, J., Segura, M., 2010.
 Thermal inertia and bolometric Bond albedo values for Mimas,
 Enceladus, Tethys, Dione, Rhea, and Iapetus as derived from Cassini/CIRS
 measurements.
 Icarus 206, 573-593.

\item
Kouchi, A., Greenberg, J.M., Yamamoto, T., Mukai, T., 1992.
Extremely low thermal conductivity of amorphous ice: 
relevance to comet evolution. 
Astrophys. J. 388, L73--L76.

\item 
Leyrat, C., Ferrari, C., Charnoz, S., Decriem, J., Spilker, L., Pilorz, S., 2008a.
Spinning particles in Saturn's C ring from mid-infrared
observations: Pre-Cassini results.
Icarus 196, 625--641.

\item
Leyrat, C., Spilker, L, J., Altobelli, N., Pilorz, S., Ferrari, C., 2008b.
Infrared observations of Saturn's rings by Cassini CIRS : Phase angle and local time dependence
Planet. Space. Sci. 56, 117--133.

\item
Marouf, E.A., Tyler, G.L., Zebker, H.A., Simpson, R.A., 
Eshleman, V.R., 1983. 
Particle size distributions in Saturn's rings from Voyager 1 radio occultation.
Icarus 54, 189--211.

\item
Morishima, R., Salo, H., 2006.
Simulations of dense planetary rings IV: 
Rotating self-gravitating particles
with size distribution. 
Icarus 181, 272--291.

\item
Morishima, R., Salo, H., Ohtsuki, K., 2009 (Paper I).
A multilayer model for thermal infrared emission of Saturn's rings:
Basic formulation and implications for Earth-based observations.  
Icarus 201, 634--654.

\item
Morishima, R., Spilker, L., Salo, H., Ohtsuki, K., Altobelli, N., Pilorz, S., 2010 (Paper II).
A multilayer model for thermal infrared emission of Saturn's rings. II:
Albedo, spins, and vertical mixing of ring particles inferred from Cassini CIRS.  
Icarus 210, 330-345. 

\item  
Morrison, D., 1974.
Infrared radiometry of the rings of Saturn. 
Icarus 22, 57-65.

\item 
Nicholson, P.D., and 15 coauthors,  2008.
A close look at Saturn's rings with Cassini VIMS.
Icarus 193, 182-212.

\item
Ohtsuki, K., 2005.
Rotation rates of particles in Saturn's rings.
Astrophys. J. 626, L61--L64.

\item
Ohtsuki, K., 2006.
Rotation rate and velocity dispersion of planetary ring particles 
with size distribution II. Numerical simulation for gravitating particles.
Icarus 183, 384--395.

\item
Poulet, F., Cruikshank, D.P., Cuzzi, J.N., Roush, T.L., French, R.G., 2003.
Composition of Saturn's rings A, B, and C from high resolution
near-infrared spectroscopic observations.
Astron. Astrophys. 412, 305--316.

\item
Press, W.H., Teukolsky, S.A., Vetterling, W.T., Flannery, B.P., 1986.
Numerical Recipes.
Cambridge Univ. Press, Cambridge, UK.

\item
Salo, H., 1987. 
Numerical simulations of collisions between rotating particles.
Icarus 70, 37--51.

\item
Salo, H., 1992. 
Numerical simulations of dense collisional systems. II - Extended distribution of particle sizes.
Icarus 96, 85--106.

\item
Salo, H., 1995. 
Simulations of dense planetary rings. III. Self-gravitating identical particles. 
Icarus 117, 287--312.

\item
Salo, H., Karjalainen, R., 2003.
Photometric modeling of Saturn's rings 
I. Monte Carlo method and the effect of nonzero volume filling factor.
Icarus 164, 428--460.

\item
Salo, H., Schmidt, J., Spahn, F., 2001. 
Viscous overstability in Saturn's B ring I. Direct simulations and measurement of transport coefficients. 
Icarus 153, 295--315.

\item
Shoshany, Y., Prialnik, D., Podolak, M.,  2002.
Monte Carlo modeling of the thermal conductivity of porous cometary ice.
Icarus 157, 219--227.

\item
Spilker, L.J., Pilorz, S.H., Edgington, S.G., Wallis, B.D.,
Brooks, S.M., Pearl, J.C., Flasar, F.M., 2005.
Cassini CIRS observations of a roll-off in Saturn ring spectra at
submillimeter wavelengths.  
Earth, Moon, and Planets 96, 149--163. 

\item
Spilker, L.J., and 11 colleagues, 2006.
Cassini thermal observations of Saturn's main rings: 
Implications for particle rotation and vertical mixing.
Planet. Space Sci. 54, 1167--1176.

\item
Spilker, L.J., Flandes, A., Morishima, R., Altobelli, N., Leyrat, C., Pilorz, S., Ferrari, C.C.,  Edgington, S.G., Brooks, S.M., 2009.
Saturn ring temperatures at equinox with Cassini CIRS.
American Geophysical Union, Fall Meeting 2009, abstract $\#$P51B-1131.

\item Stuart, S.J., Stewart, G.R., Lewis, M.C., Colwell, J.E., Srem\v{c}evi\'{c}, M.,  2010.
Estimating the masses of Saturn's A and B rings from high-optical depth $N$-body simulations and steller occultations.
Icarus 206, 431--445.

\item
Zebker, H.A., Marouf, E.A., Tyler, G.L., 1985.
Saturn's rings - Particle size distributions for thin layer model.
Icarus 64, 531--548.

\end{description}

\clearpage

\renewcommand{\baselinestretch}{1.5}

\begin{table}
\begin{center}
\scriptsize
\begin{tabular}{|ccccccccccccc|} \hline
Scan No.   &  Date      & $N_{\rm foot}$ &  $a_{\odot}$ & $B'$  & $B$   & $\alpha$ & $\phi_{\rm p}$ & $r_{\rm p}$  & $\phi_{\rm Cas}$& $r_{\rm Cas}$  & $\tau_{\rm PPS}$& $t_{\rm eclip}$    \\ 
                    &                &                            & (AU)              & (deg) &(deg) &  (deg)                    &  (deg)         & (1,000 km)   & (deg)                     &  $(r_{\rm Sat})$ && (hr)   \\ \hline
C1  & 05-156  & 1610 &9.076  & -21.48 & -18.73 &  41.38 & 261.10 (co)&  83.10 & 143.00 &  26.30 &0.108& 1.56\\
       &                 &           &            &              & -20.02 &  34.67 &  17.25         &  82.81 & 135.78 &  24.46 &&\\   
C2  & 05-176 & 1979 &  9.079 & -21.28 & -20.20 &  30.51 & 276.91 (co) &  83.11 & 157.85 &  19.19 &0.111& 1.56\\
      &                 &           &             &             & -22.30 &  20.64 &  27.82           &  82.80 & 147.23 &  17.08 &&\\
C3 & 05-178  & 1711 &  9.079  & -21.25 &   9.59  & 125.27 & 286.54 (cl) &  83.59 &  59.99 &  11.18 &0.107&1.57\\
      &                &            &             &              &   9.55 & 121.33 &   1.69           &  82.39 &  55.75 &  10.56 &&\\   
C4 & 05-194  & 1764 &  9.081  & -21.09 & -19.88 &  28.25 &  22.05 (cl)   &  83.08 & 151.57 &  21.31 &0.110& 1.57\\
       &               &           &               &              & -21.76 &  26.49 & 100.21       &  82.81 & 149.83 &  19.03 &&\\
C5  & 06-300 & 1126 &  9.164  & -15.29 & -22.66 & 112.84 & 283.54 (cl) &  83.76 &  60.10 &  12.92 &0.111&1.66\\
       &                &           &              &              & -28.19 & 107.66 &   1.37         &  82.90 &  56.61 &  10.88 &&\\
C6 &  06-337  & 2482 &  9.172 & -14.77 &  60.00 & 113.76 & 270.33 (cl) &  83.58 & 289.52 &  12.77 &0.111& 1.67\\
      &                 &           &              &              &  50.03 & 105.73 &  89.70         &  82.91 & 276.30 &   9.98 &&\\
C7 & 07-016  & 1041 &   9.181 & -14.16 & -48.86 &  45.17   & 297.13 (cl)&  83.11 & 171.86 &  13.40 &0.112& 1.67\\
      &                  &           &              &              & -53.71 &  37.09   & 119.60       &  82.42 & 146.87 &  11.96 &&\\
C8 & 07-048  & 1132 &  9.188 & -13.70 & -25.66 & 117.50 & 281.25 (cl)&  84.54 &  51.69 &  18.54 &0.114& 1.68\\
        &                 &           &              &              & -29.92 & 114.26 &   2.20         &  82.66 &  49.87 &  16.82 &&\\
C9& 07-049   & 2188 &  9.189 & -13.69 & -52.95 &  80.11   &  92.11 (co)&  83.92 & 100.70 &  16.47 &0.114&1.68\\
       &                 &            &              &              & -59.23 &  72.70   & 301.52      &  82.93 &  86.40 &  15.09 &&\\
C10 & 07-082 &  982    & 9.196  & -13.22 & -50.18 &  75.22   & 346.88 (cl)&  84.01 & 101.22 &   9.97 &0.113&1.69\\
        &                &             &             &              & -60.14 &  72.80  &  74.78        &  81.49 &  97.32 &   9.10 &&\\ \hline
\end{tabular}
\caption{Geometry data of azimuthal scans for the C ring. 
Date is year and day of year, $N_{\rm foot}$ is the number of footprints (or spectra), $a_{\odot}$ is the heliocentric distance, $B'$ is the solar elevation angle, 
$B$ is the elevation angle of the Cassini spacecraft,
$\alpha$ is the solar phase angle, $\phi_{\rm p}$ is the local hour angle (or the longitude) of ring particles around Saturn 
with the origin at the midnight, $r_{\rm p}$ is the saturnocentric distance, $\phi_{\rm Cas}$ is the longitude of the spacecraft around the footprint, $r_{\rm Cas}$ is the range of the spacecraft ($r_{\rm Sat}$ is 
the radius of Saturn),
$\tau_{\rm PPS}$ is the mean optical depth from observations taken by Voyager PPS (see the text), and $t_{\rm eclip}$ is the mean time of the eclipse. 
For $B$, $\alpha$,  $r_{\rm p}$, $\phi_{\rm Cas}$, and $r_{\rm Cas}$, 
the upper and lower limits of each parameter are shown in the upper and lower rows. For $\phi_{\rm p}$, the values of initial and last foot prints during the scans 
are shown in the upper and lower rows. The characters in the parenthesis in the upper row of $\phi_{\rm p}$ represents 
the azimuthal direction of the scan: ``co" is for a counter-clockwise scan whereas ``cl"  for a clockwise scan.}
\end{center}
\end{table}

\clearpage

\begin{table}
\begin{center}
\scriptsize
\begin{tabular}{|cccccccccccccc|} \hline
Scan No.  &  Date      & $N_{\rm foot}$ &  $a_{\odot}$ & $B'$  & $B$   & $\alpha$ & $\phi_{\rm p}$  & $r_{\rm p}$  & $\phi_{\rm Cas}$& $r_{\rm Cas}$   & $\tau_{\rm PPS}$ &$\tau_{\rm UVIS}$&  $t_{\rm eclip}$ \\ 
                    &                &                            & (AU)              & (deg) &(deg) &  (deg)                    &  (deg)                 & (1,000 km)   & (deg)          &  $(r_{\rm Sat})$ &&& (hr)   \\ \hline
B1  & 04-184    &  503   & 9.042  & -24.46 & -14.89 & 105.93 & 284.21 (cl) & 107.08 &  66.12 &  26.06 &1.660&2.491& 1.24\\
       &                  &            &             &              & -16.65 & 103.95 & 26.65          & 105.60 &  64.34 &  24.00 &&&\\
B2 & 05-158     & 1533  &   9.076 & -21.46 & -19.81 &  13.01 &  18.90 (cl)  & 106.12 & 171.23 &  15.79 &1.856&3.204& 1.42\\
      &                   &            &               &              & -24.53 &   8.19 & 139.28         & 105.21 & 166.20 &  12.25 &&&\\
B3  & 05-175   & 1372 &  9.078   & -21.29 & -18.72 &  34.25 &   0.40 (cl)     & 105.01 & 145.79 &  25.29 &1.572&2.279&1.44\\
       &                 &           &                &              & -21.15 &  31.98 & 108.99         & 104.76 & 143.57 &  22.20 &&&\\
B4& 05-229   & 1353  &  9.086  & -20.70 & -18.60 &  43.78 & 357.03 (cl)& 105.37 & 141.26 &  26.67 &1.570&2.277& 1.47\\
       &                 &             &             &              & -20.80 &  36.28 & 277.87       & 104.46 & 133.19 &  23.81 &&&\\
B5& 06-300  & 1571   &  9.164 & -15.29 & -26.29 & 107.06 &  17.21 (co)& 106.29 &  65.12 &  12.08 &1.877&3.228&1.67\\
       &                 &             &             &             & -30.13 & 103.58 & 269.10       & 104.92 &  60.30 &  10.29 &&&\\
B6 & 07-016    & 1311  & 9.181 & -14.16 & -39.76 &  43.16 & 152.84 (co)& 105.53 & 180.02 &  13.85 &1.877&3.460&1.70\\
        &                  &            &             &               & -56.28 &  26.73 & 287.33       & 104.27 & 166.05 &  11.50 &&&\\
B7 & 07-048    & 1515 & 9.188  & -13.70 & -29.06 & 114.20 &  28.36 (co) & 106.02 &  55.24 &  18.40 &1.857&3.135& 1.71\\
        &                   &           &             &              & -31.53 & 111.16 & 283.54       & 104.78 &  49.87 &  16.20 &&&\\
B8 & 07-049    &  822  &  9.189 & -13.69 & -51.02 &  86.44 &  356.27 (cl)  & 106.84 &  85.26 &  15.97 &1.974&4.014&1.72\\
        &                  &           &              &              & -56.91 &  82.08 &  82.35           & 104.97 &  75.37 &  14.99 &&&\\ \hline
\end{tabular}
\caption{Same as Table~1 but for  the B ring.  Here the mean optical depth from observations taken by Cassini UVIS, $\tau_{\rm UVIS}$, 
is also shown, in addition to $\tau_{\rm PPS}$.}
\end{center}
\end{table}

\clearpage

\begin{table}\begin{center}
\scriptsize
\begin{tabular}{|ccccccccccccc|} \hline
Scan No.   &  Date      & $N_{\rm foot}$ &  $a_{\odot}$ & $B'$  & $B$   & $\alpha$ & $\phi_{\rm p}$  & $r_{\rm p}$  & $\phi_{\rm Cas}$& $r_{\rm Cas}$  & $\tau_{\rm PPS}$&  $t_{\rm eclip}$   \\ 
                    &                &                            & (AU)              & (deg) &(deg) &  (deg)                    &  (deg)                 & (1,000 km)   & (deg)           &  $(r_{\rm Sat})$ & &(hr)  \\ \hline 
CD1 & 07-098  &  402  &  9.200 & -12.99 & -49.76 &  83.85 &  28.68 (cl) & 120.84 &  94.65 &   6.82 & 0.069&1.76\\
         &                &           &              &              & -56.03 &  77.02 &  60.91        & 119.69 &  81.68 &   6.36 &&\\
CD2 & 07-098  & 1126 &  9.200 & -12.99 & -41.76 &  70.27 &  30.93 (co)& 120.95 & 124.91 &   9.19 &0.097& 1.76\\
         &                 &           &             &              & -53.53 &  59.19 &  300.20      & 118.61 & 104.98 &   7.24 &&\\
CD3 & 07-130   & 2678 & 9.207 & -12.53 &  41.60 & 150.90 & 314.45 (cl)& 121.84 & 337.04 &   8.06 &0.149&1.77\\
         &                 &            &            &             &  23.76 & 140.37 &  98.20        & 119.81 & 324.19 &   6.76 &&\\
CD4  & 08-015   & 2364 & 9.270  &  -8.82 &  57.92 & 117.62 & 359.99 (co)& 120.04 & 300.32 &   8.67 &0.095 & 1.88\\
          &                  &          &                &           &  34.87 &  84.08 &   0.03            & 119.76 & 252.06 &   6.49 && \\
CD5  &  08-027  & 2767 &9.273   &  -8.64 &  57.37 & 109.88 & 359.97(co) & 120.18 & 289.68 &   9.55 &0.100 & 1.88\\
         &                   &          &               &           &  36.83 &  79.59 &   0.02            & 119.74 & 245.31 &   7.48 &            &  \\ 
CD6  & 08-262   & 3328 & 9.337   &  -5.04 & -31.83 &  53.11 & 293.87 (cl) & 121.79 & 158.53 &  13.75 &0.121 &1.94 \\
          &                   &         &               &            & -45.36 &  39.20 & 289.98       & 118.92 & 135.06 &  10.36   & &\\ \hline         
\end{tabular}
\caption{Same as Table~1 but for  the Cassini division. }
\end{center}
\end{table}

\begin{table}\begin{center}
\scriptsize
\begin{tabular}{|cccccccccccccc|} \hline
Scan No.   &  Date      & $N_{\rm foot}$ &  $a_{\odot}$ & $B'$  & $B$   & $\alpha$ & $\phi_{\rm p}$  & $r_{\rm p}$  & $\phi_{\rm Cas}$& $r_{\rm Cas}$  & $\tau_{\rm PPS}$ &$\tau_{\rm UVIS}$& $t_{\rm eclip}$   \\ 
                    &                &                            & (AU)              & (deg) &(deg) &  (deg)                    &  (deg)                 & (1,000 km)   & (deg)           &  $(r_{\rm Sat})$ & && (hr)   \\ \hline   
A1  & 06-300    & 2626 &  9.164  & -15.28 & -25.60 & 106.56 & 269.34 (cl)& 129.79 &  74.88 &  12.39 &0.432& 0.598& 1.64\\
       &                   &           &              &             & -45.43 &  92.41 &  90.64         & 128.80 &  55.34 &   7.83   && &\\
A2  & 07-016     & 1061 &  9.181  & -14.15 & -35.16 &  28.72 & 311.93 (cl)& 129.58 & 194.27 &  14.31 &0.445& 0.618&1.71\\
       &                   &           &               &              & -40.40 &  21.09 & 134.59       & 128.06 & 168.26 &  11.75 &&&\\
A3 & 07-048    & 2619 &   9.188   & -13.70 & -24.38  & 118.18 & 260.41 (cl)& 130.72 &  59.09 &  19.44 &0.441&0.596&1.73\\
      &                    &          &               &              & -38.54  & 106.19 &  87.43        & 128.55 &  47.15 &  15.47 &&&\\ \hline
\end{tabular}
\caption{Same as Table~2 but for  the A ring. }
\end{center}
\end{table}

\clearpage
\begin{landscape}
\begin{table}\begin{center}
\scriptsize 
\begin{tabular}{|cccccccccccccc|} \hline
Ring         & $\tau$             & $h_{\rm r}$& Bouncing &  $r_{\rm fast}$ & $\Gamma_{\rm slow}$   & $\Gamma_{\rm fast}$     & $A_{\rm V}$ & $f_{\rm fast}$ & $\chi^2_1$  & $\Gamma_0$       & $A_{\rm V,0}$                   & $f_{\rm fast,0}$                & $\chi^2_0$     \\ \hline     
C ring      &  $\tau_{\rm PPS}$ &          3       & no             &	  10 cm               &${\bf 8.5^{+22.8}_{-8.5}}$    & ${\bf 33.5^{+71.6}_{-15.5}}$ & ${\bf 0.14^{+0.06}_{-0.09}}$ &  ${\bf 0.68^{+0.11}_{-0.10}}$ &  {\bf 1.58}  &  ${\bf 20.0^{+10.1}_{-5.7}}$     &  ${\bf 0.12^{+0.06}_{-0.12}}$ & ${\bf 0.68^{+0.10}_{-0.10}}$ &  {\bf 2.25}  \\ 
                  &                 &                    &                  &	  1 cm                 &$15.1^{+29.9}_{-11.9}$ & $80.0^{+120.0}_{-65.9}$ & $0.17^{+0.22}_{-0.11}$ &  $0.58^{+0.42}_{-0.15}$ &  2.51  &     &   &  &    \\
                   &                &                    &                   &	  5 mm                &$20.0^{+80.0}_{-7.0}$             & $80.0^{+120.0}_{-70.4}$ & $0.14^{+0.10}_{-0.14}$ &  $0.58^{+0.42}_{-0.13}$ &  5.82  &     &   &  &    \\ \hline                                  
B ring (all)& $\tau_{\rm PPS}$ &	1          &yes	   & 	10 cm               &${\bf  5.0^{+4.7}_{-3.1}}$  &	${\bf  45.0^{+143.0}_{-25.1}}$  &  ${\bf  0.58^{+0.01}_{-0.01}}$  & ${\bf 0.49^{+0.14}_{-0.13}}$   &   {\bf 18.62}  &  ${\bf 13.0^{+2.1}_{-2.1}}$     &  ${\bf  0.58^{+0.01}_{-0.01}}$ & ${\bf  0.35^{+0.13}_{-0.10}}$ &  {\bf  19.66}  \\  
B ring (high $B'$)& $\tau_{\rm PPS}$ &	1          &yes	   & 	10 cm                &${\bf 7.4^{+4.9}_{-2.9}}$&	${\bf 35.0^{+92.7}_{-16.5}}$   &  ${\bf 0.56^{+0.01}_{-0.01}}$ &  ${\bf 0.48^{+0.12}_{-0.08}}$ & {\bf 7.91}&  ${\bf 14.3^{+1.8}_{-2.5}}$     &  ${\bf 0.56^{+0.01}_{-0.01}}$ & ${\bf 0.42^{+0.11}_{-0.07}}$ &  {\bf 9.38}  \\ 
                              &              &	            &	            & 	1 cm                  &$8.0^{+6.1}_{-2.8}$	     &	$80.0^{+120.0}_{-80.0}$   &  $0.56^{+0.02}_{-0.01}$ &  $0.42^{+0.08}_{-0.17}$ & 8.36&     &   & &   \\ 
                   
	              &            &	            &	            & 	5 mm                &$14.0^{+6.6}_{-3.3}$	     &	$80.0^{+120.0}_{-70.8}$   &  $0.56^{+0.02}_{-0.01}$ &  $0.34^{+0.09}_{-0.15}$ & 10.33&     &   &  &  \\ 	
    &$\tau_{\rm PPS}$  &	3                  & yes	   &	10 cm      &$6.1^{+5.3}_{-1.4}$	     &  $150.0^{+50.0}_{-130.1}$  & $0.57^{+0.01}_{-0.01}$  &  $0.20^{+0.04}_{-0.05}$& 8.27  &  $13.9^{+1.6}_{-1.8}$     &  $0.56^{+0.01}_{-0.01}$ & $0.17^{+0.04}_{-0.04}$ &  9.72  \\
     &$\tau_{\rm UVIS}$ &	1          &yes	   & 	10 cm               &$8.4^{+4.5}_{-3.8}$	     &	$20.0^{+46.5}_{-7.1}$    & $0.60^{+0.01}_{-0.01}$  & $0.42^{+0.13}_{-0.12}$ & 5.96     &   $12.3^{+1.9}_{-1.5}$  &  $0.60^{+0.01}_{-0.01}$  &  $0.37^{+0.09}_{-0.09}$ &  6.65  \\ 
   &$\tau_{\rm PPS}$    &	1                   &no	   & 	10 cm               &$5.0^{+8.1}_{-1.2}$  &	$39.5^{+56.0}_{-6.6}$	& $0.29^{+0.01}_{-0.02}$  & $0.75^{+0.04}_{-0.12}$  & 10.13  &  $21.3^{+5.3}_{-2.8}$     &  $0.31^{+0.01}_{-0.01}$ & $0.58^{+0.08}_{-0.08}$ &  12.16  \\ 
B ring (low $B'$)& $\tau_{\rm PPS}$ &	1          &yes	   & 	10 cm               &${\bf  9.1^{+8.9}_{-4.4}}$  &	${\bf  19.0^{+181.0}_{-19.0}}$  &  ${\bf  0.64^{+0.02}_{-0.01}}$  & ${\bf 0.27^{+0.22}_{-0.27}}$   &   {\bf 1.72}  &  ${\bf 11.7^{+3.6}_{-2.5}}$     &  ${\bf  0.64^{+0.01}_{-0.02}}$ & ${\bf  0.23^{+0.20}_{-0.21}}$ &  {\bf  1.80}  \\ 
                               &      &                    &          	   & 	1 mm               &$18.4^{+22.7}_{-6.3}$  &	$35.0^{+165.0}_{-35.0}$                   &  $0.62^{+0.02}_{-0.03}$  & $0.21^{+0.26}_{-0.25}$   &  2.51  &       &   &  &   \\
                               &     &                    &          	   & 	0.5 mm             &$19.0^{+81.0}_{-6.2}$  &	$55.0^{+145.0}_{-55.0}$                   &  $0.62^{+0.03}_{-0.02}$  & $0.18^{+0.24}_{-0.18}$   &  2.85  &       &   &  &   \\ 	
 & $\tau_{\rm PPS}$    &	3                  & yes	   &	10 cm               &$10.1^{+5.7}_{-4.5}$   &$20.0^{+180.0}_{-20.0}$ & $0.65^{+0.02}_{-0.01}$ & $0.06^{+0.11}_{-0.06}$	 & 1.66  &  $11.7^{+3.5}_{-2.5}$  &  $0.65^{+0.01}_{-0.02}$ & $0.05^{+0.10}_{-0.05}$ &  1.79  \\
  &$\tau_{\rm UVIS}$  &	1          &yes	   & 	10 cm               &$8.4^{+3.8}_{-3.6}$         &	$8.5^{+191.5}_{-8.5}$   &  $0.71^{+0.02}_{-0.02}$  &  $0.02^{+0.33}_{-0.02}$    &  2.39&    $8.7^{+1.8}_{-1.9}$  &  $0.70^{+0.02}_{-0.02}$ & $0.03^{+0.28}_{-0.03}$  & 2.16   \\ 
  & $\tau_{\rm PPS}$    &	1          &no	   & 	10 cm               &$15.9^{+84.1}_{-10.5}$       &	$11.5^{+40.3}_{-11.5}$  &    $0.50^{+0.02}_{-0.04}$   &   $0.38^{+0.29}_{-0.20}$ & 2.22 &  $14.6^{+5.0}_{-3.6}$     &  $0.49^{+0.02}_{-0.03}$ & $0.42^{+0.19}_{-0.19}$ &  2.38  \\ \hline
Cassini div.   & $\tau_{\rm PPS}$ &	3          &no	   &	10 cm               &$4.7^{+36.4}_{-4.7}$	     &  $55.0^{+145.0}_{-55.0}$ &    $0.43^{+0.08}_{-0.09}$   &	  $0.46^{+0.26}_{-0.22}$     &  1.26   &  $11.0^{+67.7}_{-6.2}$    & $0.42^{+0.08}_{-0.08}$    & $0.50^{+0.24}_{-0.25}$ &   1.46 \\ 
                                &  &	        &	   &	2 mm                &$15.0^{+85.0}_{-13.8}$	     &  $5.0^{+195.0}_{-5.0}$ &    $0.41^{+0.12}_{-0.17}$   &	  $0.53^{+0.47}_{-0.50}$     &  2.10   &  &  &  &    \\ 
                                &  &	        &	   &	1 mm                &$15.0^{+85.0}_{-11.2}$	     &  $5.0^{+195.0}_{-5.0}$ &    $0.41^{+0.12}_{-0.17}$   &	  $0.45^{+0.55}_{-0.45}$     &  3.02   &  &  &  &    \\ \hline
A ring  &$\tau_{\rm PPS} $  &	3          &no	   & 	10 cm               &${\bf 7.8^{+18.9}_{-4.2}}$	      &	${\bf  76.5^{+123.5}_{-65.8}}$   &  ${\bf  0.49^{+0.03}_{-0.03}}$  &  ${\bf  0.62^{+0.12}_{-0.11}}$  & {\bf 1.54}  &  ${\bf 16.2^{+10.3}_{-5.1}}$     &  ${\bf 0.49^{+0.03}_{-0.03}}$ & ${\bf 0.62^{+0.11}_{-0.11}}$ &{\bf 1.93}  \\ 
	    &                     &	                     &       	   & 	1 cm                 &$10.0^{+49.9}_{-7.5}$	      &	$35.0^{+165.0}_{-30.0}$   &  $0.49^{+0.04}_{-0.04}$  &  $0.61^{+0.16}_{-0.19}$  & 1.98  &       &  &  &    \\ 
	     &                    &	                     &       	   & 	5 mm                 &$20.0^{+80.0}_{-15.8}$	      &	$10.0^{+190.0}_{-10.0}$   &  $0.50^{+0.03}_{-0.04}$  &  $0.58^{+0.17}_{-0.16}$  & 2.99  &       &  &  &    \\ 	
	 &$\tau_{\rm PPS}$    &	1                  & no	   &	10 cm               &$9.0^{+19.4}_{-5.7}$     &  $35.0^{+165.0}_{-23.6}$   &   $0.49^{+0.03}_{-0.03}$  & $0.69^{+0.09}_{-0.11}$ & 1.52   &  $16.2^{+10.3}_{-4.7}$     &  $0.49^{+0.03}_{-0.03}$ & $0.68^{+0.10}_{-0.12}$ &  1.86  \\             
           &$\tau_{\rm UVIS}$     &	3          &no	   &	10 cm               &$9.4^{+15.5}_{-5.1}$	      &  $42.5^{+157.5}_{-34.7}$   &   $0.46^{+0.03}_{-0.03}$ & $0.52^{+0.22}_{-0.14}$ & 1.68   &   $15.3^{+10.8}_{-4.6}$    &   $0.46^{+0.03}_{-0.03}$ & $0.52^{+0.12}_{-0.15}$ & 2.02  \\ \hline
\end{tabular}
\caption{Input and estimated parameters. The base-line case for each ring is shown with bold numbers. 
The unit for $\Gamma_{\rm slow}$, $\Gamma_{\rm fast}$, and $\Gamma_{0}$ is Jm$^{-2}$K$^{-1}$s$^{-1/2}$.
The upper limits of $\Gamma_{\rm slow}$ and $\Gamma_{\rm fast}$ are set to be 100 and 200 Jm$^{-2}$K$^{-1}$s$^{-1/2}$, respectively,
 due to insufficient coverages in simulations. 
 The reduced chi-squares, $\chi^2_1$ and $\chi^2_0$, represent the minimum values in the parameter spaces.}
\end{center}
\end{table}
\end{landscape}
\clearpage

\begin{table}
\begin{center}
\scriptsize
\begin{tabular}{|ccc|ccc|} \hline
No.   & $\alpha$ &$\Gamma_{\rm ind}$ &  No.  & $\alpha$  & $\Gamma_{\rm ind}$       \\ 
        &  (degs) &  (Jm$^{-2}$K$^{-1}$s$^{-1/2}$)  &        &(degs)         &  (Jm$^{-2}$K$^{-1}$s$^{-1/2}$)    \\ \hline 
C2  &  25.56   & $13.6^{+6.5}_{-4.2}$  & B2  & 9.63   & $6.8^{+2.1}_{-0.9}$ ($6.3^{+1.9}_{-0.8}$)\\
C4  &  26.95   & $11.9^{+3.6}_{-2.5}$  & B6  & 31.59 & $10.0^{+2.3}_{-1.6}$ ($7.0^{+1.5}_{-0.8})$\\
C1  & 37.68    & $15.7^{+5.9}_{-3.9}$  & B3  & 32.62  & $10.8^{+1.6}_{-1.1}$ $(11.5^{+1.9}_{-1.3}$) \\
C7  & 39.23    & $15.4^{+9.4}_{-5.0}$ &B4  & 40.05  & $12.2^{+2.1}_{-1.7}$  $(11.9^{+2.1}_{-1.5}$) \\
C10& 73.56    & $17.8^{+11.5}_{-6.1}$ & B8 & 83.37  & $12.9^{+2.6}_{-2.3}$ ($7.1^{+1.3}_{-0.5}$) \\  
C9  &  74.62   & $16.6^{+12.9}_{-6.0}$ & B1  & 104.17& $16.5^{+2.6}_{-1.7}$ ($21.1^{+3.7}_{-2.7}$)\\
C6  &  106.23 & $24.5^{+56.3}_{-12.2}$ & B5 &104.71& $15.7^{+7.9}_{-3.9}$ ($9.2^{+1.8}_{-1.8}$) \\
C5  &  109.61 & $27.3^{+50.1}_{-11.5}$ & B7 &111.93 & $11.5^{+3.4}_{-2.0}$ ($6.9^{+1.5}_{-0.7}$) \\    
C8  &  115.37 & $21.6^{+26.2}_{-8.0}$ &&&\\
C3  &  122.47 &  $29.2^{+10.6}_{-5.6}$ &&&\\ \hline
CD6 & 46.26  & $5.1^{+13.8}_{-3.3}$  &A2 & 24.08 & $13.9^{+5.7}_{-3.5}$\\
CD1 and CD2 &70.61 & $9.7^{+17.4}_{-4.9}$  & A1 & 96.95 & $26.4^{+73.6}_{-12.2}$ \\
CD5  & 96.39  & $10.2^{+89.8}_{-7.2}$ & A3 &108.52 & $26.1^{+58.0}_{-11.2}$\\
CD4 & 102.67 &$10.5^{+89.5}_{-7.2}$  &&&\\
CD3 &  142.19& $20.2^{+65.2}_{-10.4}$ &&&\\ \hline
\end{tabular}
\caption{Thermal inertia of an individual scan, $\Gamma_{\rm ind}$, for the base-line cases, 
derived with fixed $A_{\rm V,0}$ and $f_{\rm fast,0}$.  The scans are on the order of the mean value of the phase angle, $\alpha$.
The upper limit of $\Gamma_{\rm ind}$ is set to be 100 Jm$^{-2}$K$^{-1}$s$^{-1/2}$ due to an insufficient coverage in simulations.
The different combinations of $A_{\rm V,0}$ and $f_{\rm fast,0}$ for the high $|B'|$ case (B1-B4) and the low $|B'|$ case (B5-B8) are used. 
For comparison, the values of $\Gamma_{\rm ind}$ estimated using a single combination of 
$A_{\rm V,0}$ and $f_{\rm fast,0}$ for all the B ring data are shown in parenthesis.}
\end{center}
\end{table}

\begin{table}\begin{center}
\scriptsize
\begin{tabular}{|cccc|} \hline
Ring     &  $r_{\rm min}$  & $r_{\rm max}$ & $q$   \\   
	     &    (cm)                &  (m)                  &      \\ \hline
C ring &    1              & 10  & 3.1  \\
B ring &    30              & 20  & 2.75  \\
Cassini division  &    0.1              & 20  & 2.75       \\ 
A ring (inner) &    30              & 20  & 2.75       \\ \hline
\end{tabular}
\caption{The lower and upper cut-off sizes and the power-law index  of ring particles estimated in French and Nicholson (2000). 
The inner A ring is the inside the Encke division.}
\end{center}
\end{table}

\clearpage

\renewcommand{\baselinestretch}{1.8}

\section*{Figure captions}
\begin{description}

\item
Fig.~1. Azimuthal temperature variations for the C ring. 
The title of each panel is the scan number, which is on the order of the scan date, whereas the 
panels are placed on the order of the phase angle, $\alpha$.
Green crosses are observed temperatures.
Curves are modeled temperatures with best-fit parameters for four different cases: 
(1) a single value of the thermal inertia ($\Gamma_0$) is used for all particles in all scans (black solid curves), 
(2)  different scans can have different values of the thermal inertia $\Gamma_{\rm ind}$ while $A_{\rm V}$ and $f_{\rm fast}$
are fixed for all scans (blue dashed curves),
(3)  the thermal inertias for slow and fast rotators have different values,  $\Gamma_{\rm slow}$ and $\Gamma_{\rm fast}$, 
whose values are shared in all scans (red solid curves), and (4) the same as (3) except the size of fast rotators, $r_{\rm fast}$, 
is 5 mm (purple dotted curves). 
See Table~5 for other fitted parameters.
The values of $\Gamma$,  $\Gamma_{\rm slow}$, and $\Gamma_{\rm fast}$ shown in panels are in units of 
Jm$^{-2}$K$^{-1}$s$^{-1/2}$. The values for the reduced chi-square, $\chi^2$, for each scan are also shown in parentheses.
The mean solar phase angle, $\alpha$, is shown on the top left of each panel.

\item
Fig.~2. Same as Fig.~1 but for the B ring. 
The panels are placed on the order of the phase angle but separately for 
the high $B'$ scans (B1-B4) and the low $B'$ scans (B5-B8).
The purple dotted curve is the case of  $r_{\rm fast} =$ 5 mm for 
the high $B'$ scans and $r_{\rm fast} =$ 1 mm for the low $B'$ scans.

\item
Fig.~3. Azimuthal variations of the temperature (top) and the optical depth averaged over a footprint (bottom) 
for the Cassini division. The purple dotted curve is the case of  $r_{\rm fast} =$ 1 mm.

\item  
Fig.~4. Same as Fig.~1, but for the A ring. The purple dotted curve is the case of  $r_{\rm fast} =$ 5 mm.

\item  
Fig.~5. Bolometric Bond albedo, fraction of fast rotators, and thermal inertia for the Saturn's main rings for the base-line cases.
The same thermal inertia value for all particles is assumed in the derivation of 
 $A_{\rm V,0}$, $f_{\rm fast,0}$, and $\Gamma_0$ (black crosses), whereas
  different thermal inertias for slow and fast rotators are assumed in the derivation of 
 $A_{\rm V}$ (red diamonds), $f_{\rm fast}$ (red diamonds), $\Gamma_{\rm slow}$ (blue diamonds), and  $\Gamma_{\rm fast}$ (red triangles).
 See also Table~5. 

\item
Fig.~6. 
Contour of the reduced chi-square, $\Delta{\chi}_0^2$, sliced on the $A_{\rm V}$ vs. $\Gamma$ plane with the best-fit value of 
the fraction of fast rotators, $f_{\rm fast,0}$ (from Table~5). Note that the error bars are estimated using  the surface of $\Delta{\chi}_0^2=1.0$
in the three dimensional parameter space 
and displaying these sliced contours is only for the purpose of giving some insights on error bars at other confidence levels (the surfaces of 
$\Delta{\chi}_0^2=1.00, 2.71$, and $6.63$ give confidence levels of 68.3 $\%$, 90$\%$, and 99$\%$).

\item
Fig.~7. Thermal inertia $\Gamma_{\rm ind}$ vs.  phase angle $\alpha$ for the base-line cases. 
In the parameter fits, $A_{,\rm V}$ (= $A_{\rm V,0}$) and $f_{\rm fast} (= f_{\rm fast,0})$ are fixed for all scans in each ring,
but the thermal inertia is treated as a free parameter.  The solid lines represent linear fits of $\Gamma_{\rm ind}$ from all scans. 
The values of $\Gamma_{\rm ind}$ for the C and B rings
at $\alpha = 4.7-6.2^{\circ}$ are from Ferrari et al. (2005) (the offset in $\alpha$ between these two data is 
for avoiding overlaps of the marks but ground-based observations for both rings were done simultaneously in two different campaigns). 
The values of $\Gamma_{\rm ind}$ are also shown in Table~6. 

\item
Fig.~8. Contour of ring physical temperature on the plane of the vertical and azimuthal locations. 
The vertical coordinate $z$ normalized by the scale height $h$ is positive on the lit face.
Three cases with different combinations of $\tau$ and type of vertical motion are shown. 
The used parameters are as follows: the heliocentric distance is 9.1 AU, $B'=21^{\circ}$, $r_{\rm p}$ = 105,000 km, 
$A_{\rm V} = 0.6$, $f_{\rm fast} = 1.0$, and $\Gamma = 13$ Jm$^{-2}$K$^{-1}$s$^{-1/2}$. 
The vertical dashed lines represent shadow lines.

\item
Fig.~9. Contour of the normalized reduced chi-square, $\Delta{\chi}_1^2$, sliced on the $\Gamma_{\rm slow}$ vs. $\Gamma_{\rm fast}$ 
plane with the best-fit values of $A_{\rm V}$ and $f_{\rm fast}$ (from Table~5). Note that the error bars are estimated  
in the four dimensional parameter space and displaying these sliced contours is only for the purpose of giving some insights on error 
bars at various confidence levels.

\item
Fig.~10. Lower limit of size of fast rotators $r_{\rm fast,min}$ vs. lower limit of thermal inertia of fast rotators $\Gamma_{\rm fast,min}$.
The size $r_{\rm fast}$ is fixed to be 10 cm for the estimation of $\Gamma_{\rm fast,min}$.
The lines represent theoretical predictions (Eq.~(\ref{eq:r2gm})) 
for three different illumination cycles, $2\pi/\omega$. 
The values of $\Gamma_{\rm fast,min}$ for the B ring at low $B'$ and the Cassini division are not well estimated and assumed to be 
1 Jm$^{-2}$K$^{-1}$s$^{-1/2}$.

\item
Fig.~11. Comparison of fitted parameters between this work and Paper II. The solid curves are from Paper II (dotted curves are error sizes) 
and diamonds are from this work. The scale hight ratio $h_{\rm r}$ is assumed to be 3 except $h_{\rm r}=1$ for the B ring.
The thermal inertia of all particles is assumed to be the same.
The red lines and marks are for the case with standard sinusoidal vertical motion without bouncing whereas the black lines and marks
are for the case with cycloidal motion which assumes particle bouncing at the midplane. 
The values for the B ring from this work are the mean values of high $B'$ and low $B'$. The radial offset between 
the red and black diamonds for the B ring is for avoiding overlaps of the marks.

\item
Fig.~12. Thermal inertia as a function of fraction of bouncing particles $f_{\rm b}$ for the mid B ring. 
The solid lines are  the thermal inertia estimated from this study, $\Gamma_0$, and the dashed 
lines are the thermal inertia estimated in Paper II, $\Gamma_{\rm PII}$. Two cases of 
optical depth profiles, $\tau_{\rm PPS}$ and $\tau_{\rm UVIS}$, are considered in both works.
 The value of $\Gamma_{\rm PII}$ for the case with $\tau_{\rm UVIS}$ and $f_{\rm b} = 1$ is uncertain due to the limited coverage 
 in parameter fits in Paper II,  but at least larger than 40 Jm$^{-2}$K$^{-1}$s$^{-1/2}$.  
The offsets in $f_{\rm b}$ are for avoiding overlaps of the marks: the simulations were made for either $f_{\rm b} = 0$ or 1.

\item
Fig.~13. Thermal inertia vs. fraction of fast rotators for the Saturn's main rings. 
Black bars represent the range of these parameters directly estimated in model fits to observed data ($\Gamma_0$ and $f_{\rm fast,0}$).
The red and blue curves are relations between these two parameters predicted from Eqs.~(\ref{eq:tspin}) - (\ref{eq:ff2}) with the size distribution 
from French and Nicholson (2000) (Table~7).  

\end{description}

\clearpage

\begin{figure}
\begin{center}
\includegraphics[width=.49\textwidth]{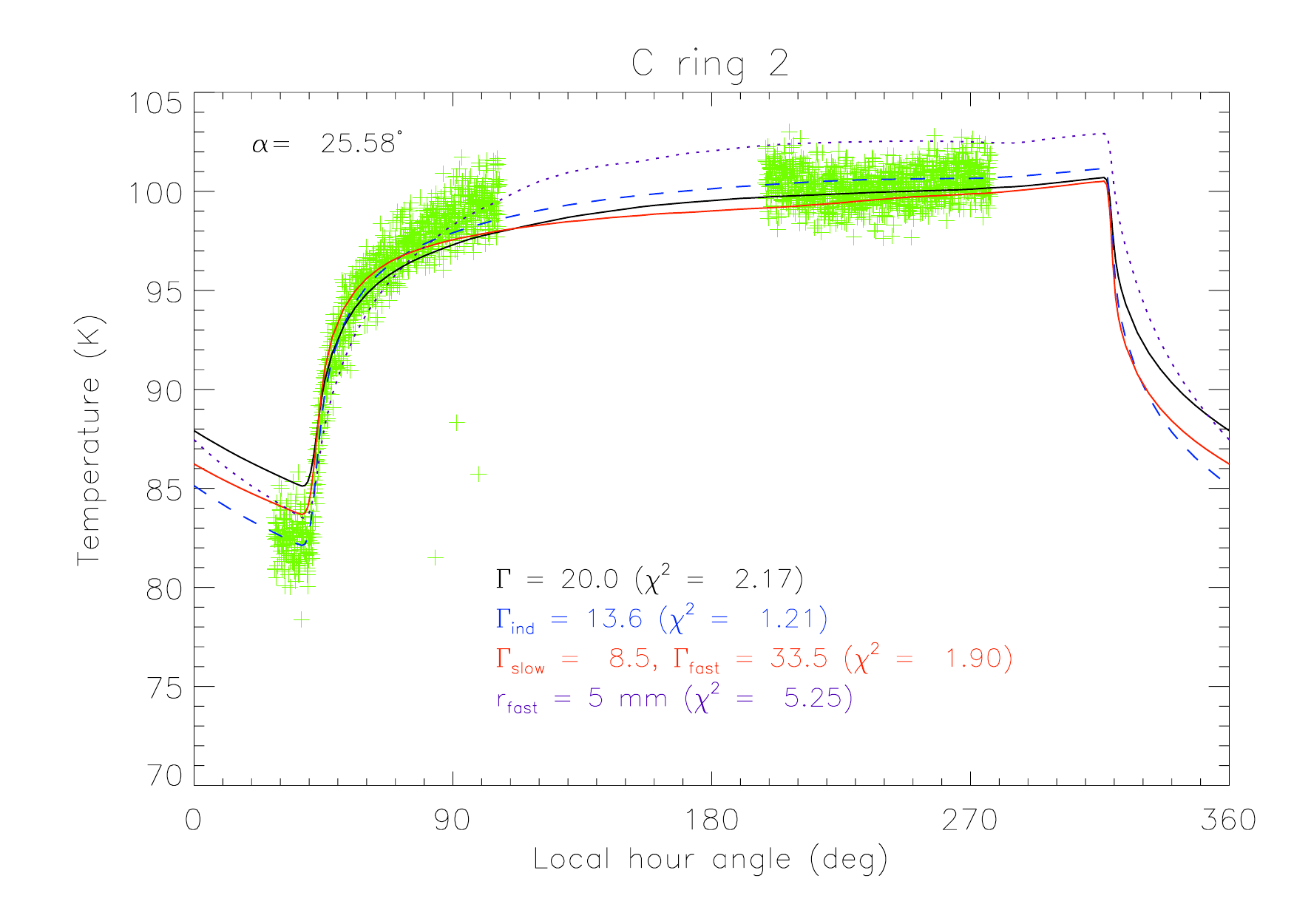}\includegraphics[width=.49\textwidth]{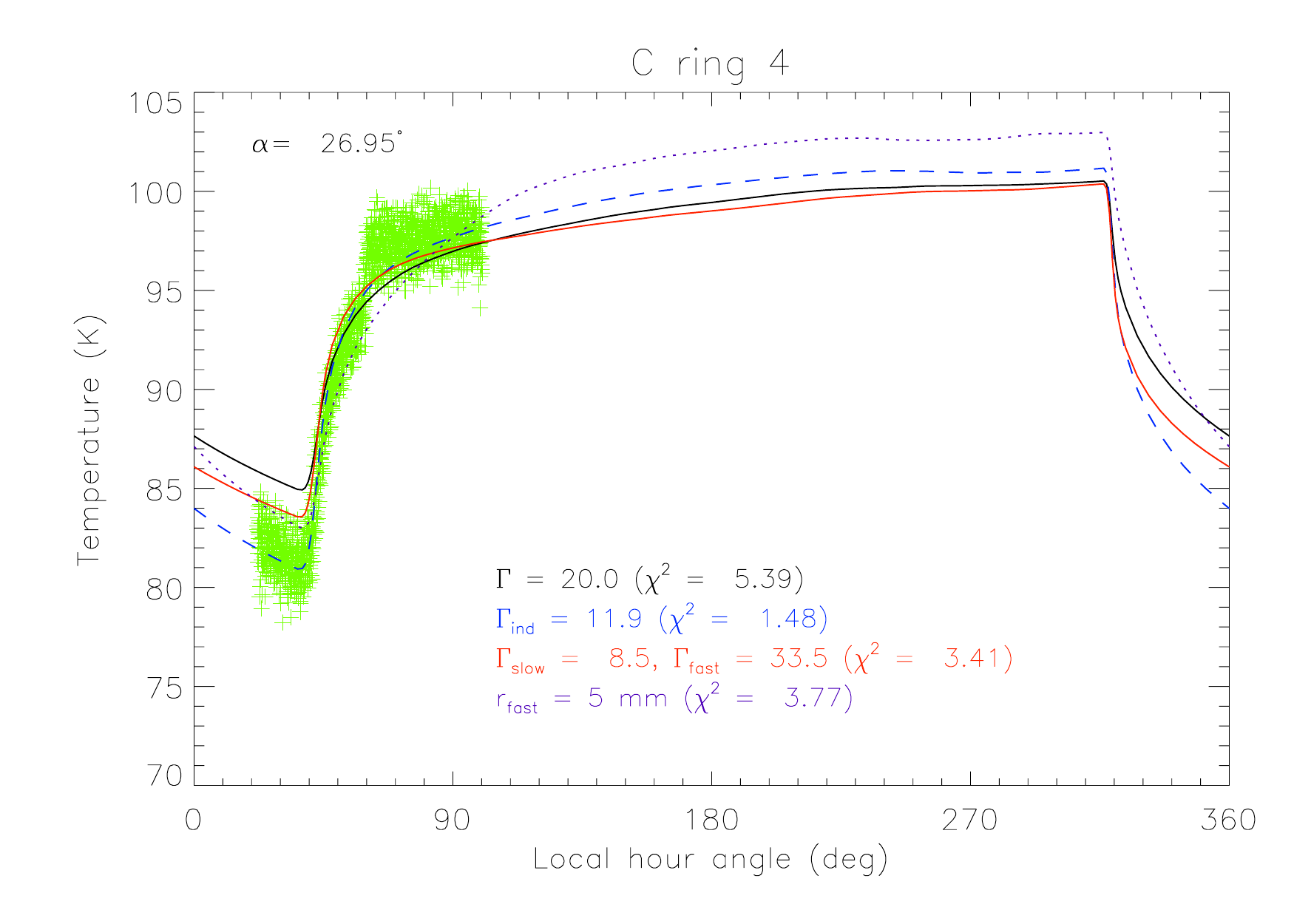}

\includegraphics[width=.49\textwidth]{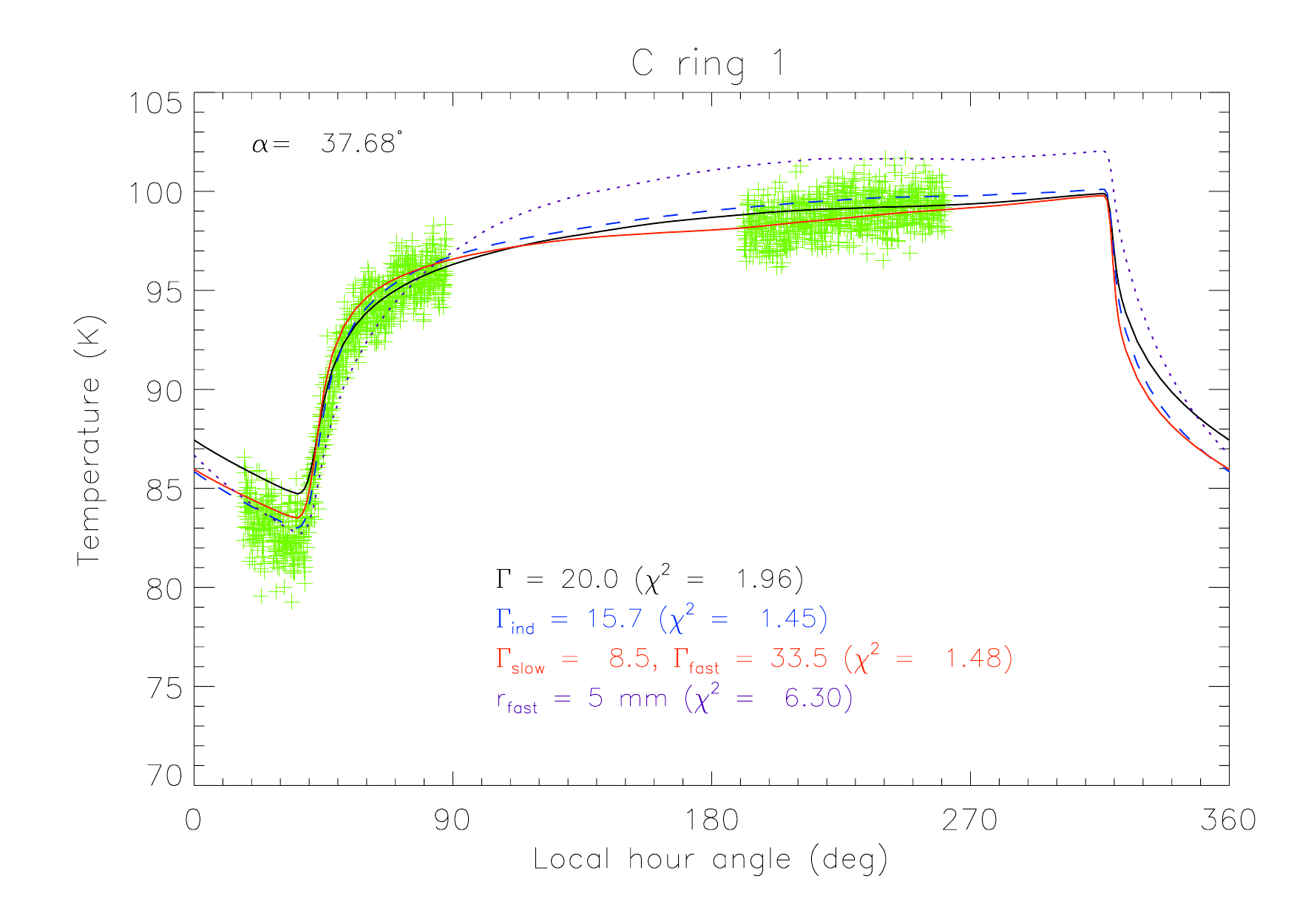}\includegraphics[width=.49\textwidth]{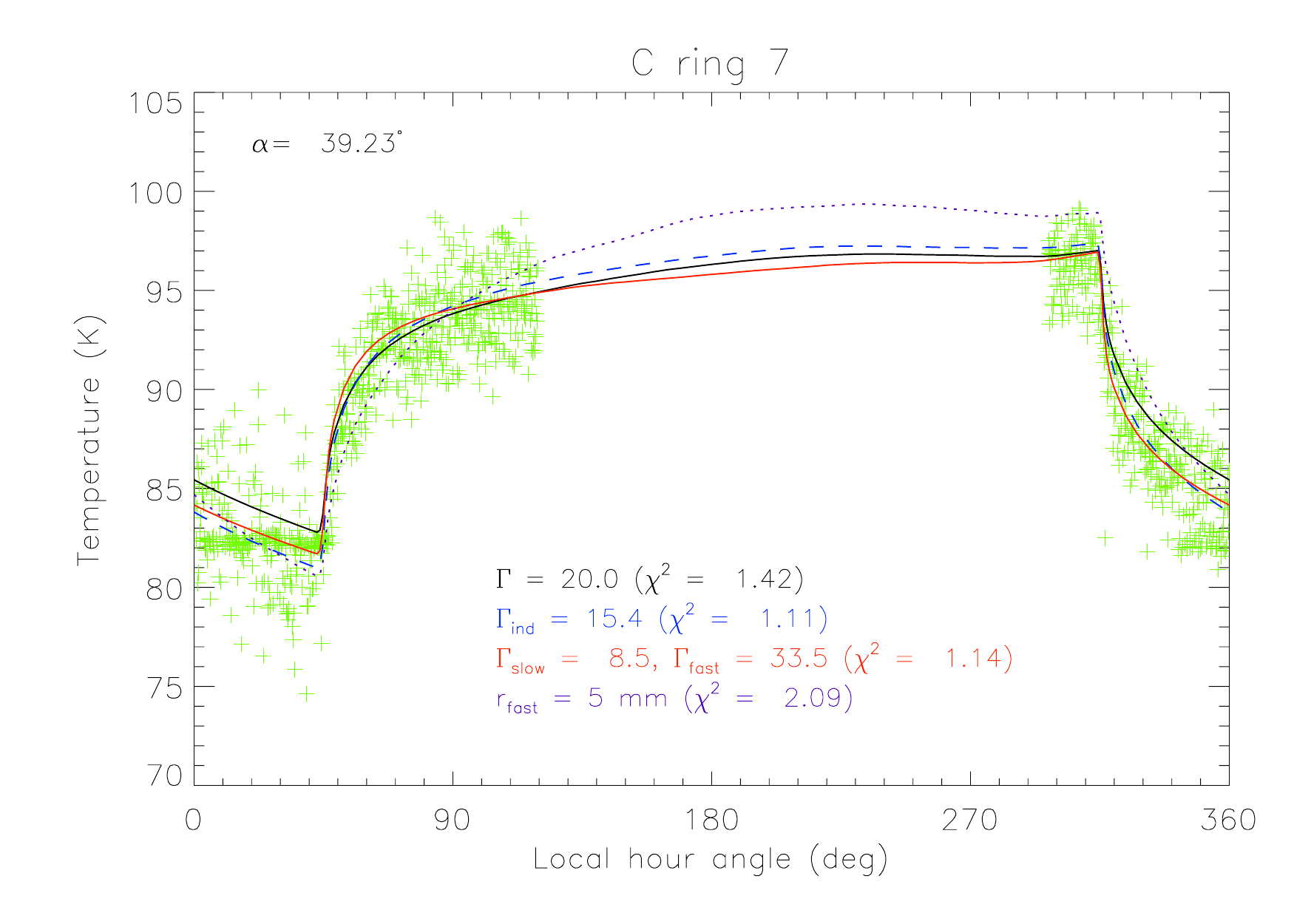}

\end{center}
Fig.~1. Morishima et al. 
\end{figure}

\clearpage

\begin{figure}
\includegraphics[width=.49\textwidth]{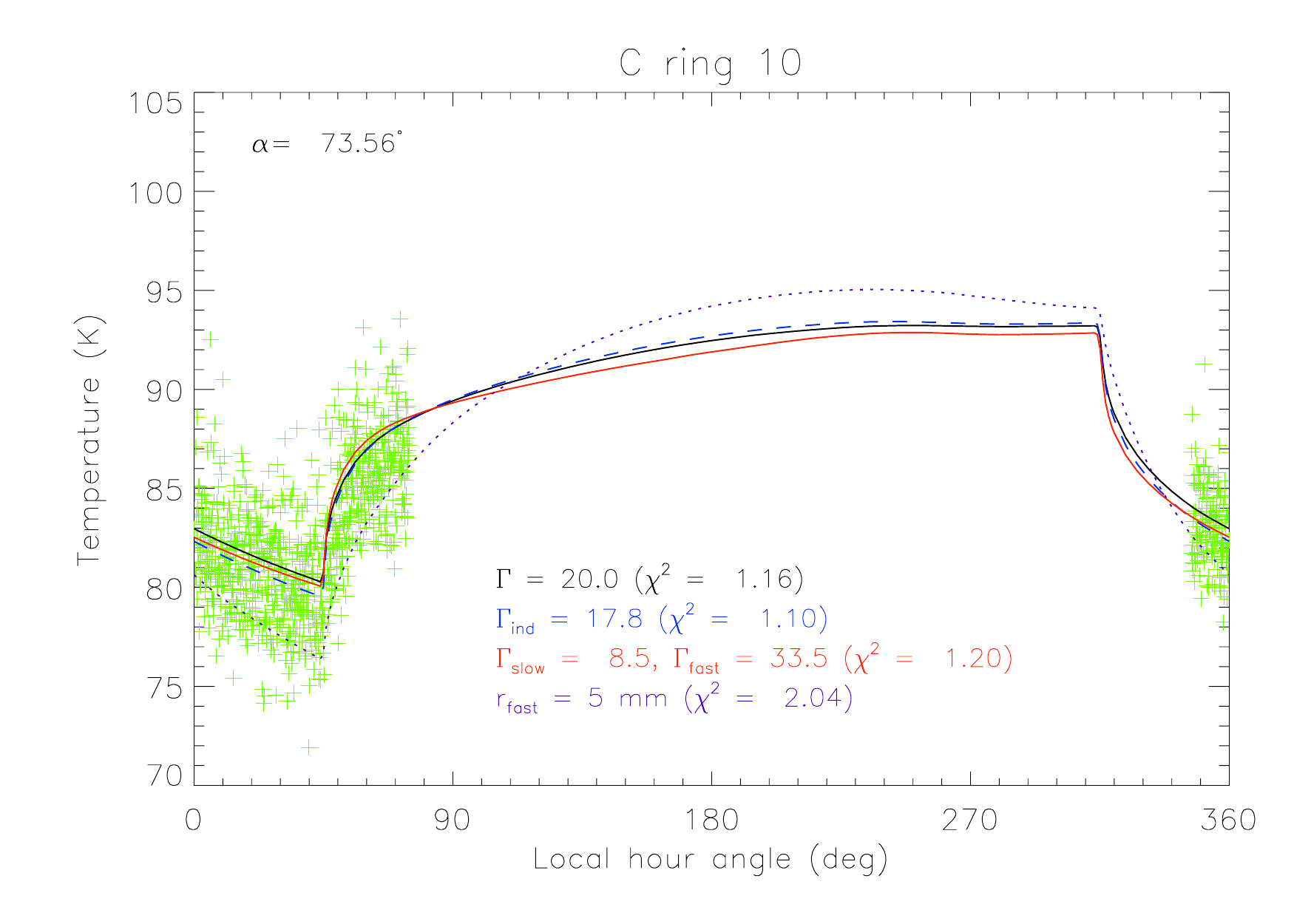}\includegraphics[width=.49\textwidth]{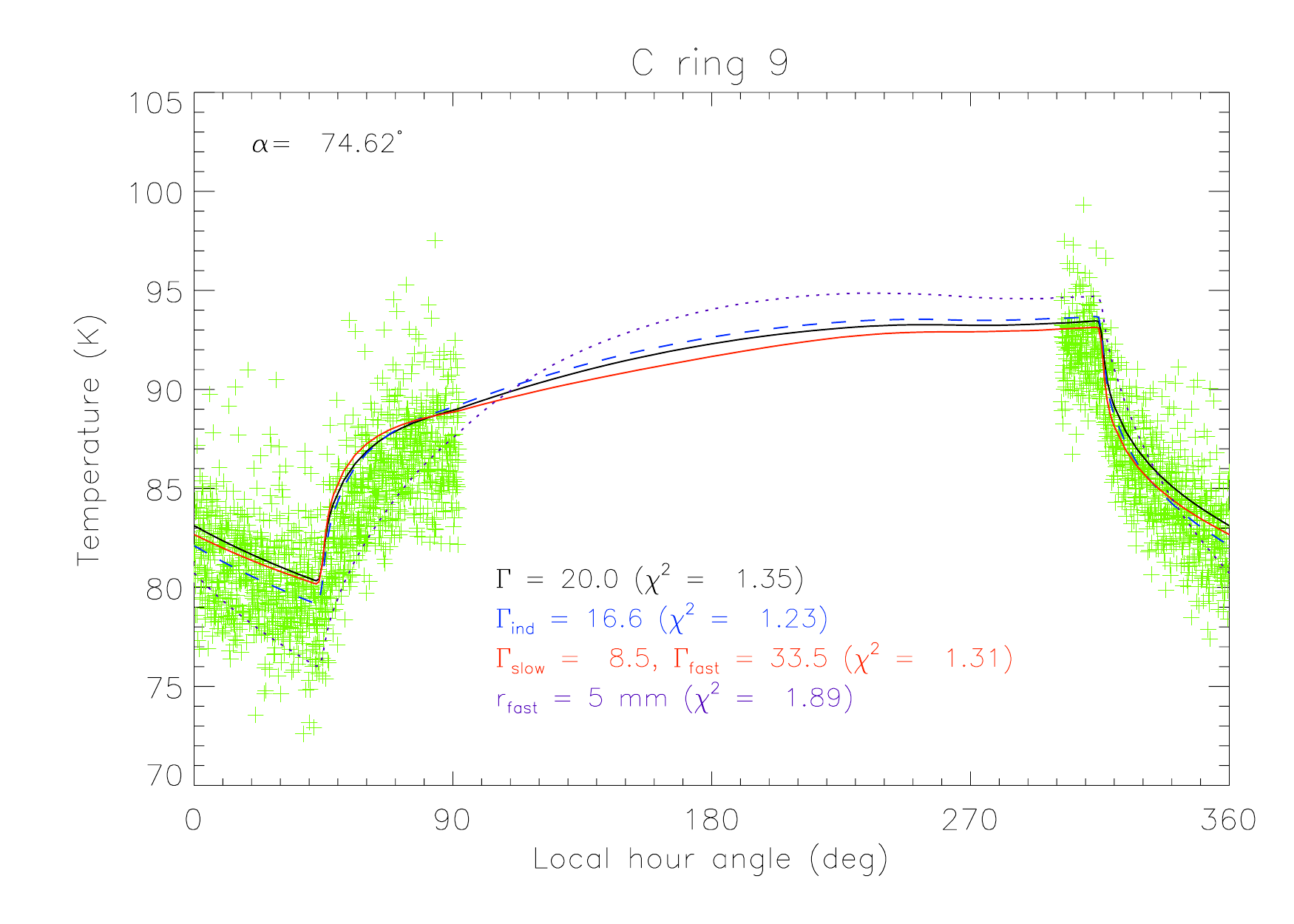}

\includegraphics[width=.49\textwidth]{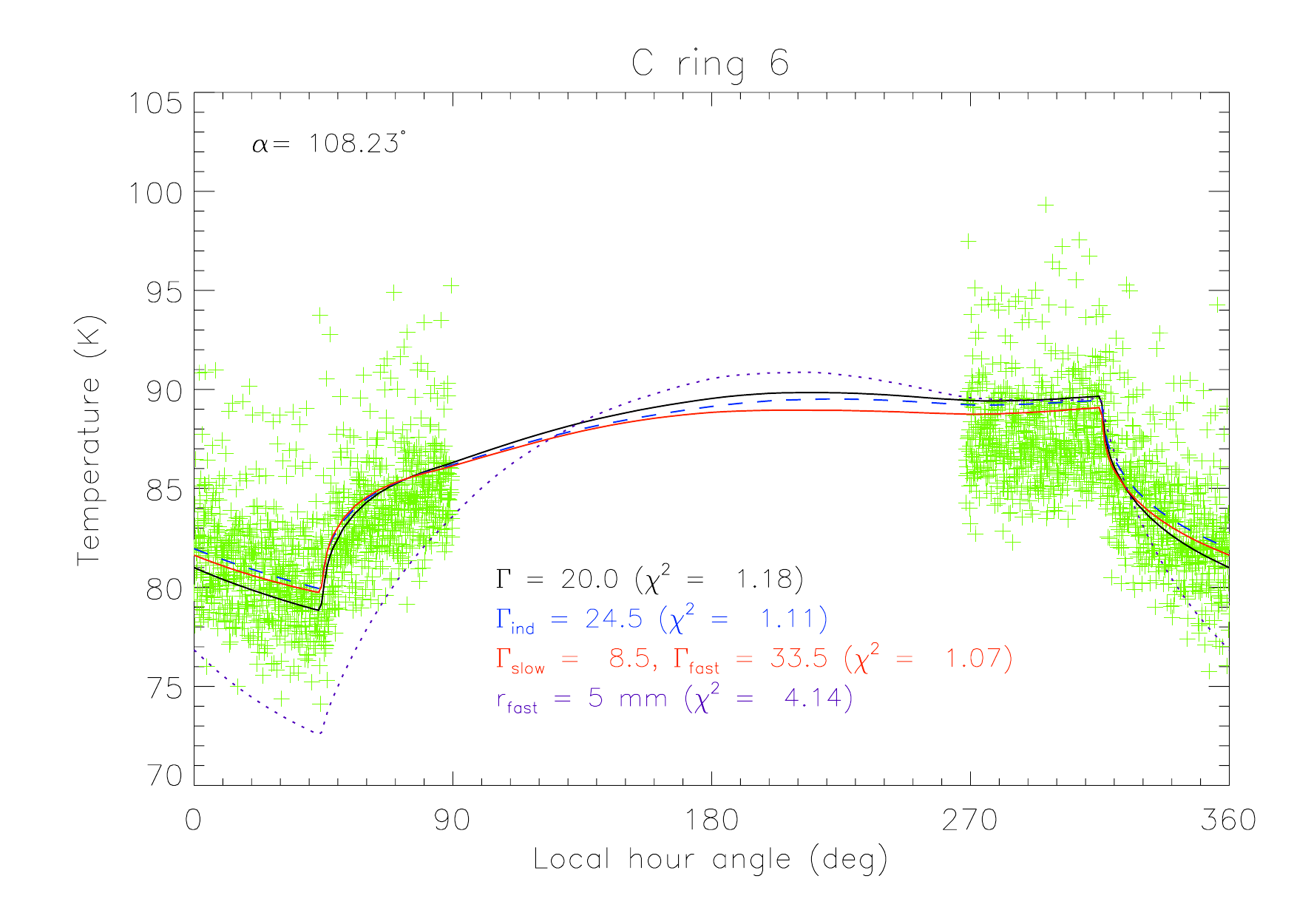}\includegraphics[width=.49\textwidth]{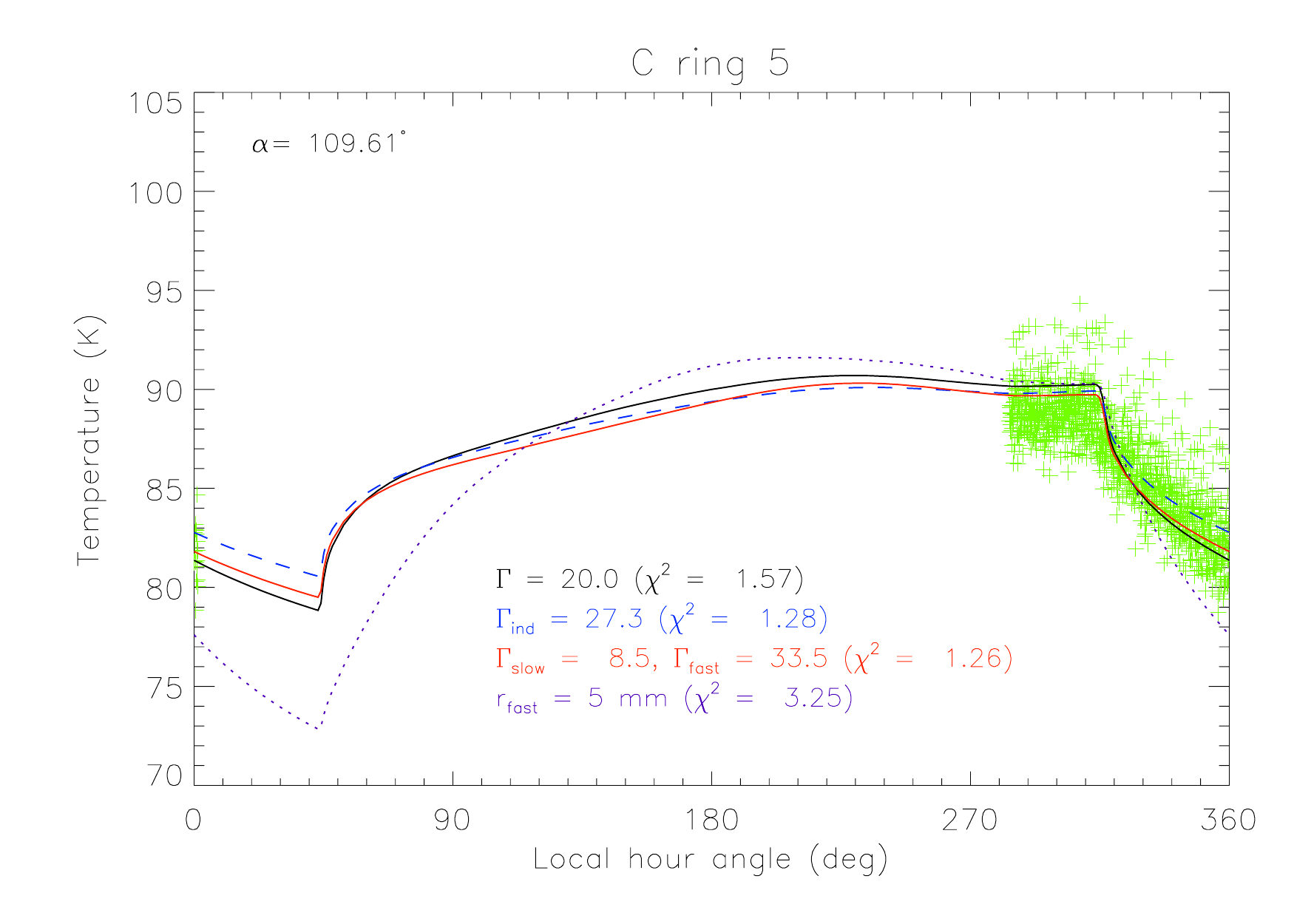}

\includegraphics[width=.49\textwidth]{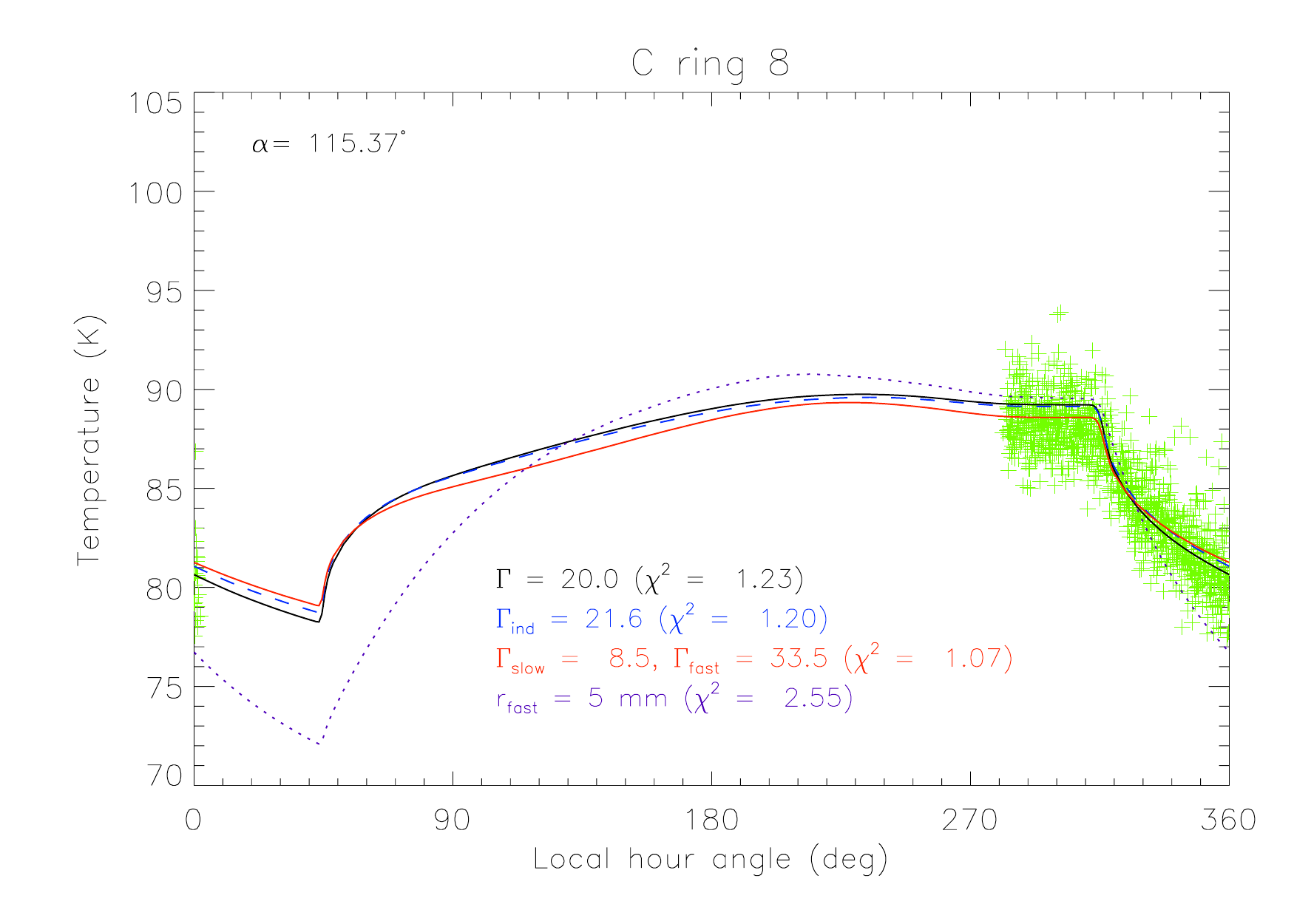}\includegraphics[width=.49\textwidth]{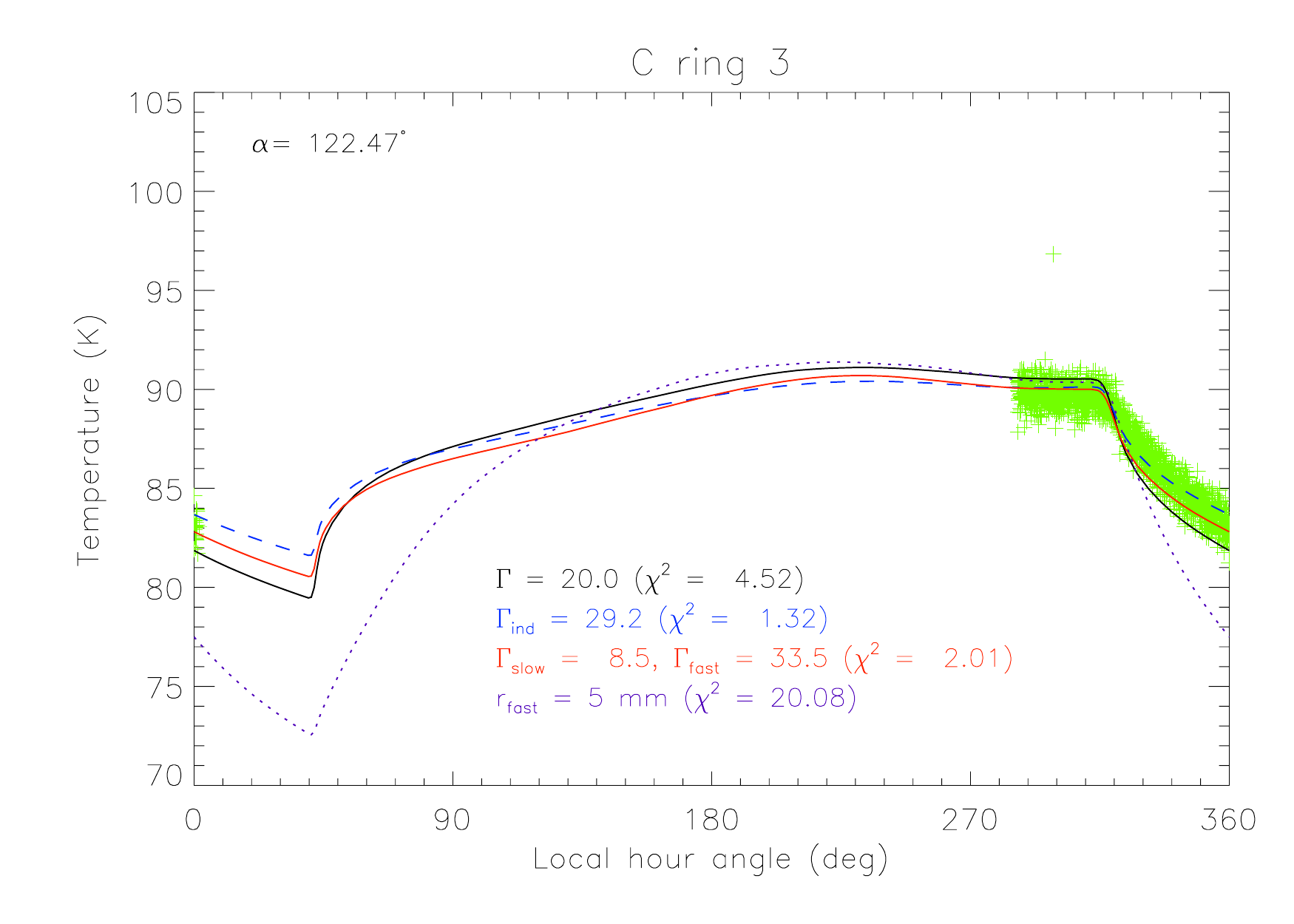}

Fig.~1-continued.

\end{figure}

\clearpage

\begin{figure}
\includegraphics[width=.49\textwidth]{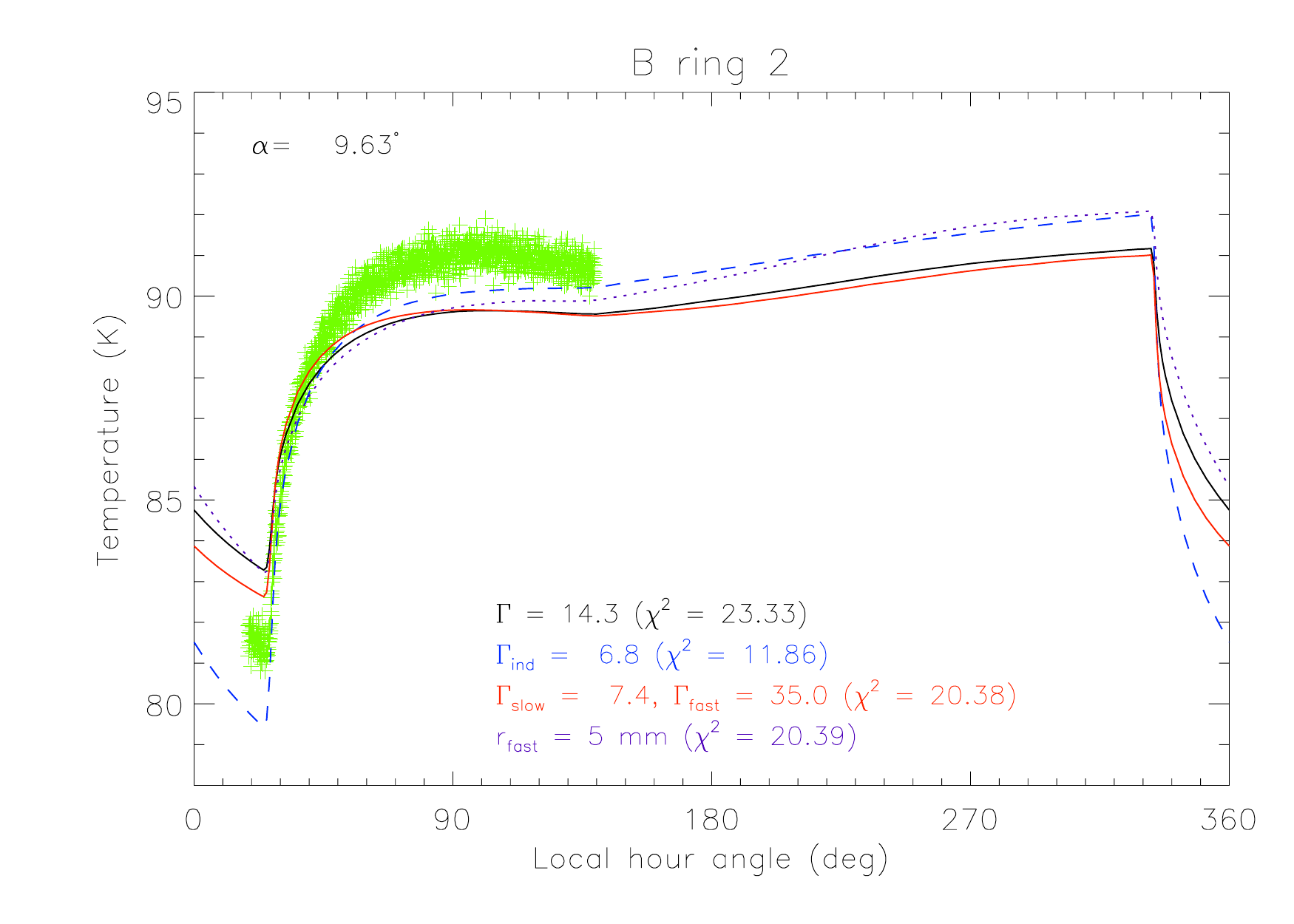}\includegraphics[width=.49\textwidth]{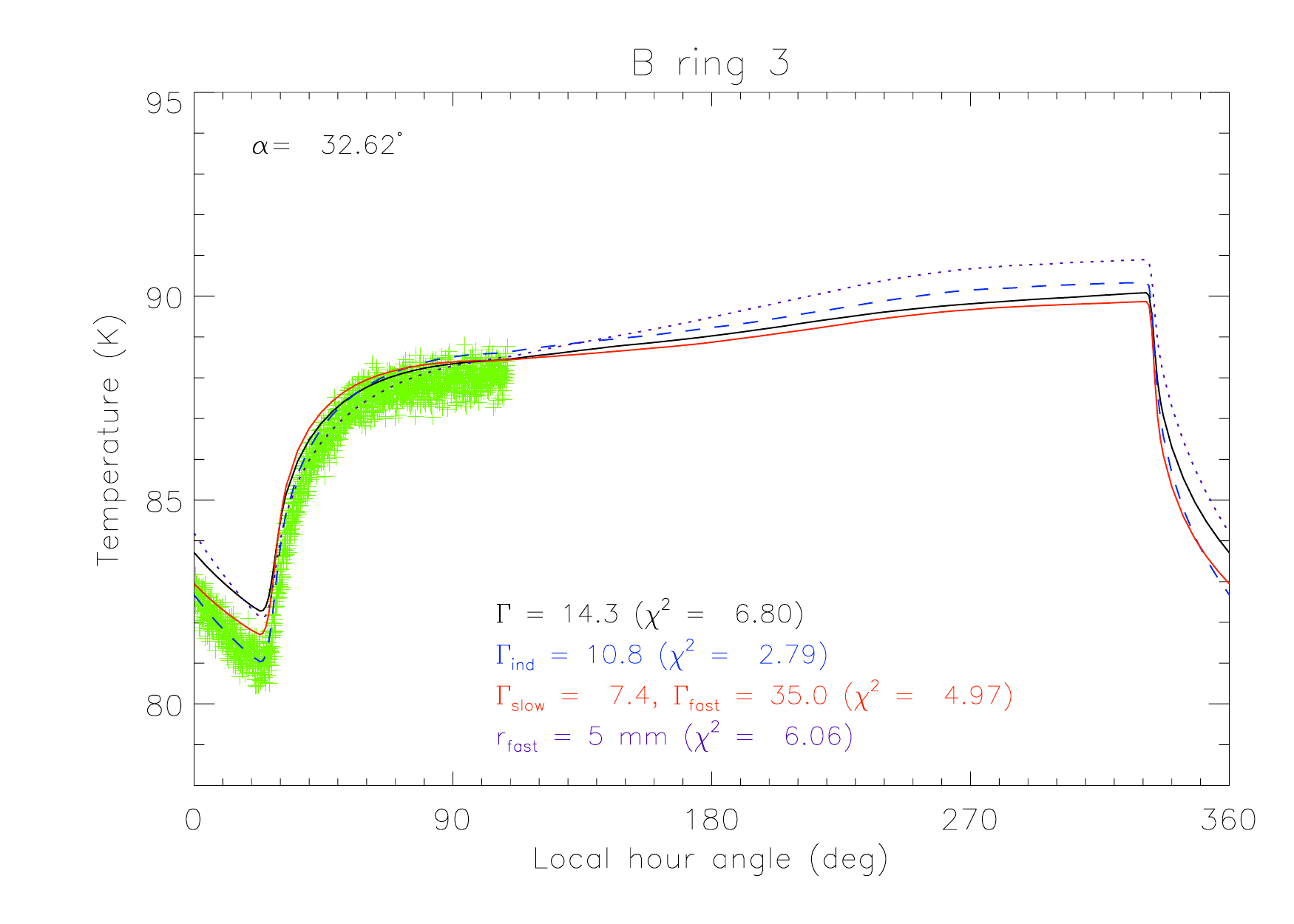}

\includegraphics[width=.49\textwidth]{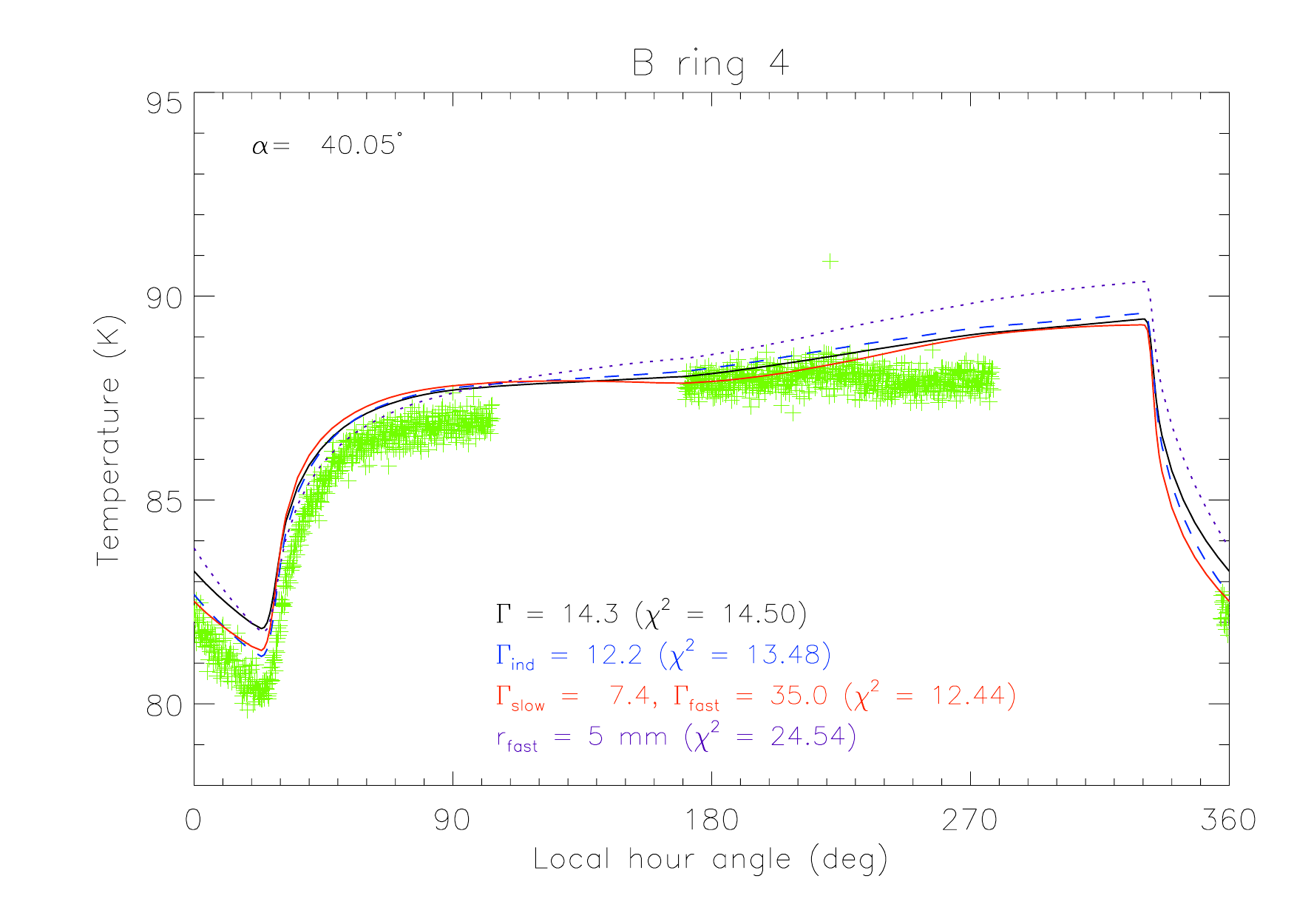}\includegraphics[width=.49\textwidth]{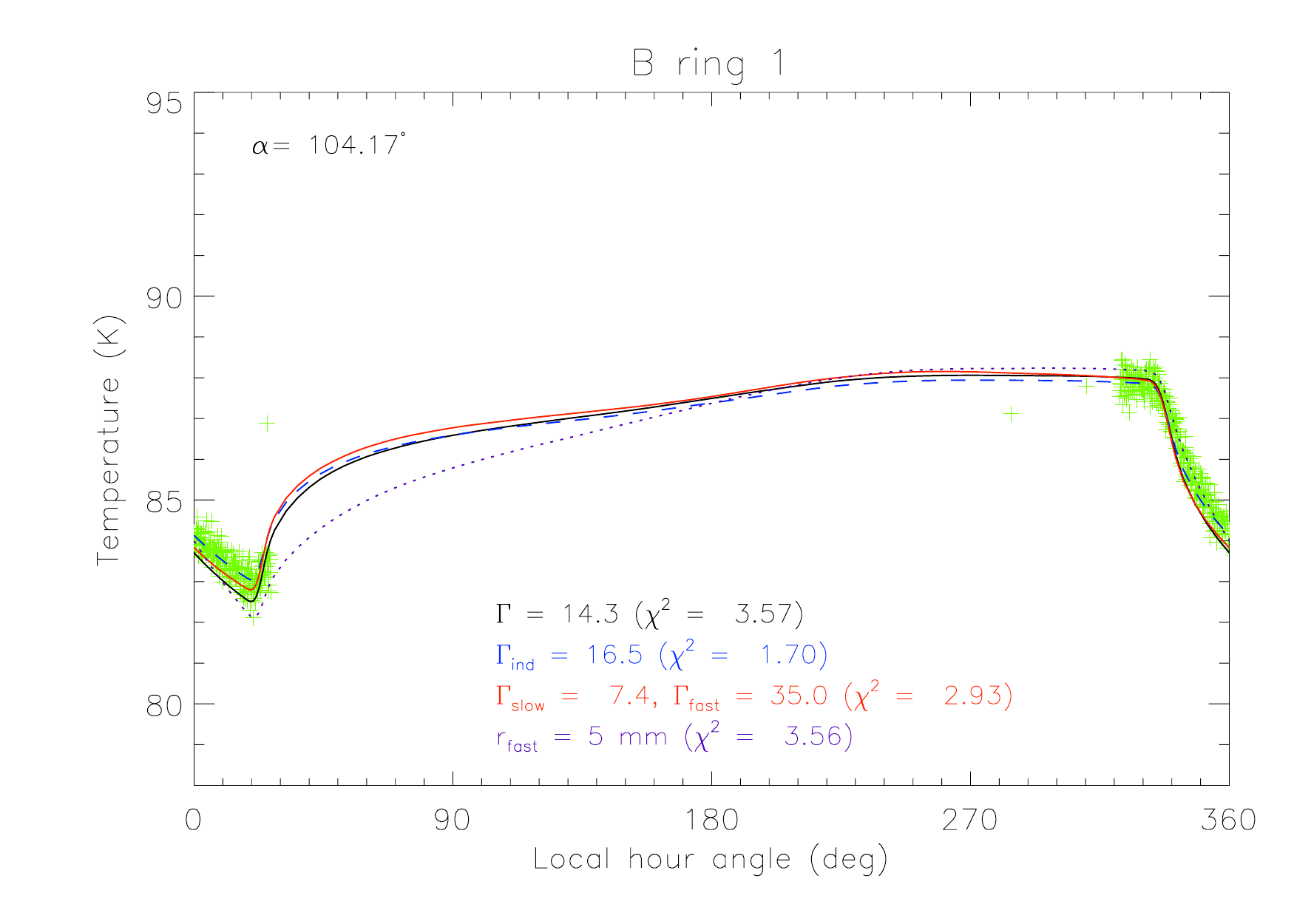}

Fig.~2. Morishima et al.

\end{figure}

\clearpage

\begin{figure}
\includegraphics[width=.49\textwidth]{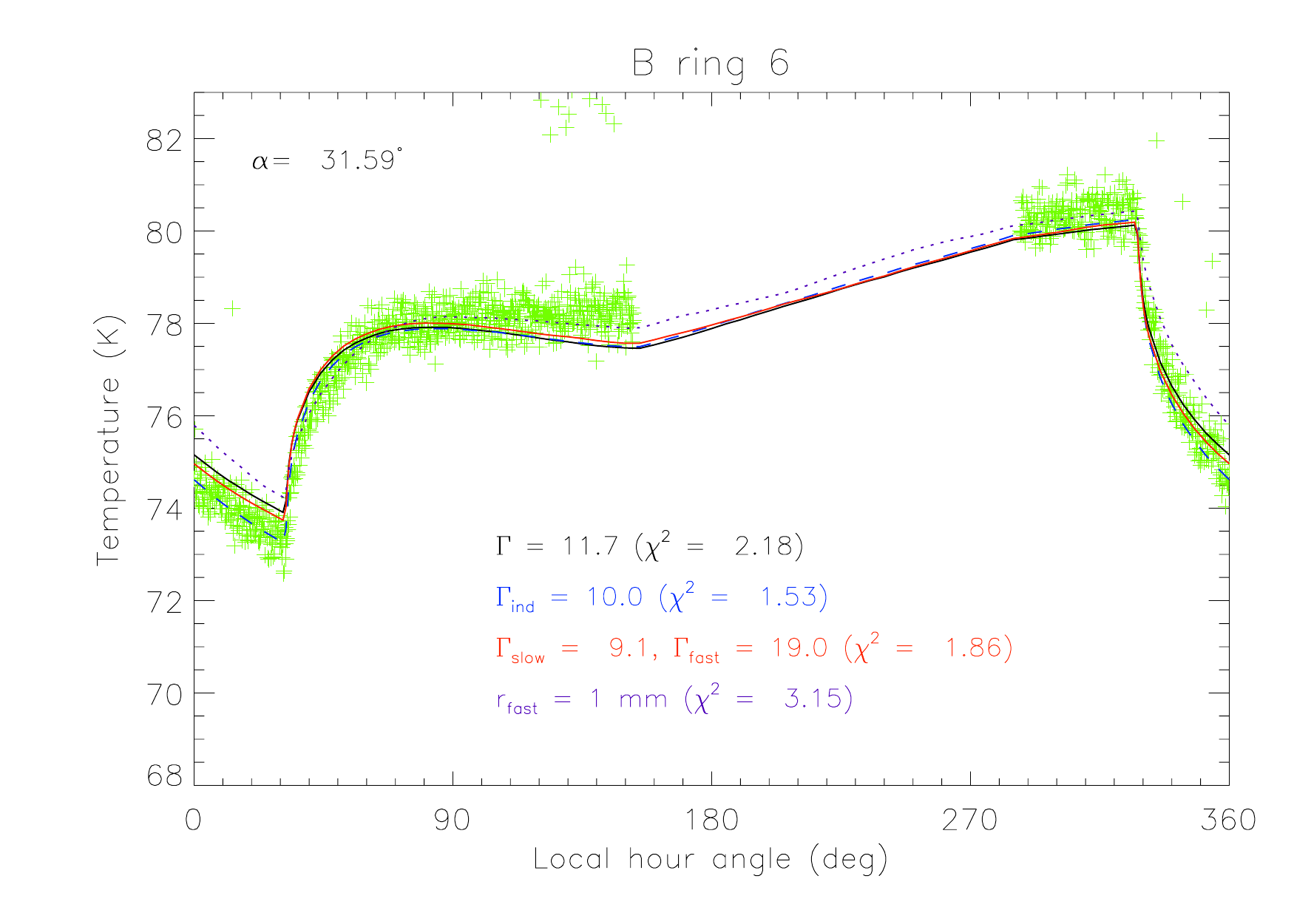}\includegraphics[width=.49\textwidth]{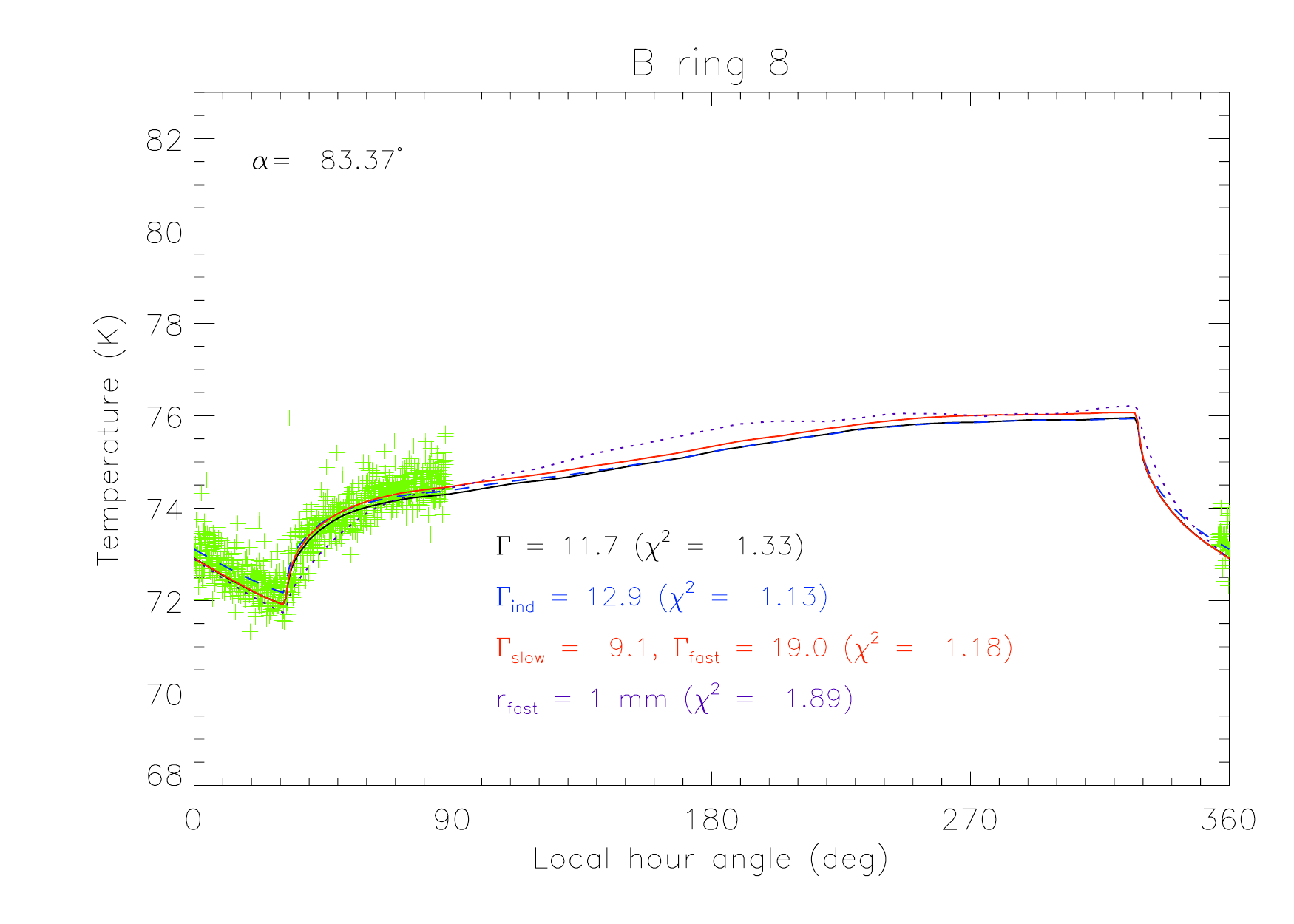}

\includegraphics[width=.49\textwidth]{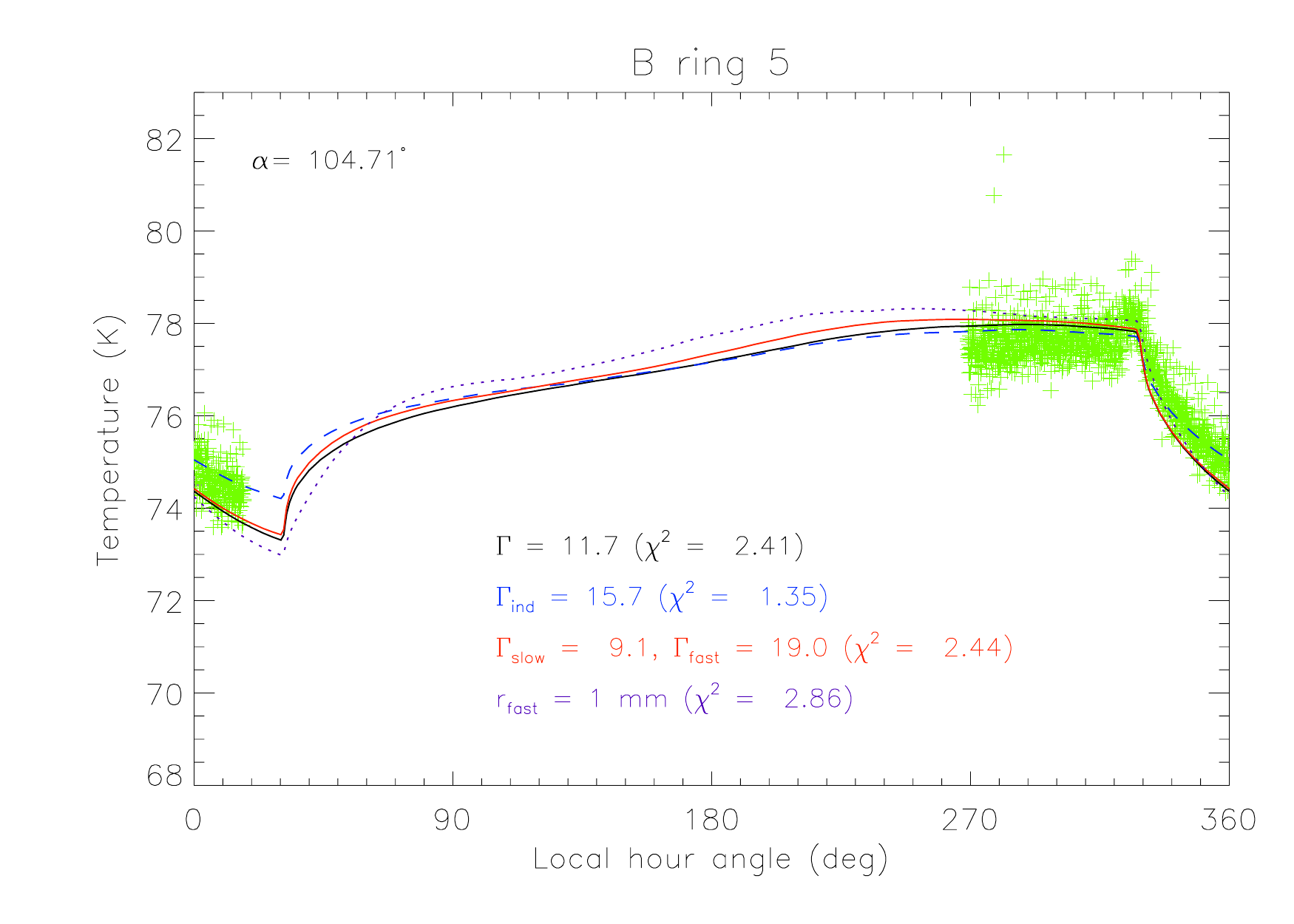}\includegraphics[width=.49\textwidth]{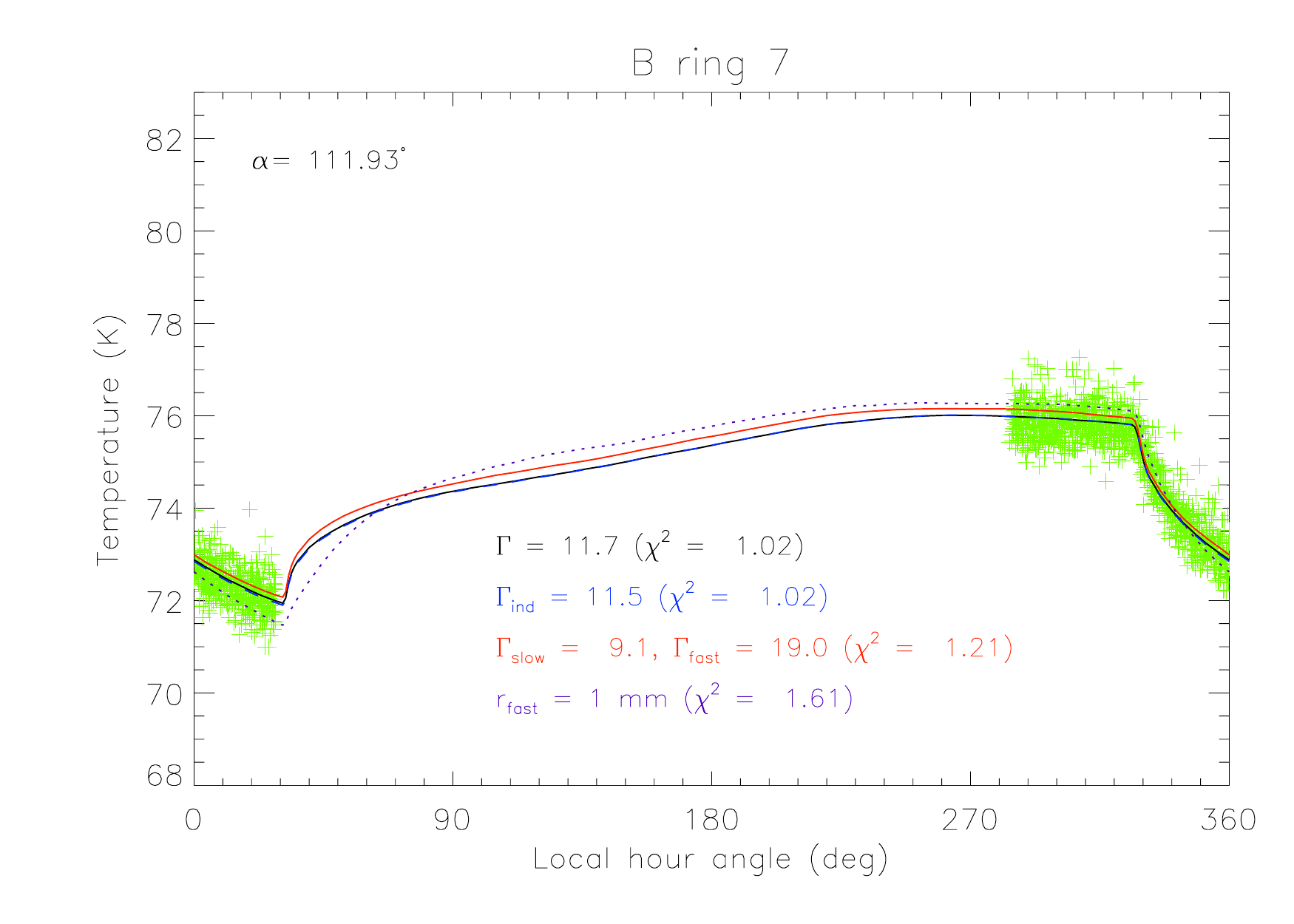}

Fig.~2-continued.

\end{figure}

\clearpage

\begin{figure}
\includegraphics[width=.49\textwidth]{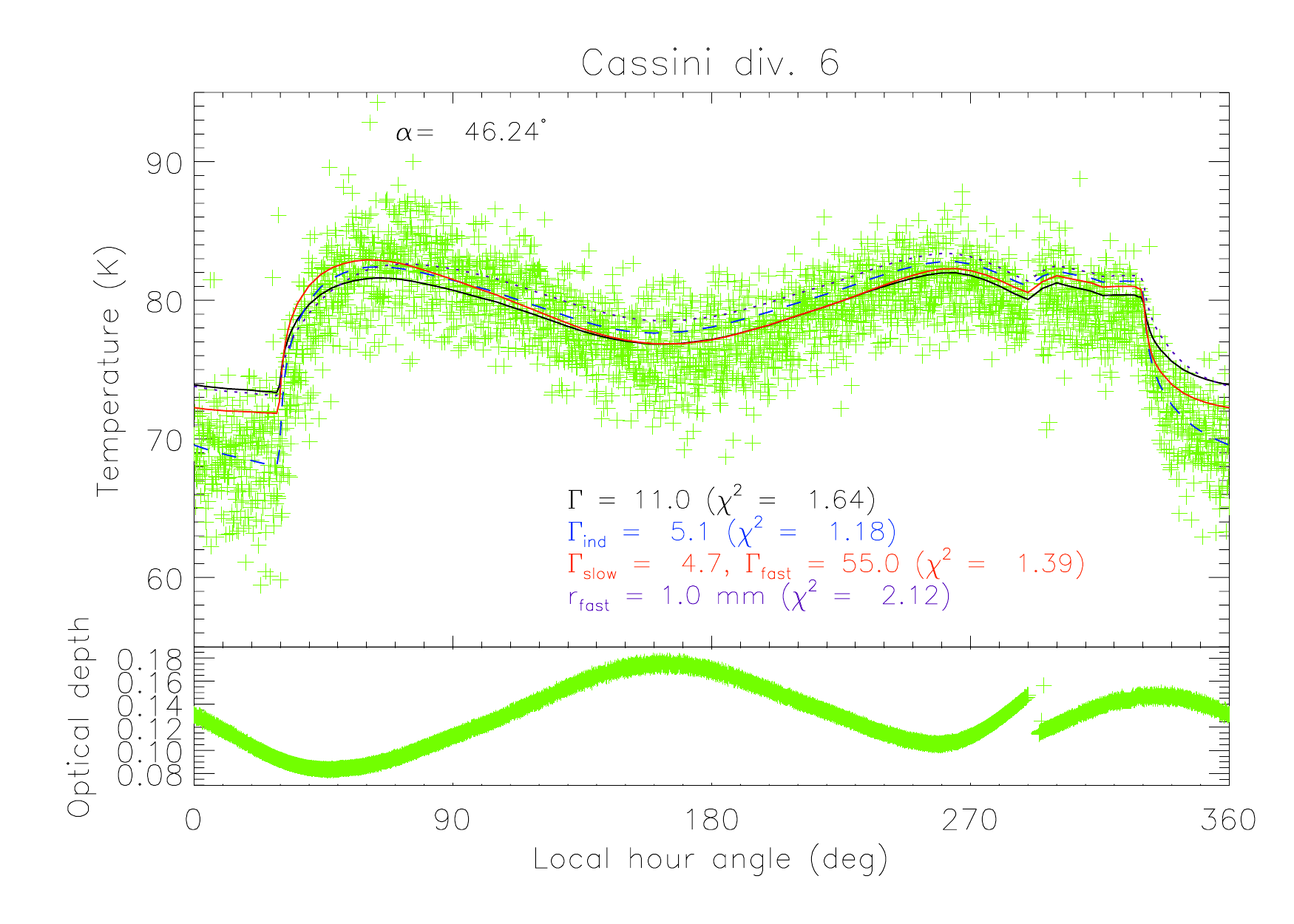}\includegraphics[width=.49\textwidth]{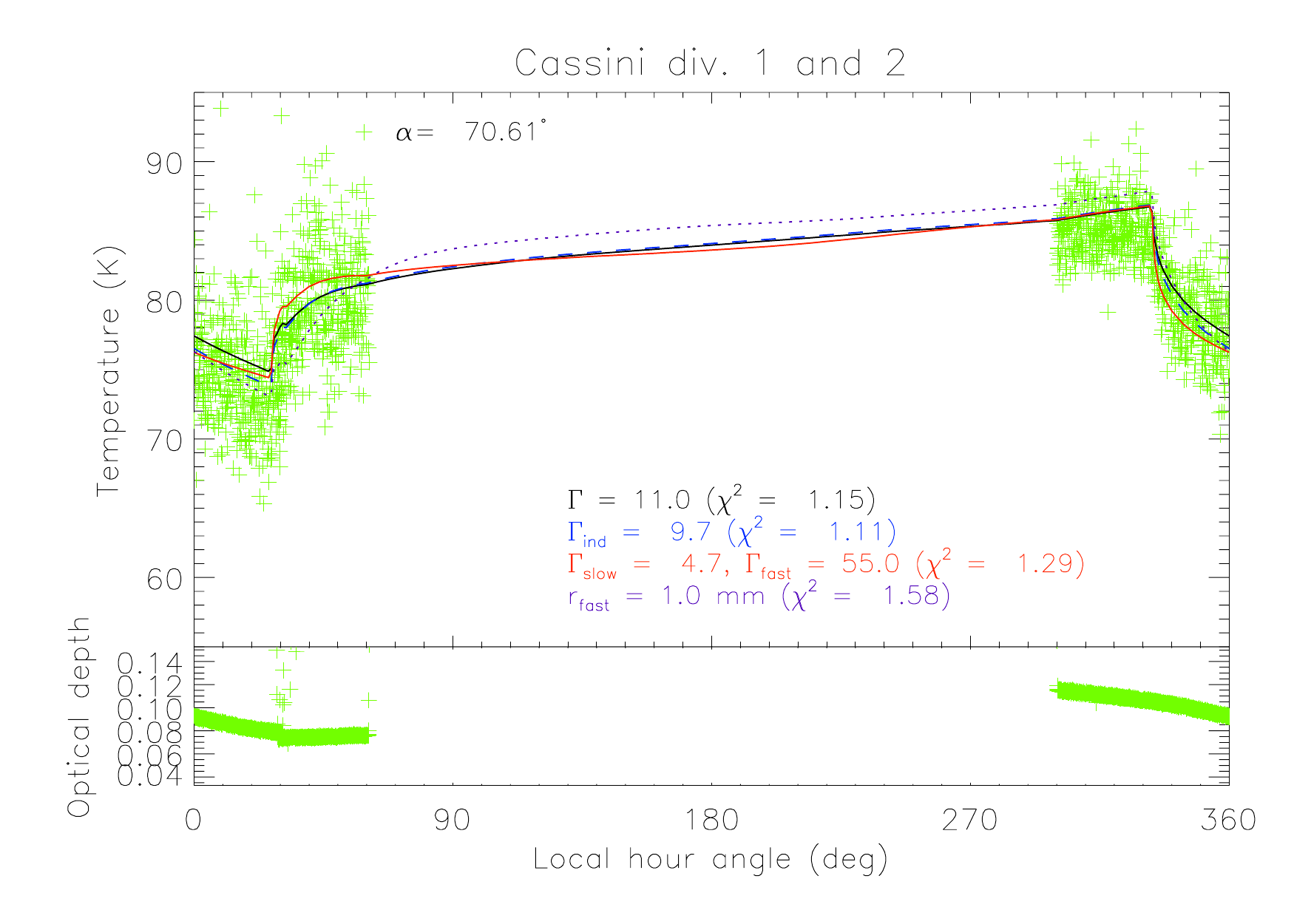}

\includegraphics[width=.49\textwidth]{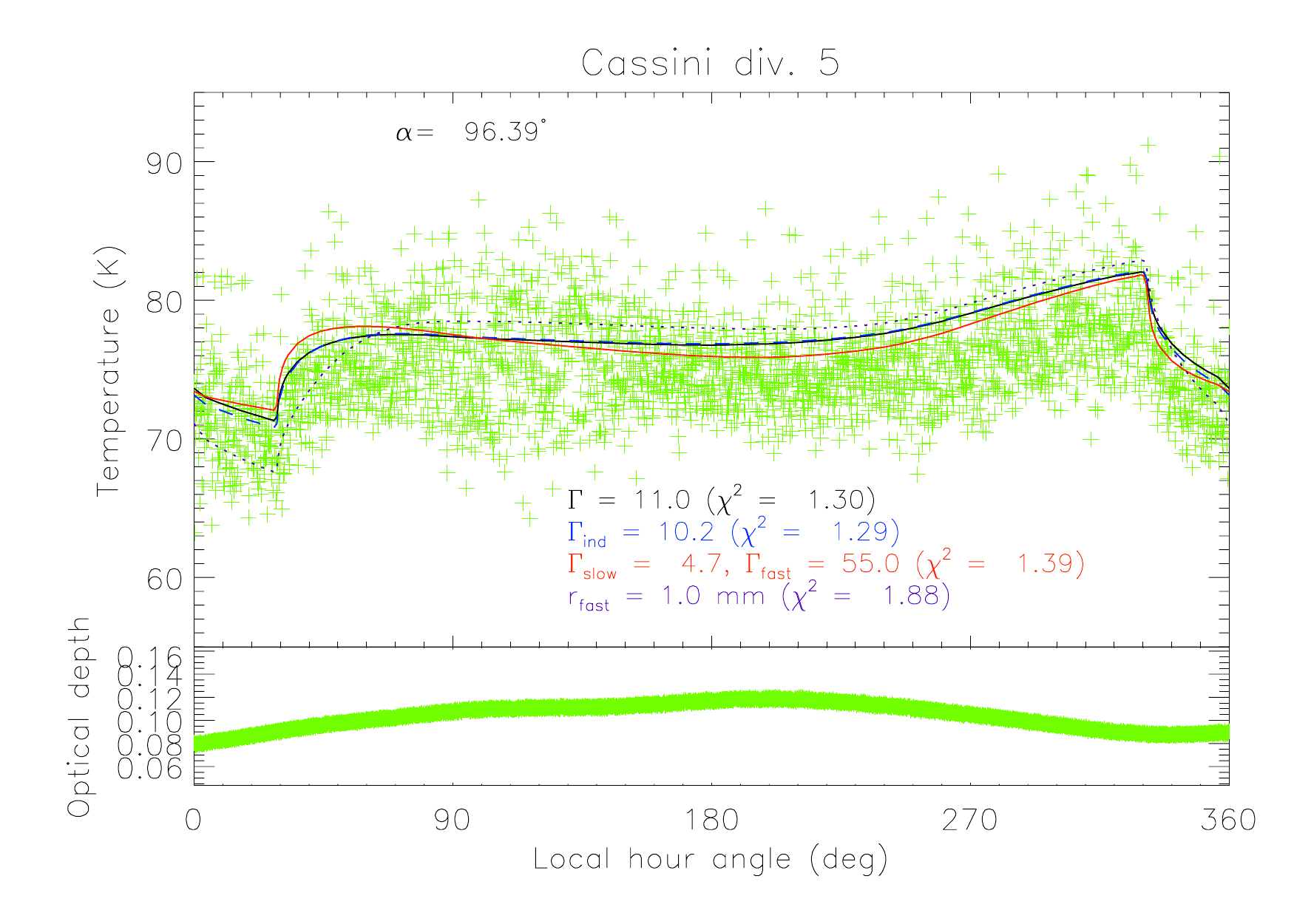}\includegraphics[width=.49\textwidth]{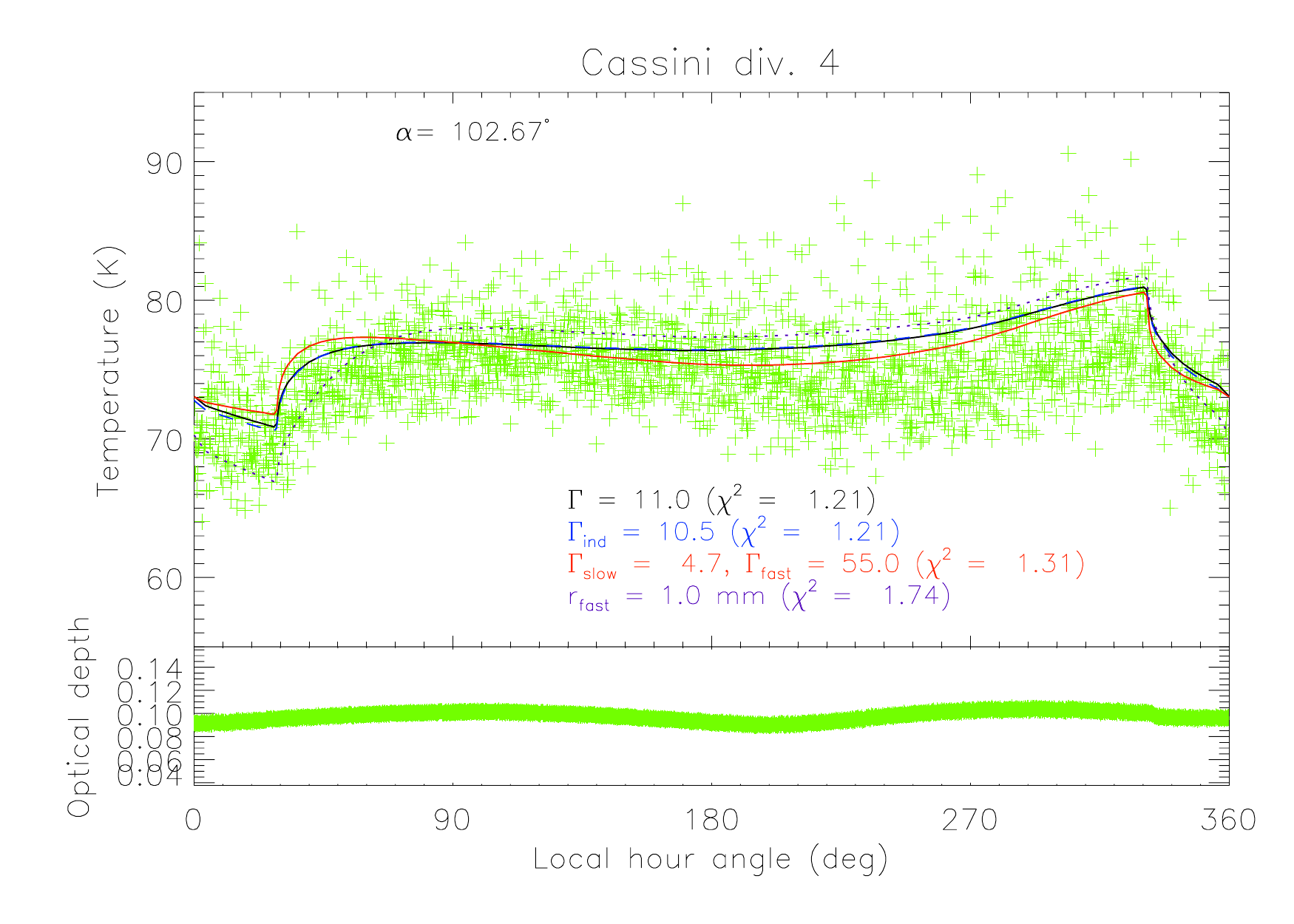}

\includegraphics[width=.49\textwidth]{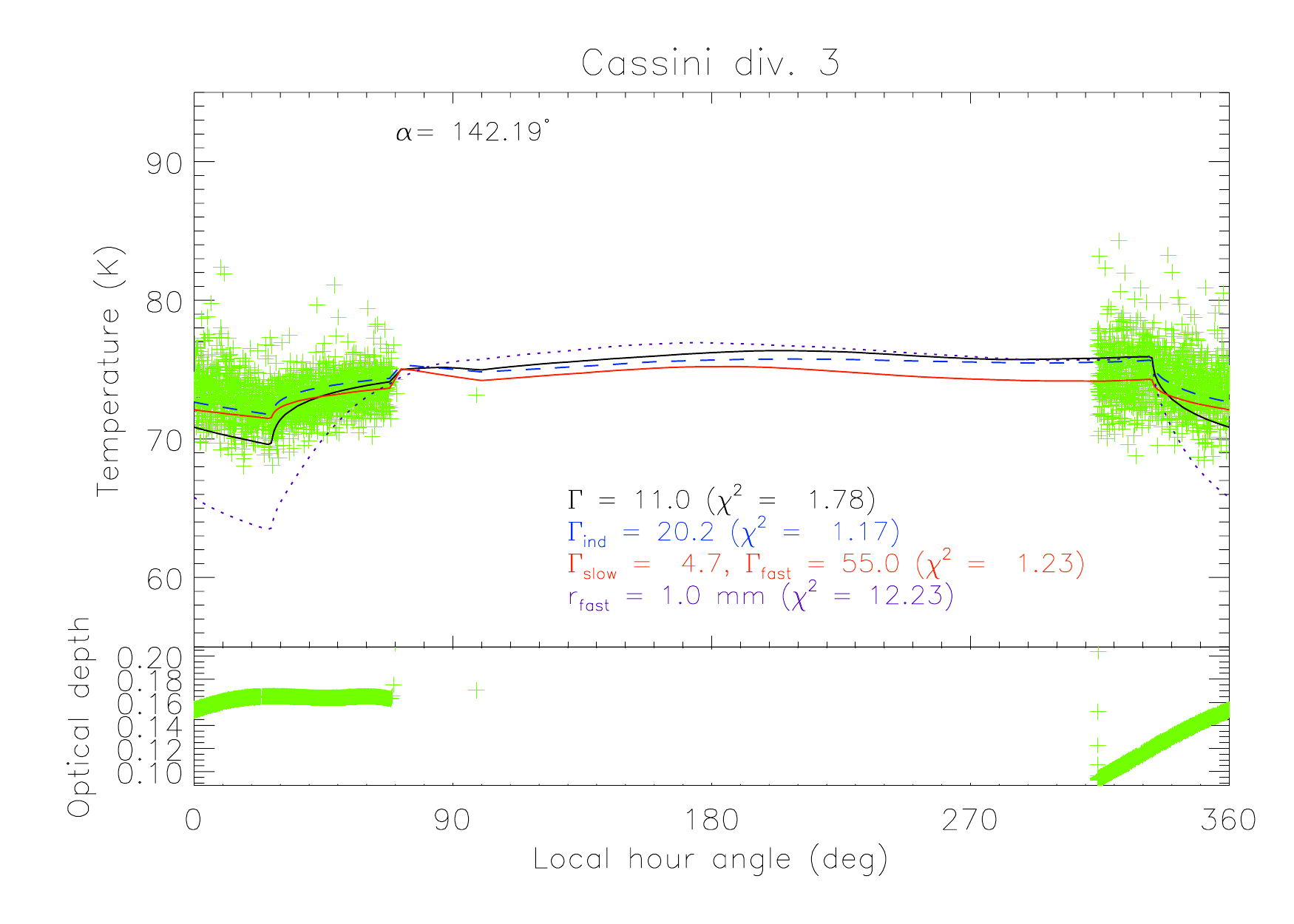}

Fig.~3. Morishima et al.

\end{figure}

\clearpage

\begin{figure}

\includegraphics[width=.49\textwidth]{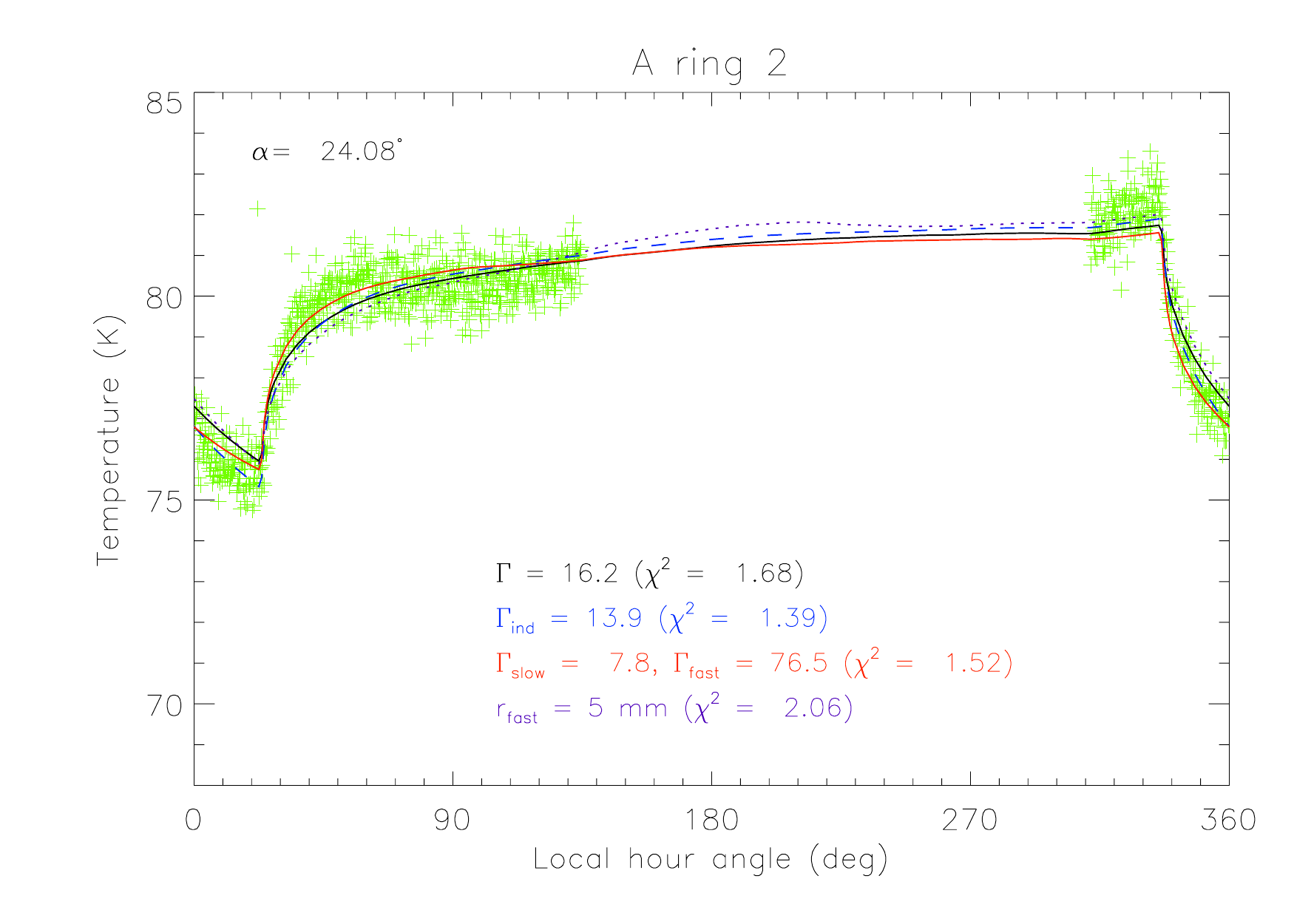}\includegraphics[width=.49\textwidth]{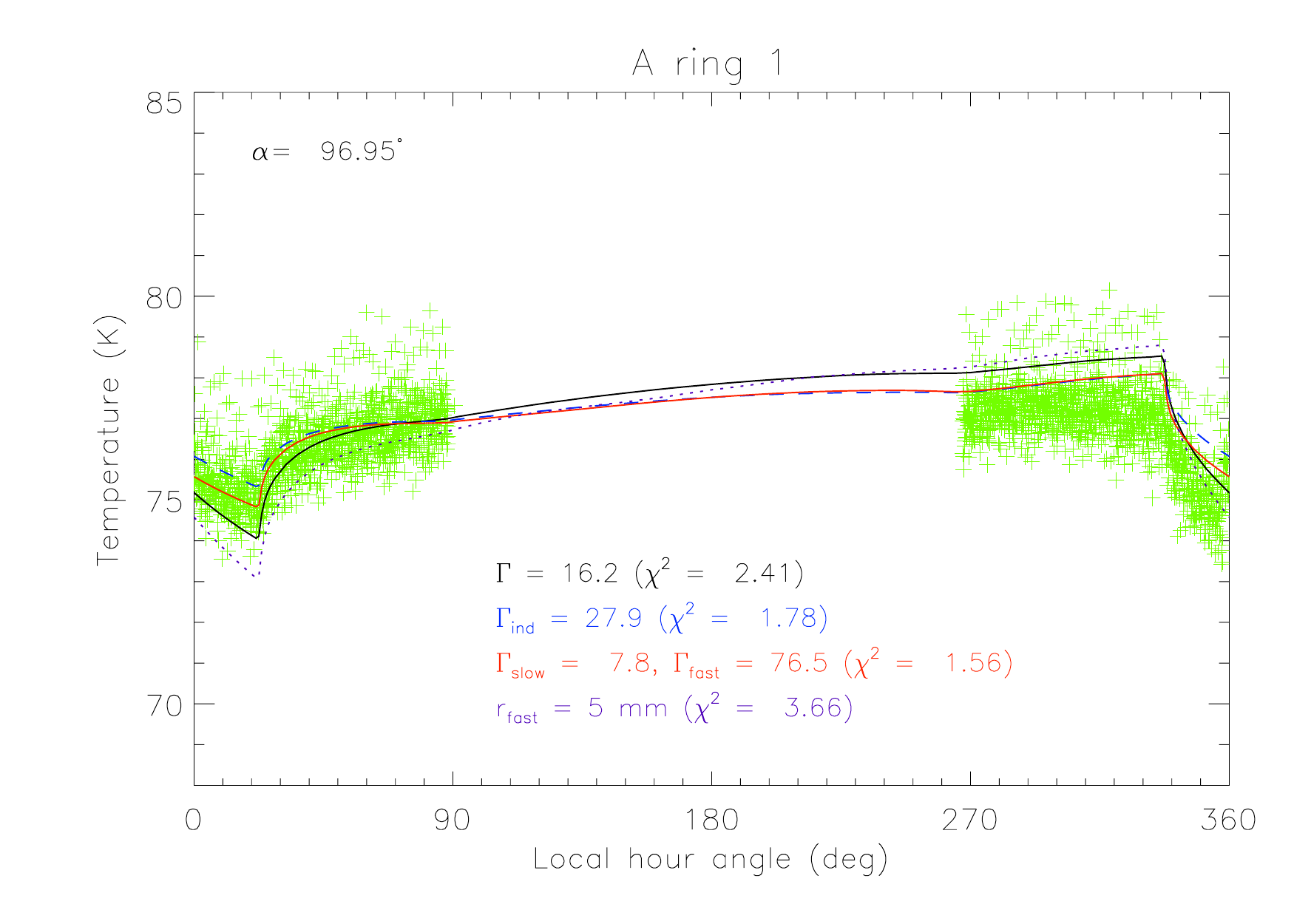}

\includegraphics[width=.49\textwidth]{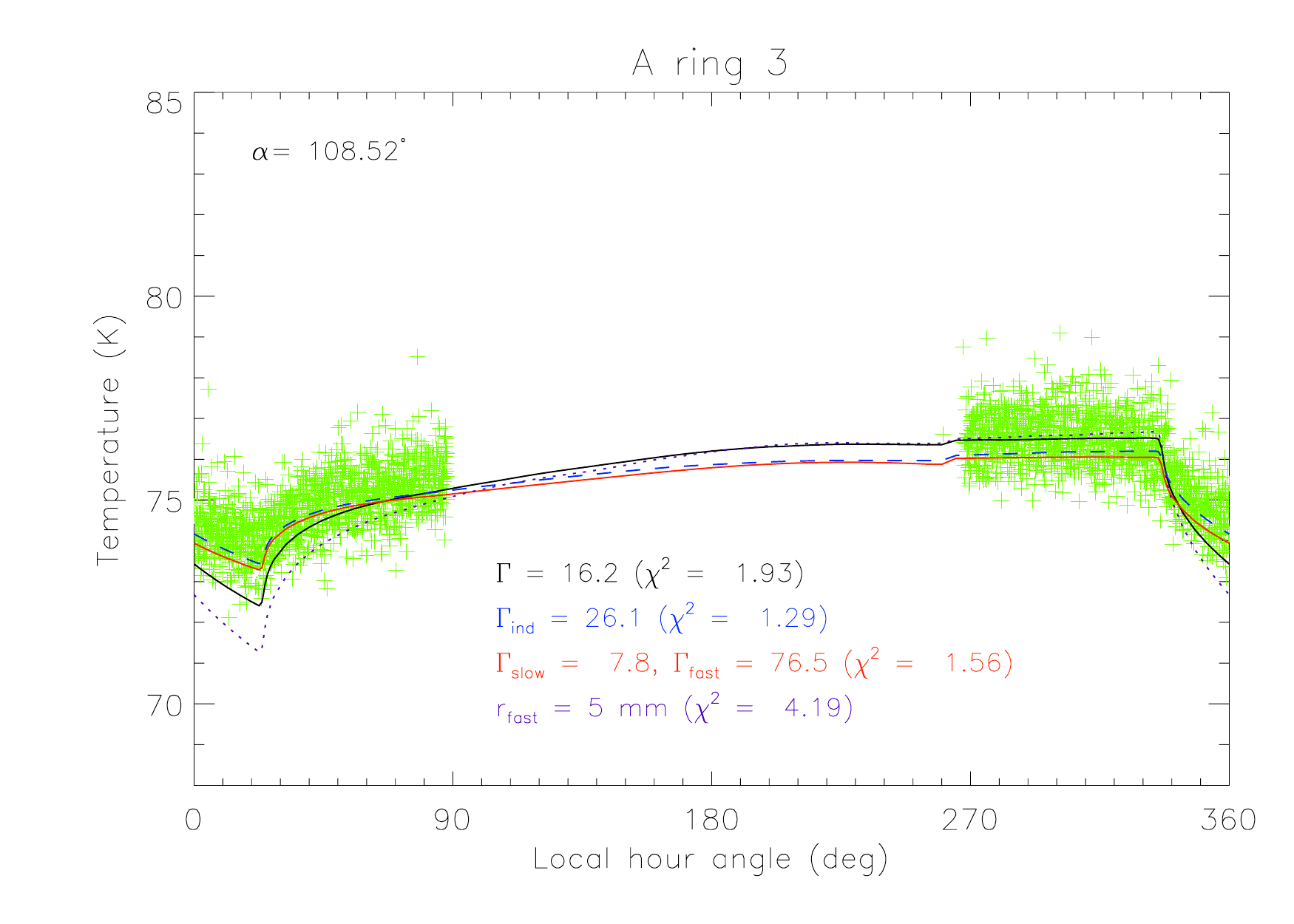}

Fig.~4. Morishima et al.
\end{figure}

\clearpage

\begin{figure}

\begin{center}
\includegraphics[width=.8\textwidth]{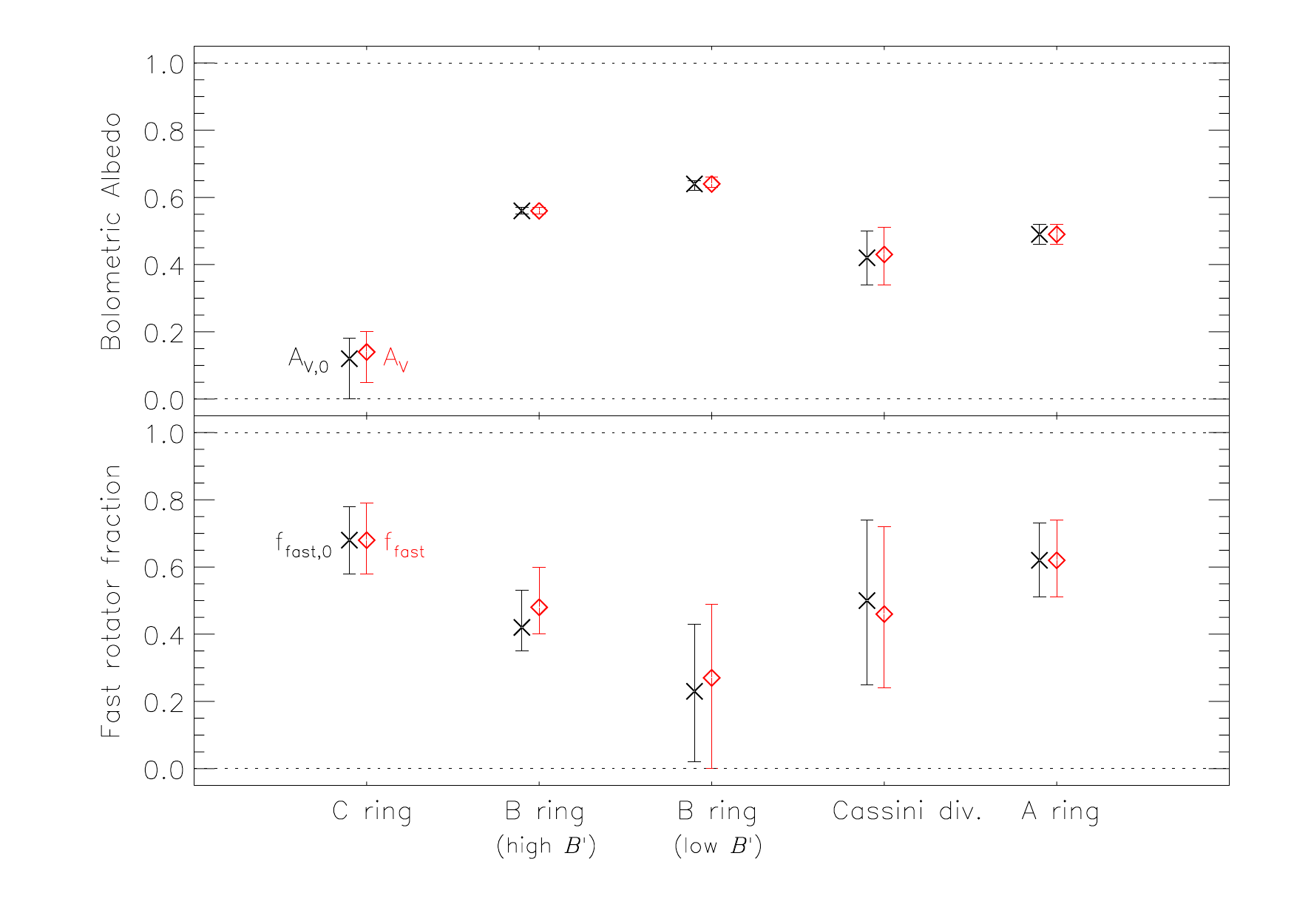}

\includegraphics[width=.8\textwidth]{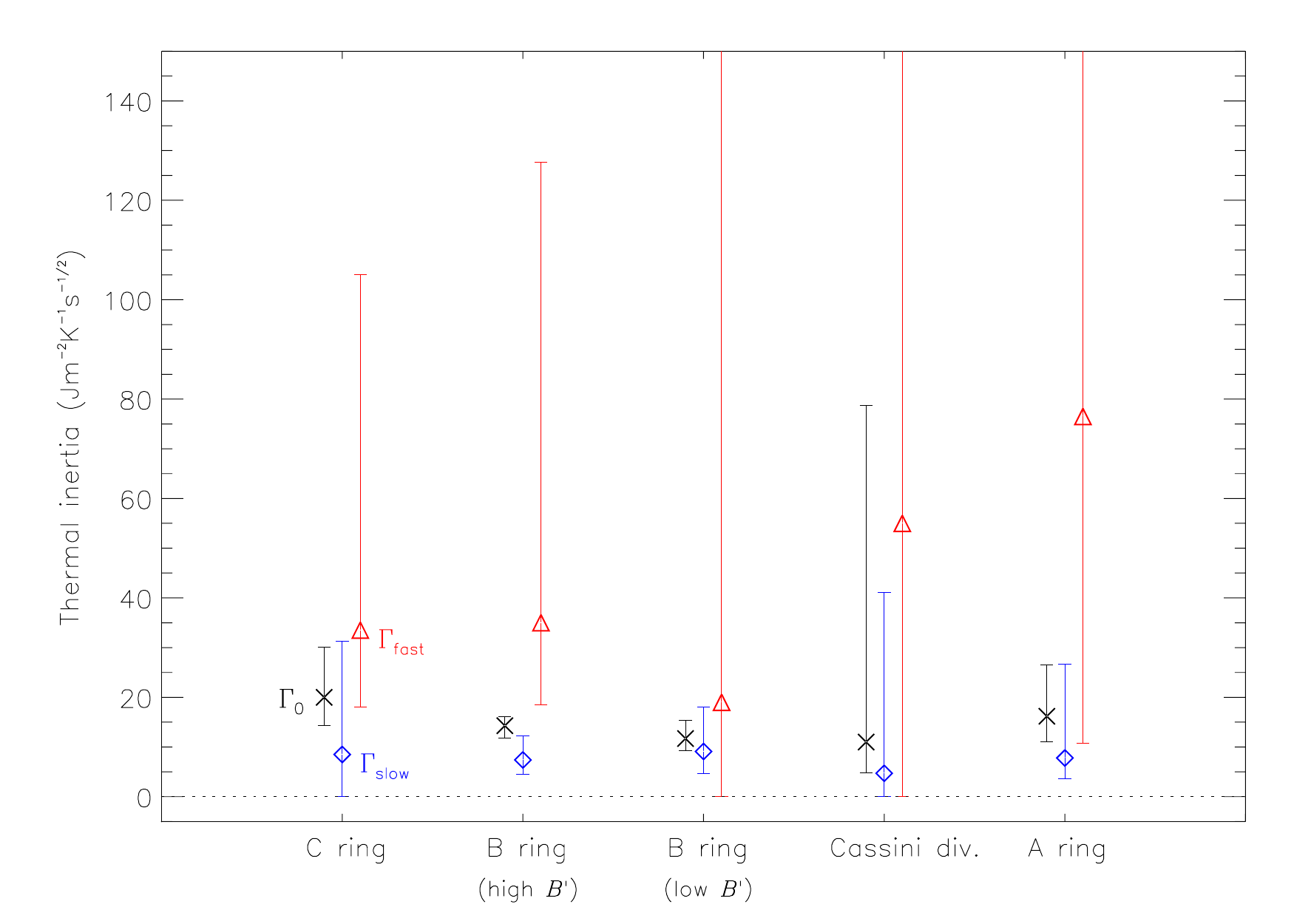}
\end{center}

Fig.~5. Morishima et al.

\end{figure}

\clearpage

\begin{figure}

\begin{center}
\includegraphics[width=.5\textwidth]{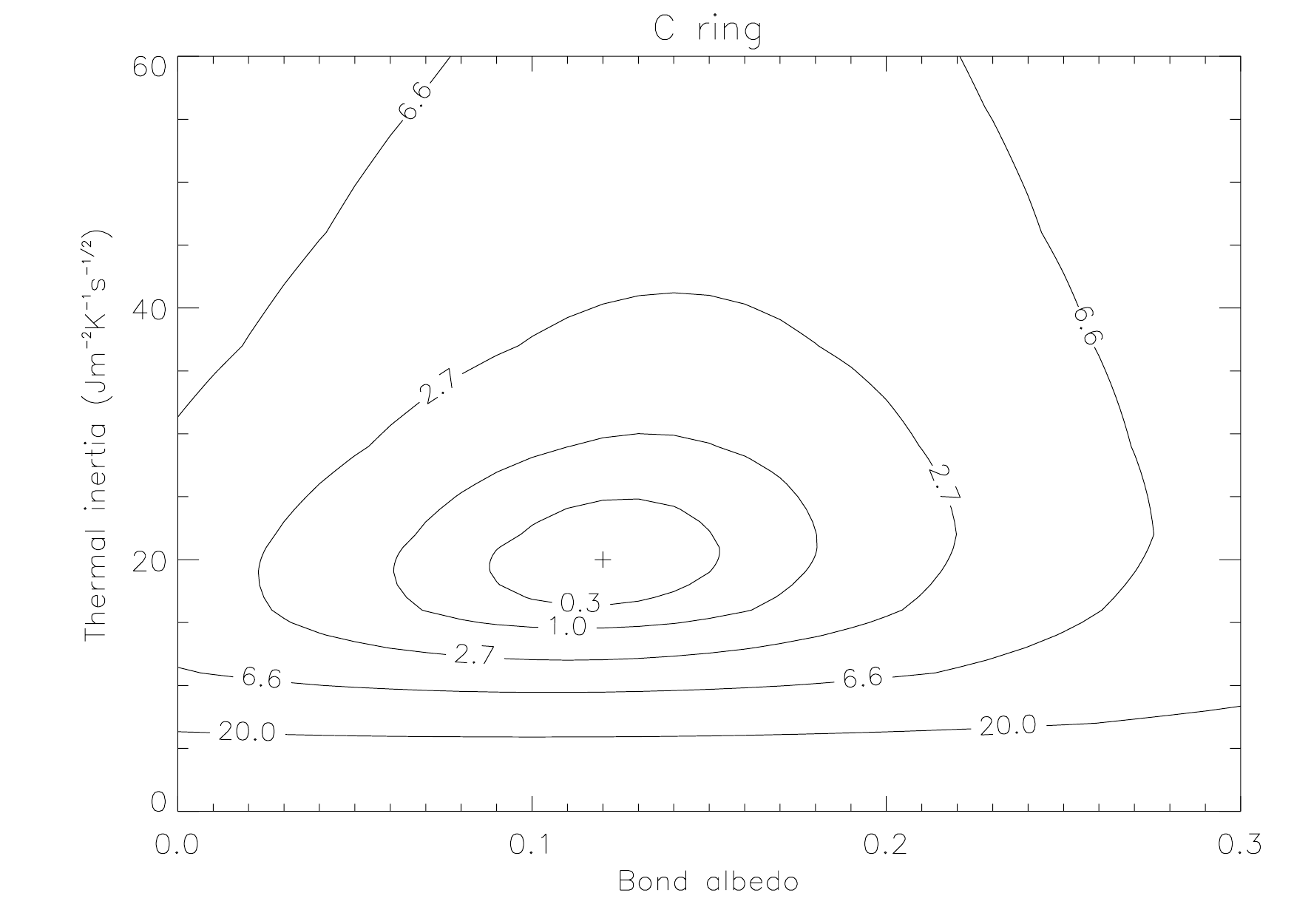}\includegraphics[width=.5\textwidth]{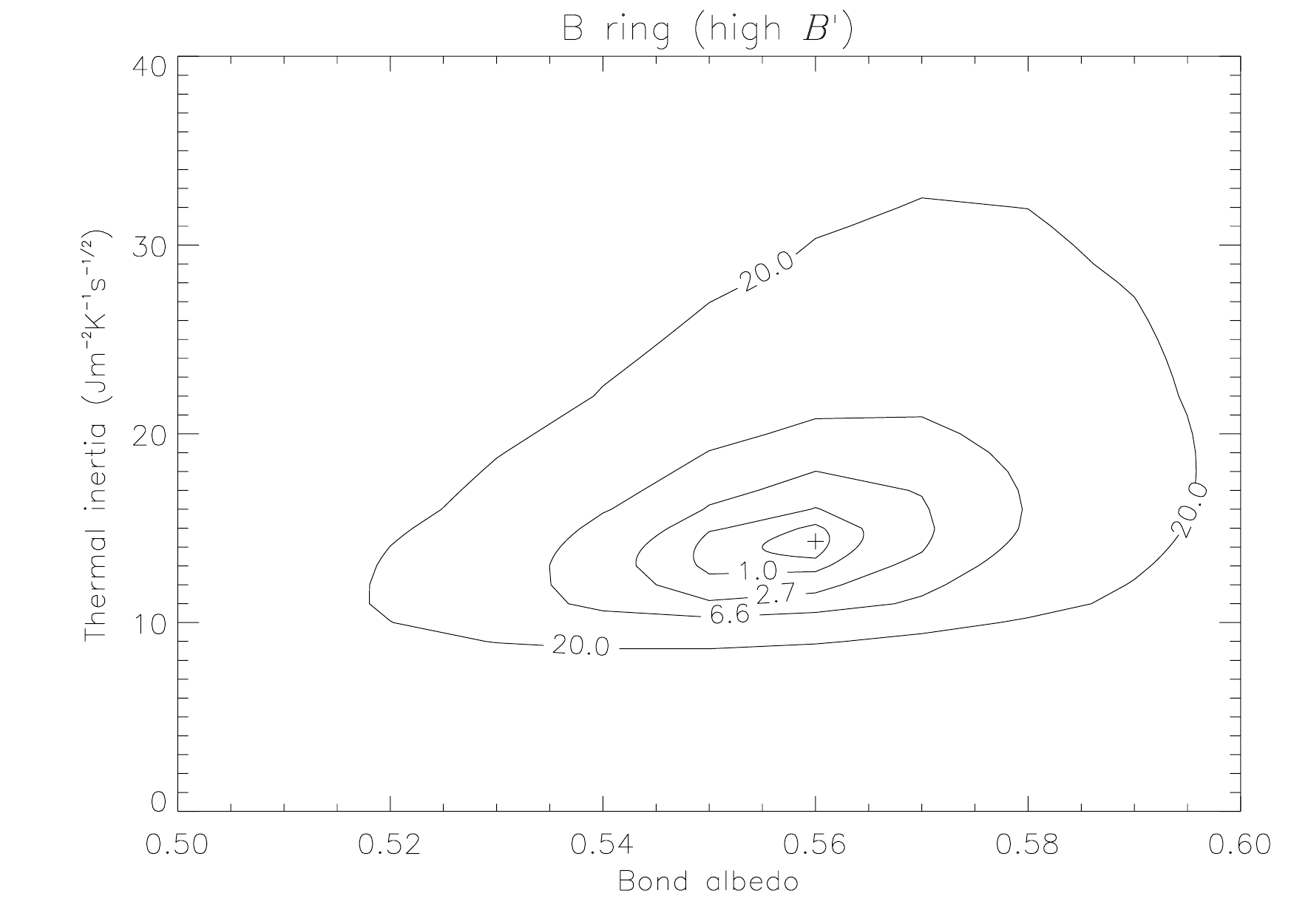}

\includegraphics[width=.5\textwidth]{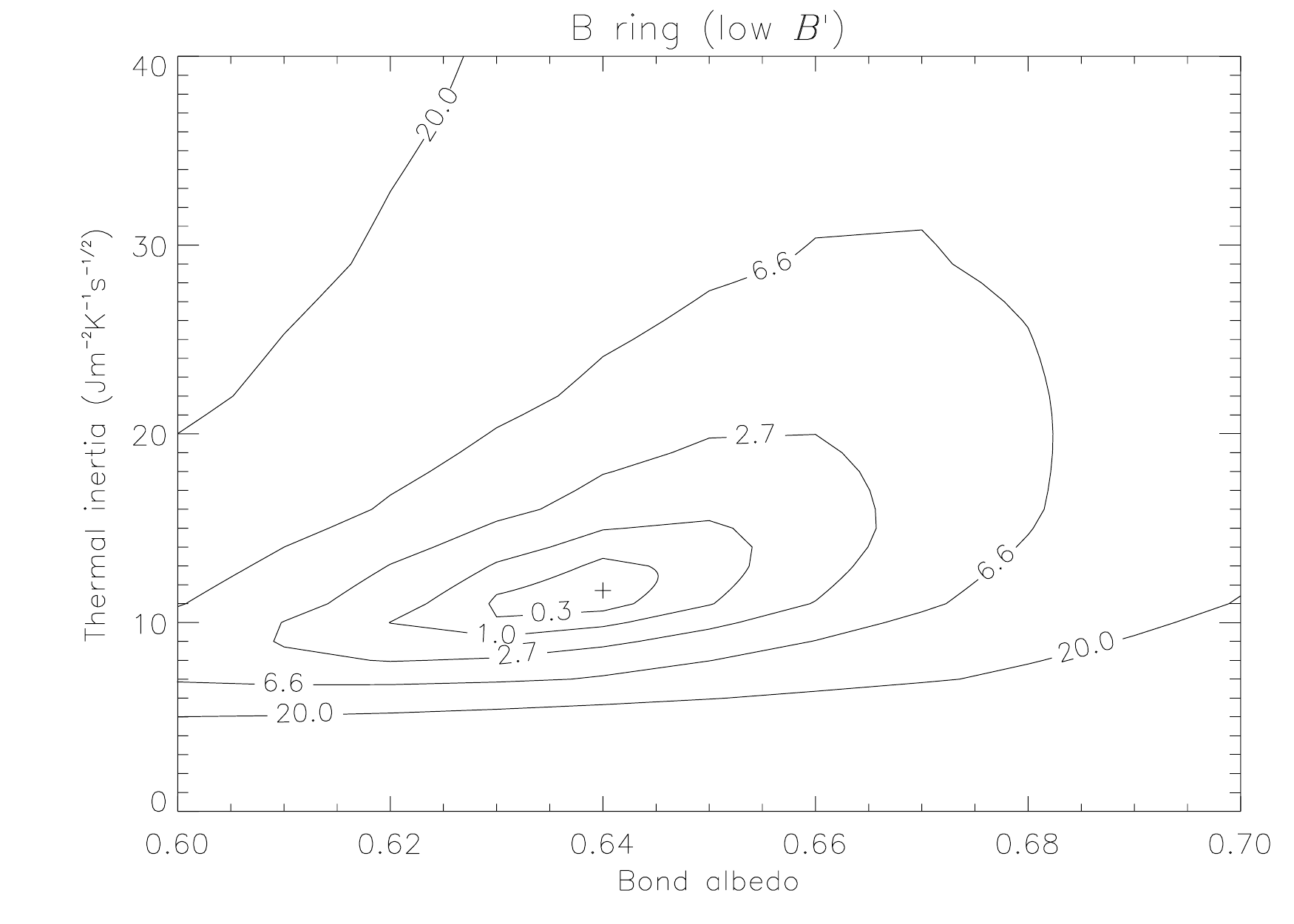}\includegraphics[width=.5\textwidth]{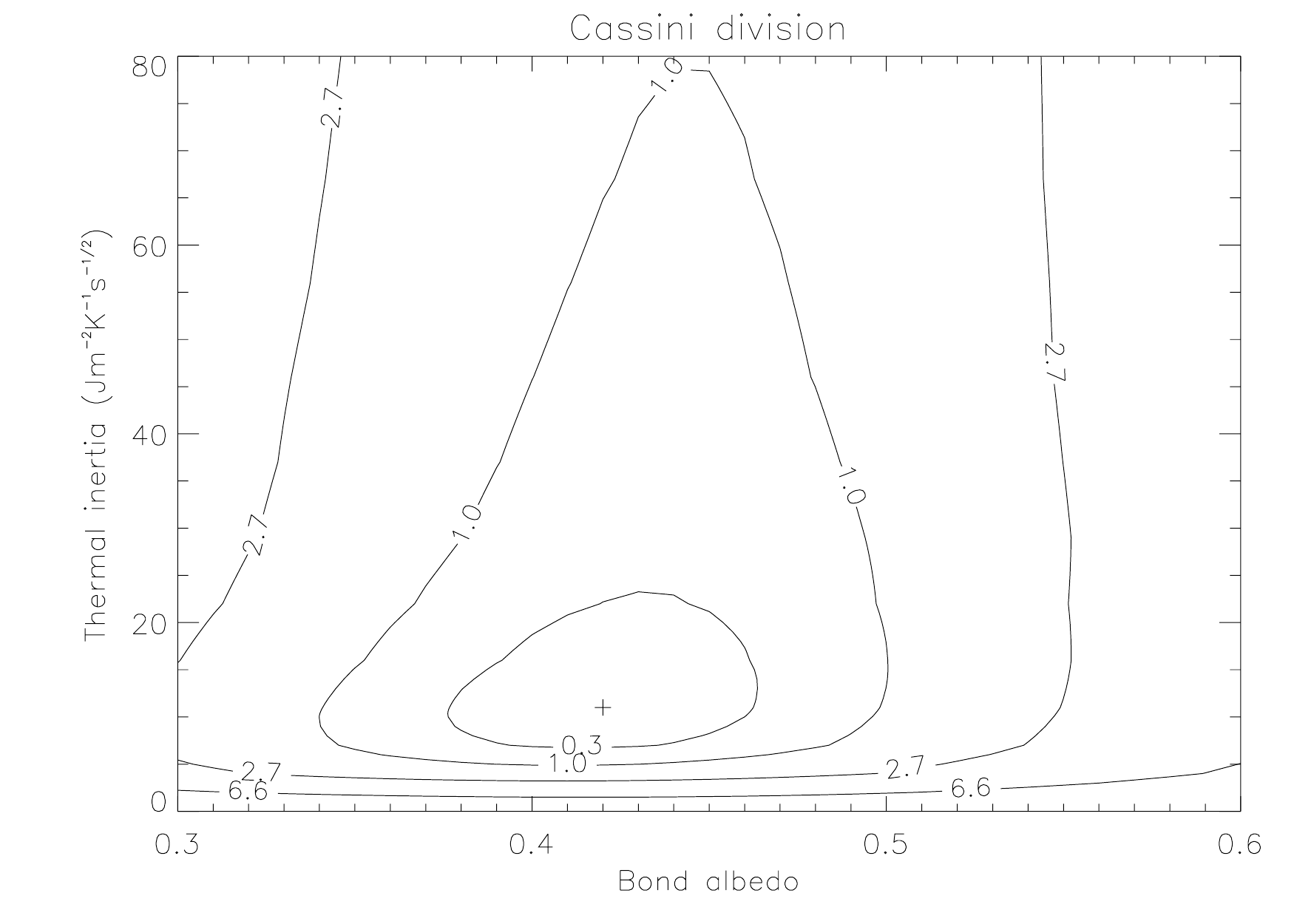}

\includegraphics[width=.5\textwidth]{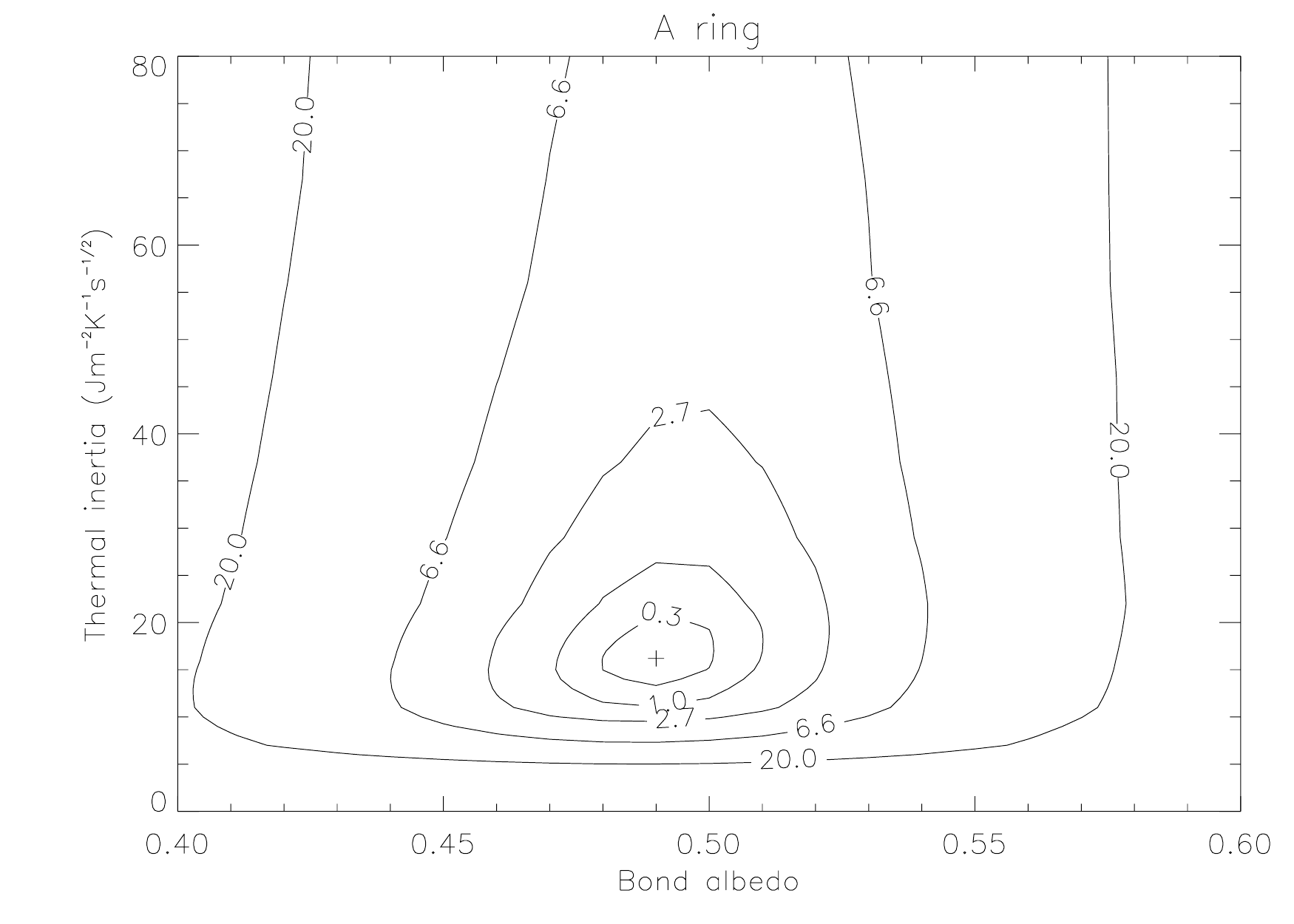}
\end{center}

Fig.~6. Morishima et al.

\end{figure}

\clearpage
\begin{figure}

\begin{center}
\includegraphics[width=.85\textwidth]{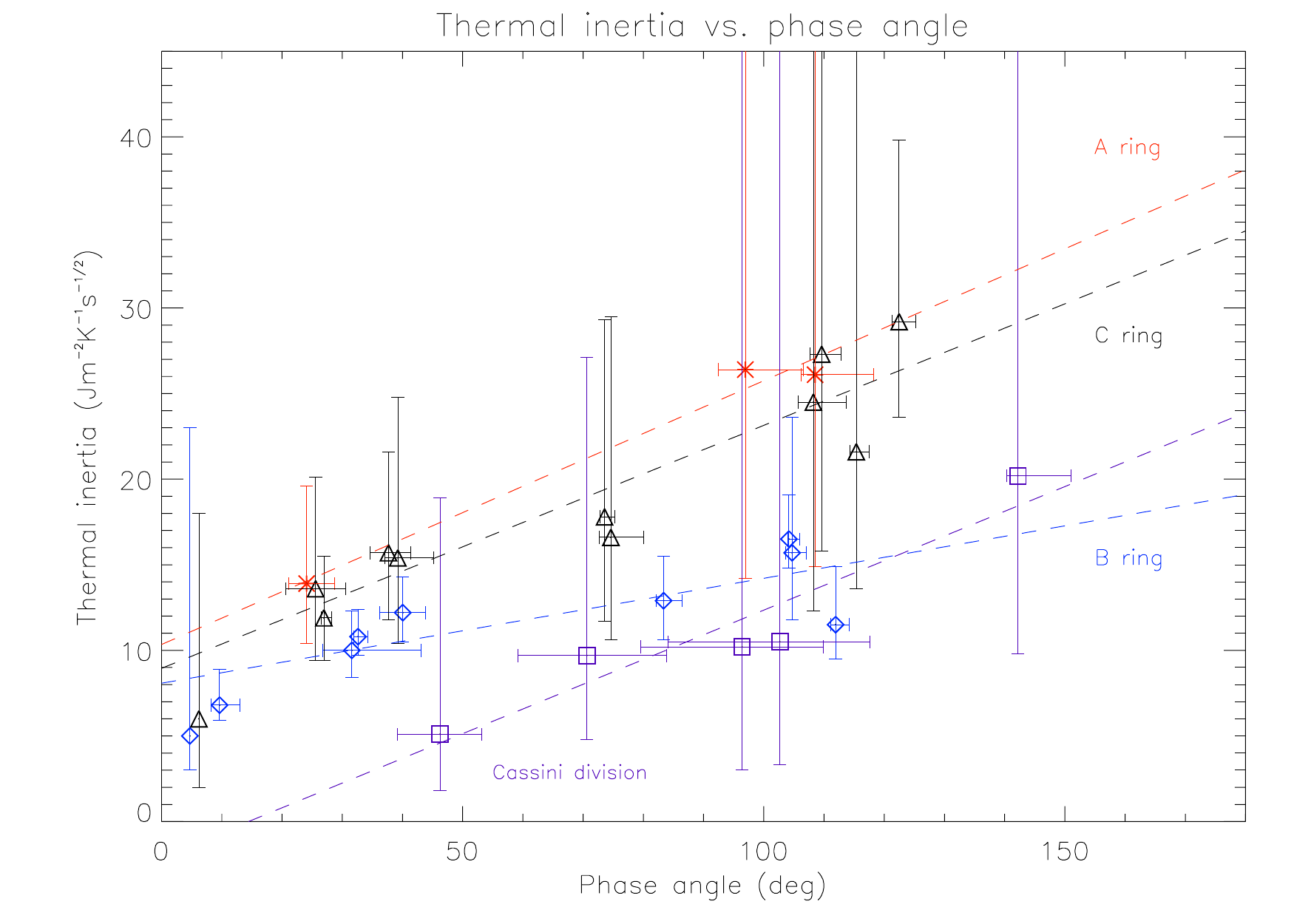}
\end{center}

Fig.~7. Morishima et al.

\end{figure}
\clearpage

\begin{figure}

\begin{center}
\includegraphics[width=.5\textwidth]{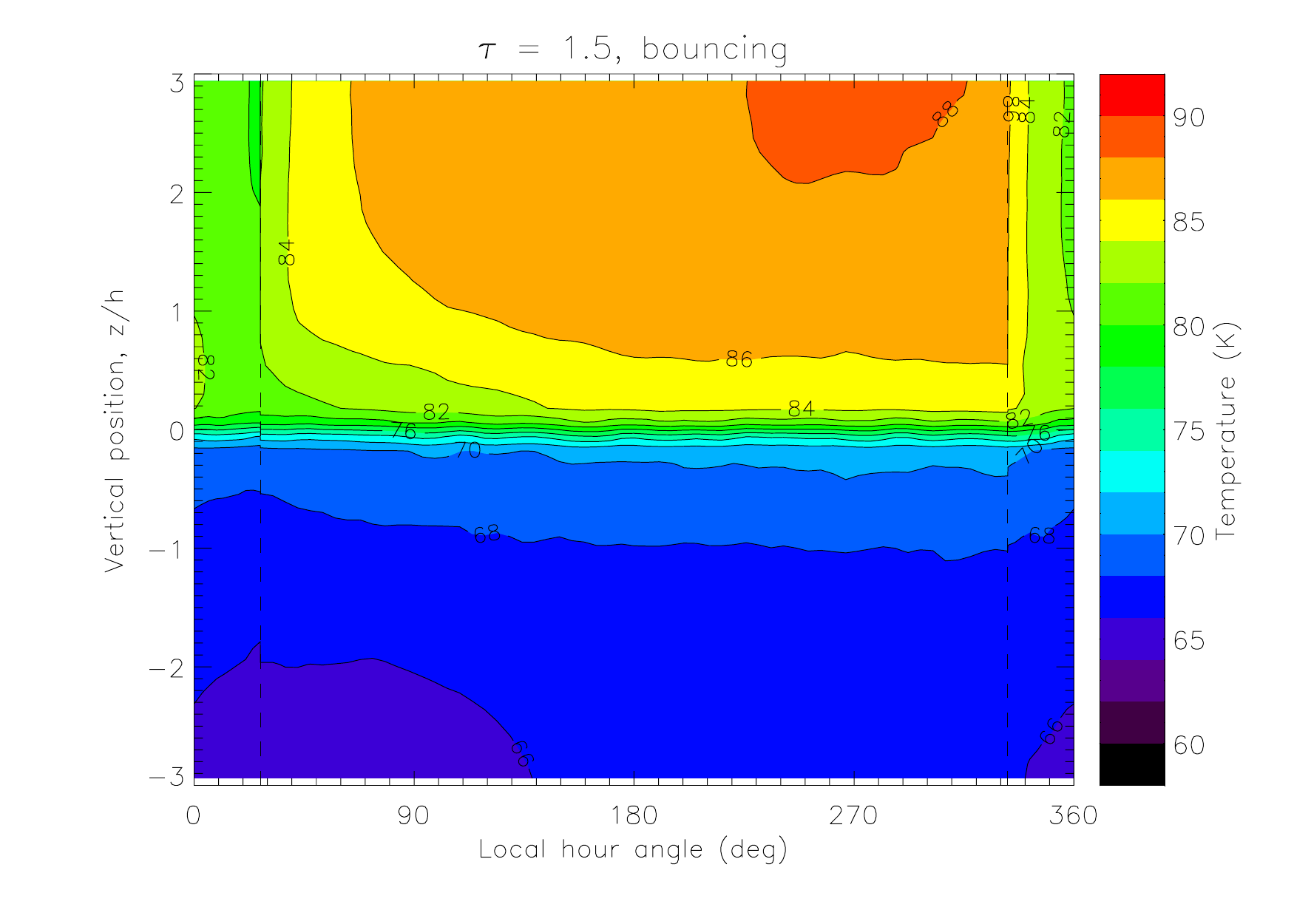}

\includegraphics[width=.5\textwidth]{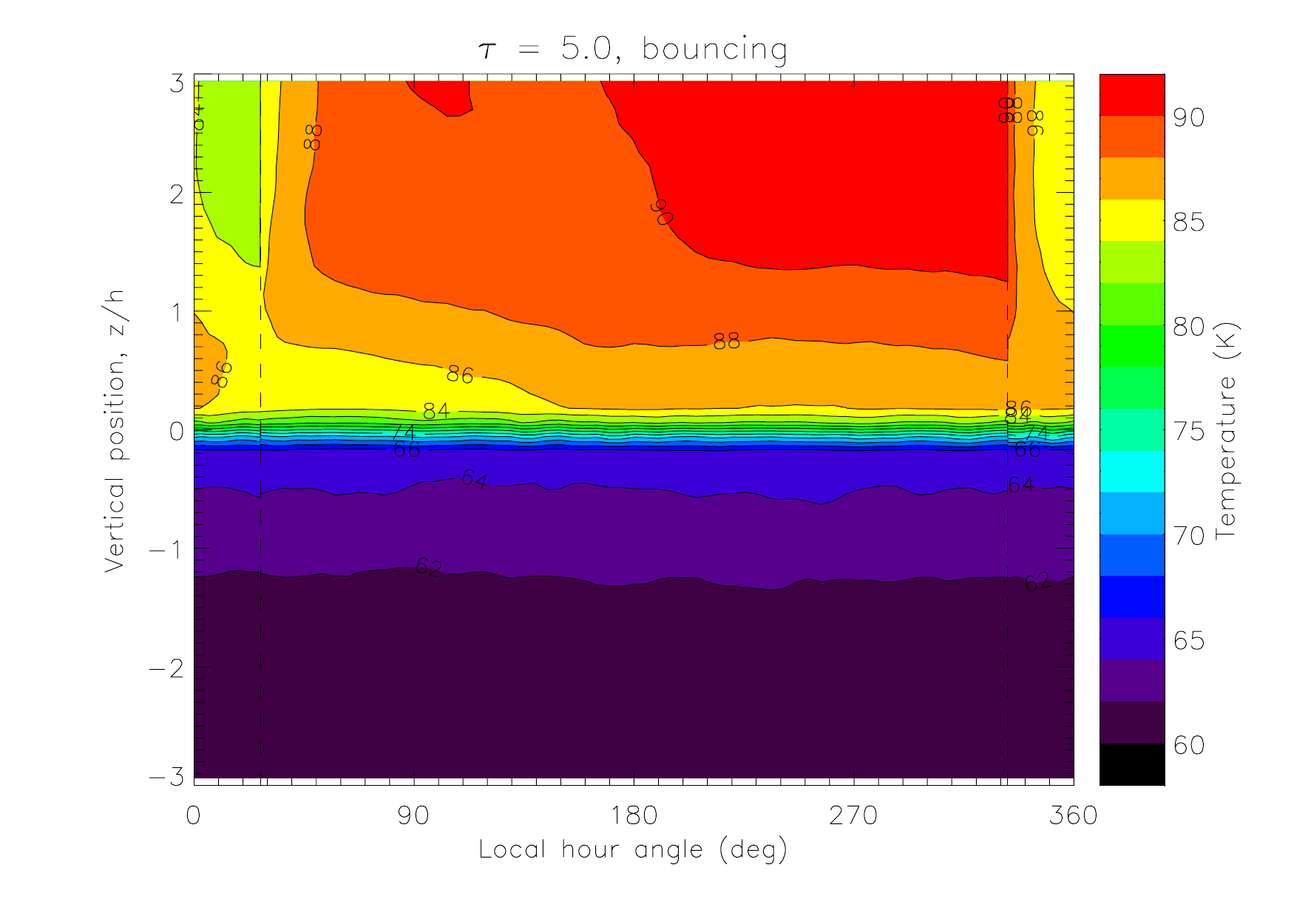}

\includegraphics[width=.5\textwidth]{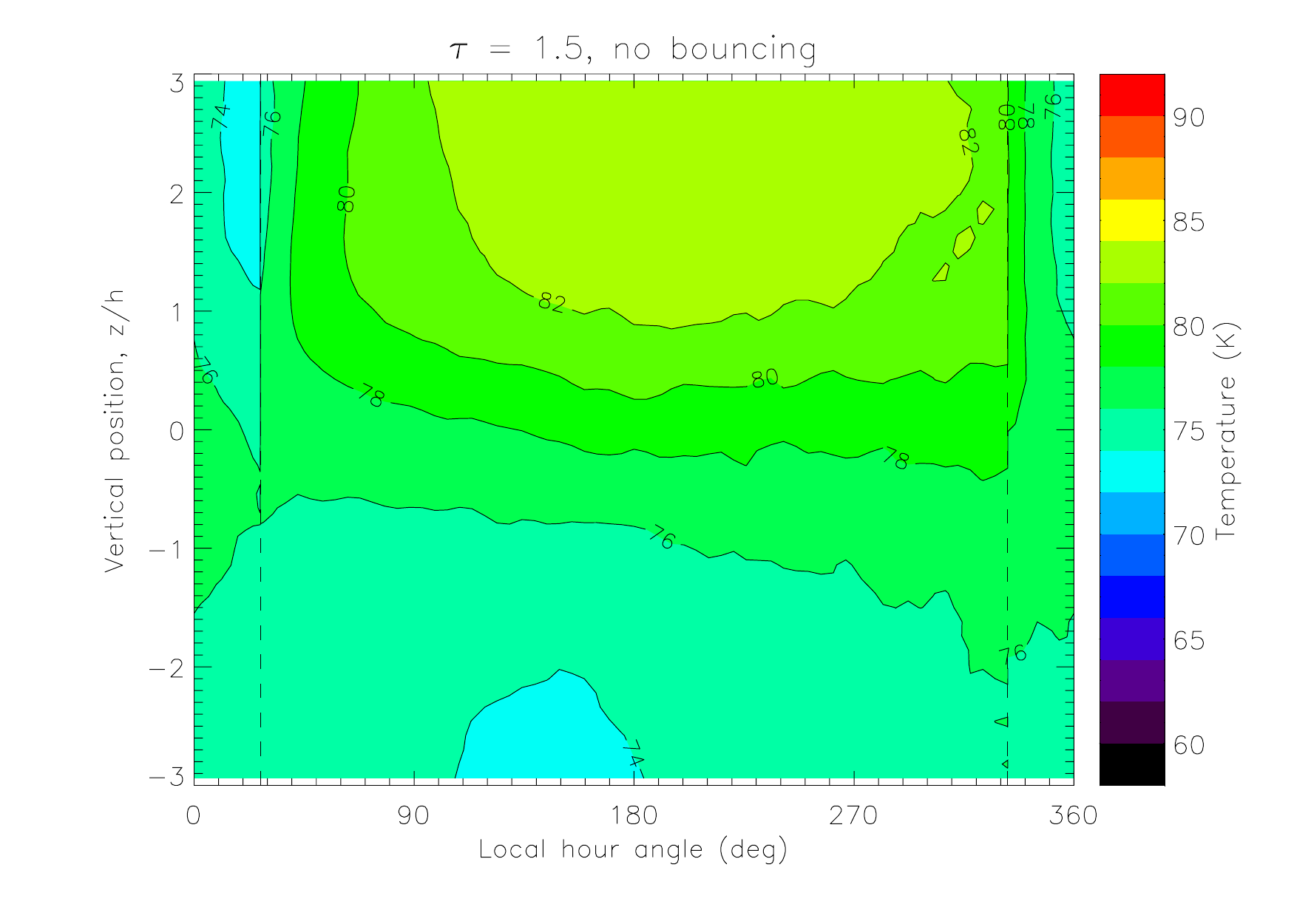}
\end{center}

Fig.~8. Morishima et al.

\end{figure}

\clearpage

\begin{figure}

\begin{center}
\includegraphics[width=.5\textwidth]{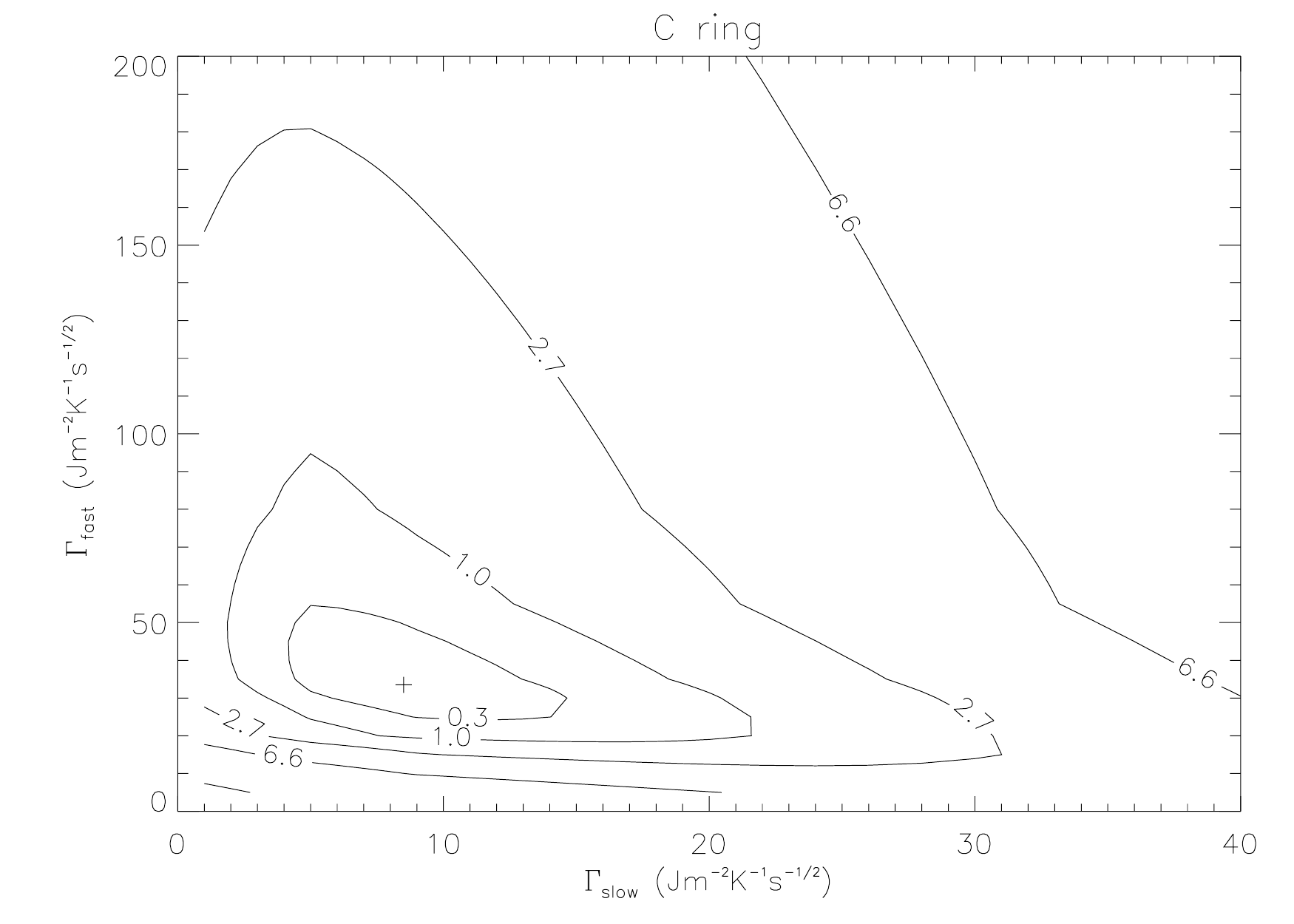}\includegraphics[width=.5\textwidth]{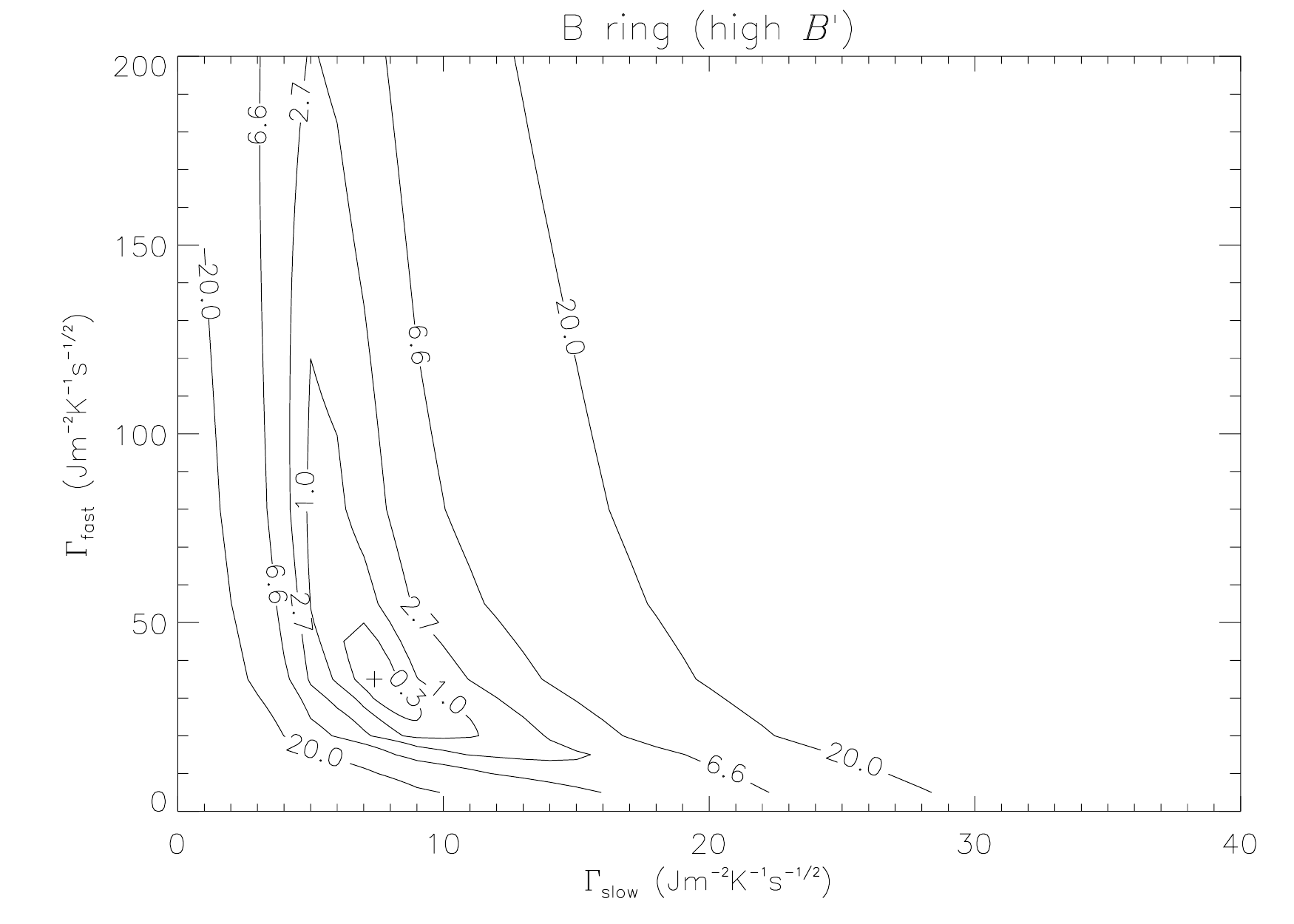}

\includegraphics[width=.5\textwidth]{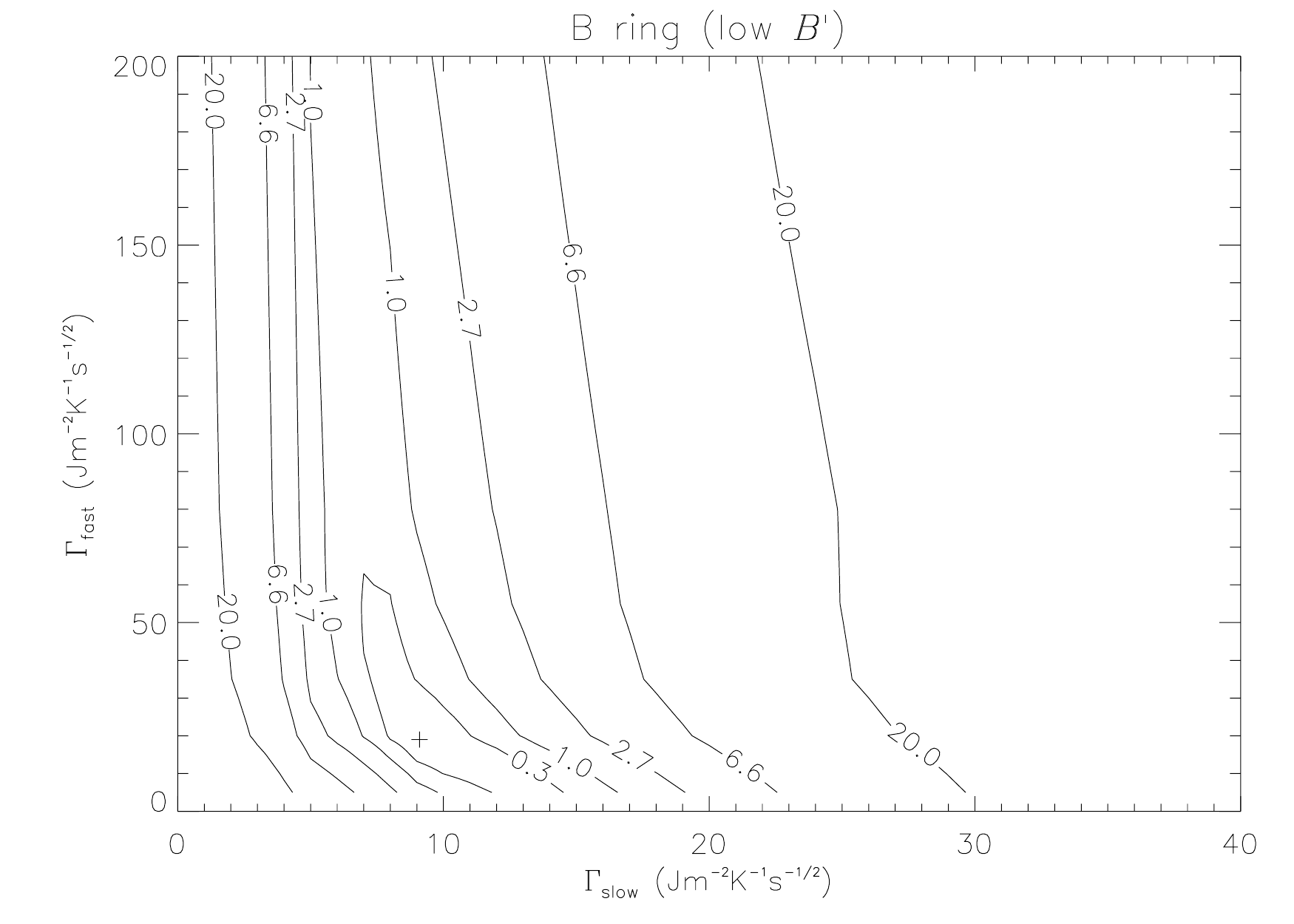}\includegraphics[width=.5\textwidth]{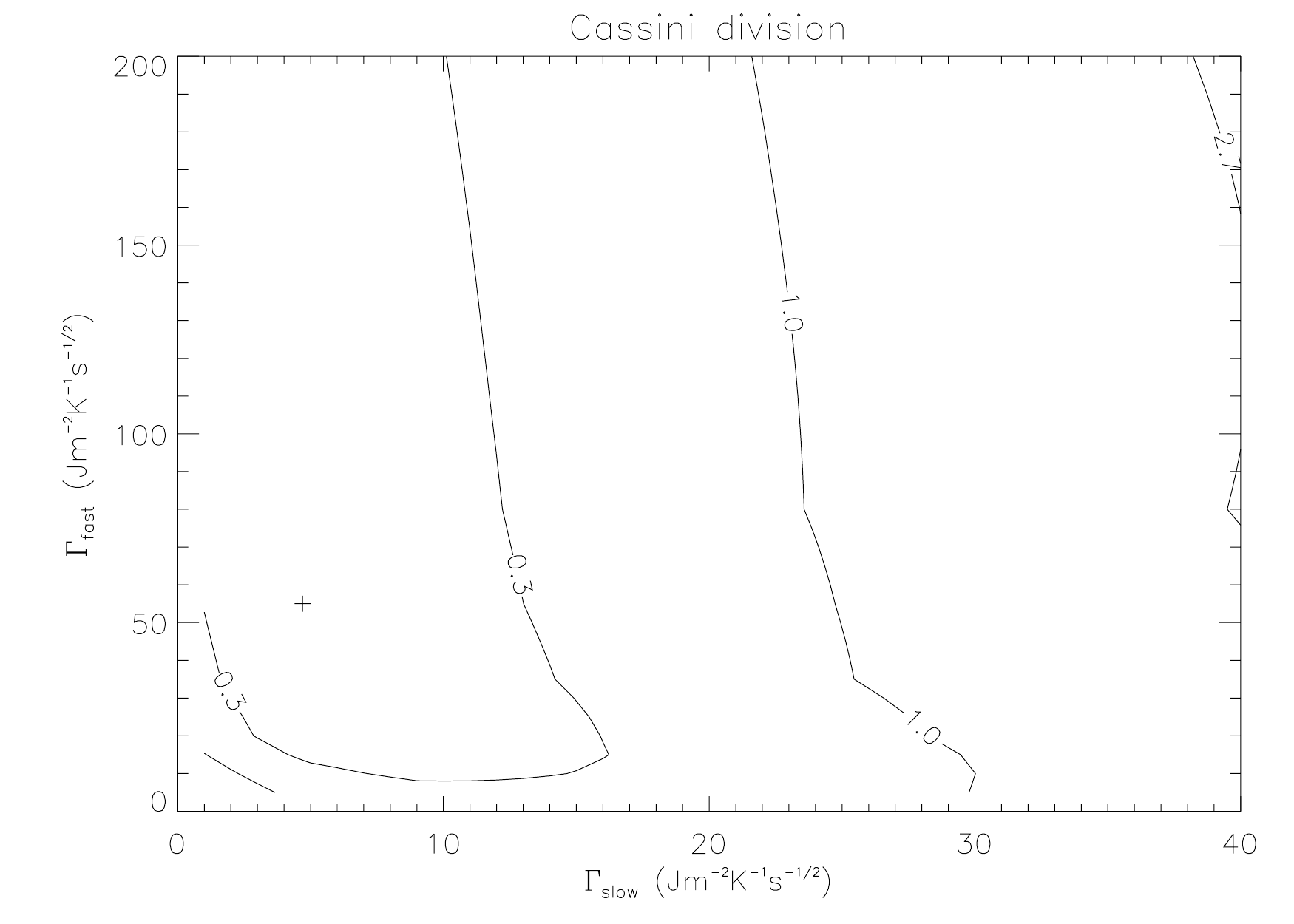}

\includegraphics[width=.5\textwidth]{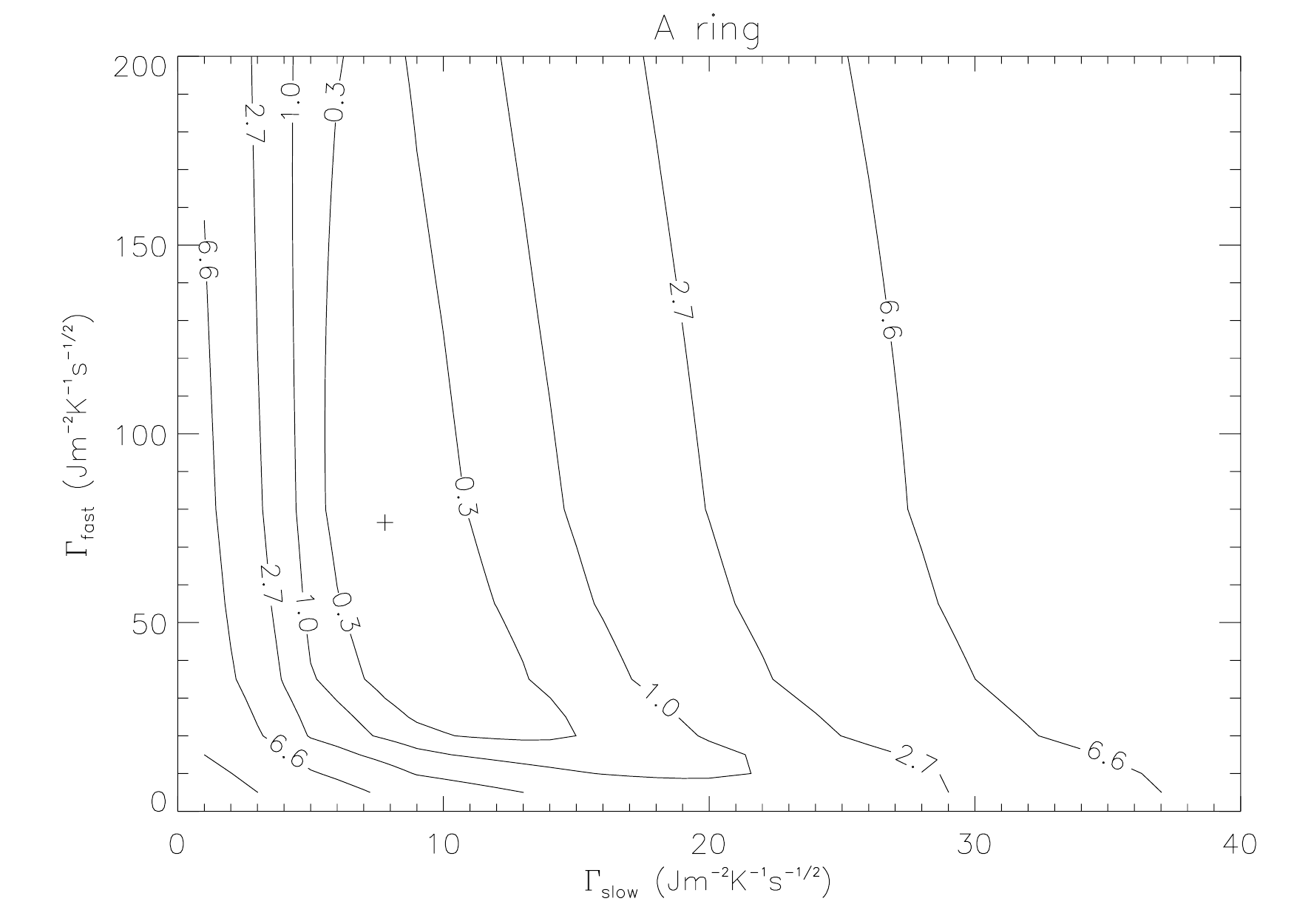}
\end{center}

Fig.~9. Morishima et al.

\end{figure}

\clearpage

\begin{figure}

\begin{center}
\includegraphics[width=.8\textwidth]{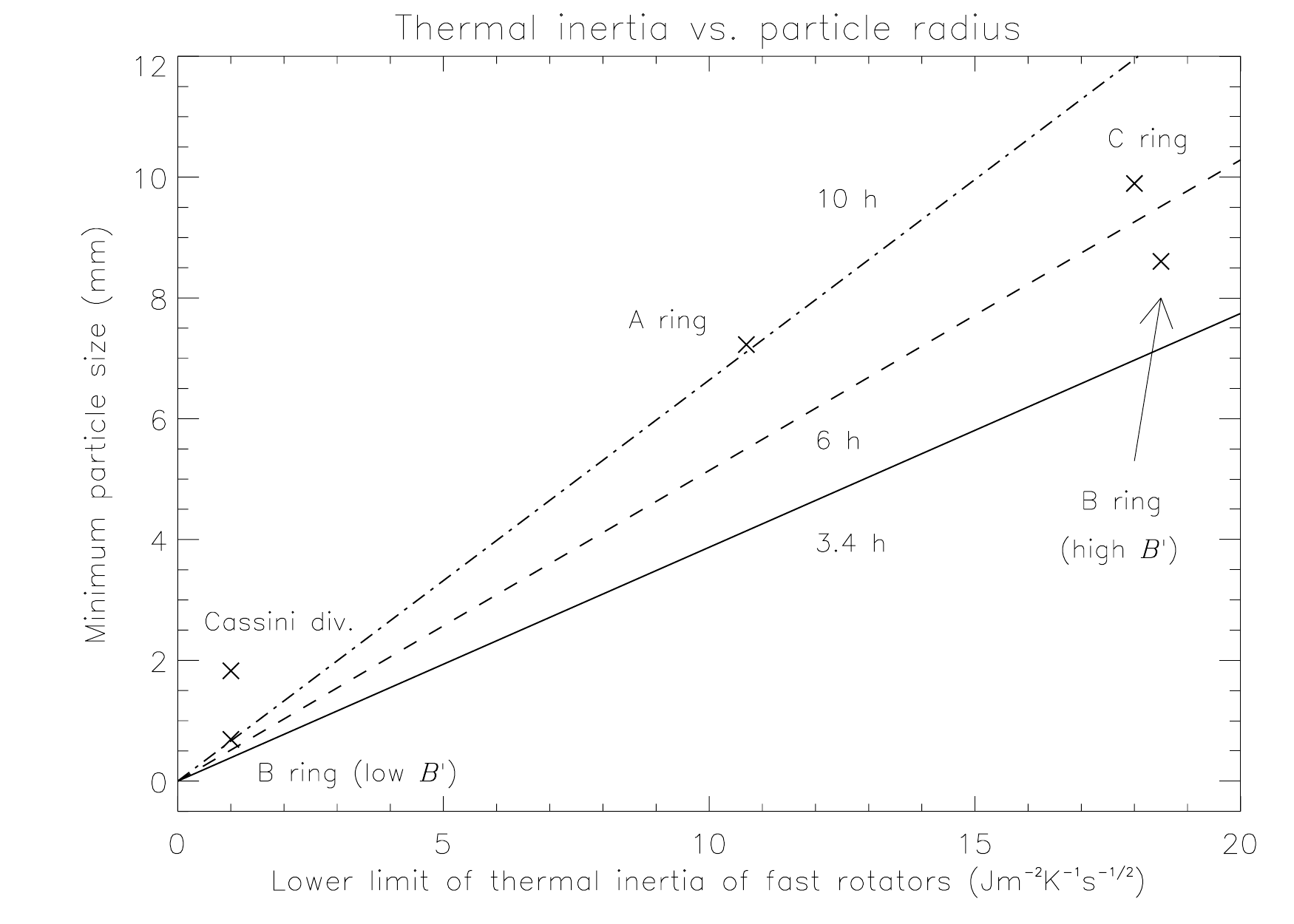}

\end{center}

Fig.~10. Morishima et al.

\end{figure}

\clearpage

\begin{figure}

\begin{center}
\includegraphics[width=.85\textwidth]{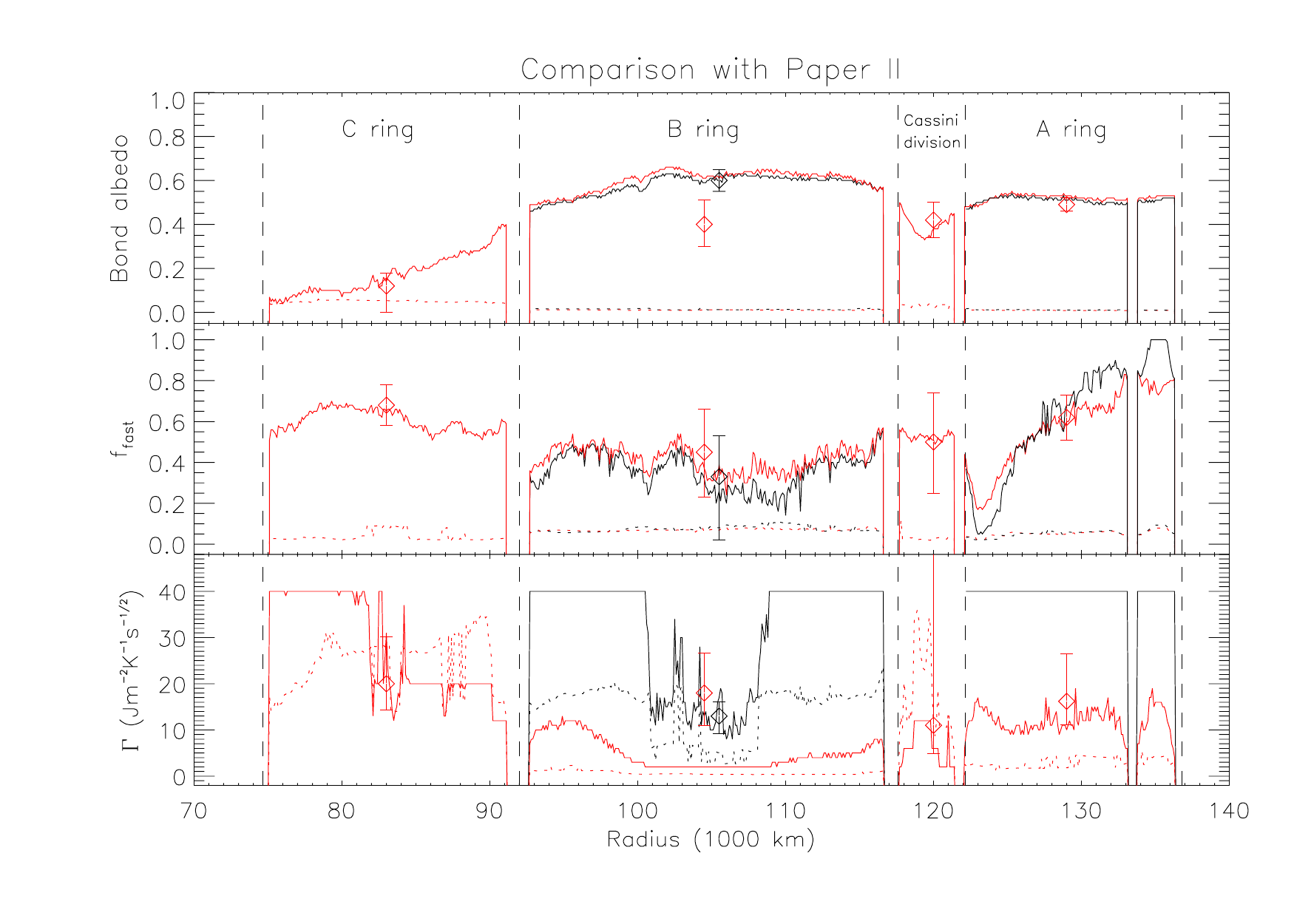}

\end{center}

Fig.~11. Morishima et al.

\end{figure}

\clearpage

\begin{figure}

\begin{center}
\includegraphics[width=.85\textwidth]{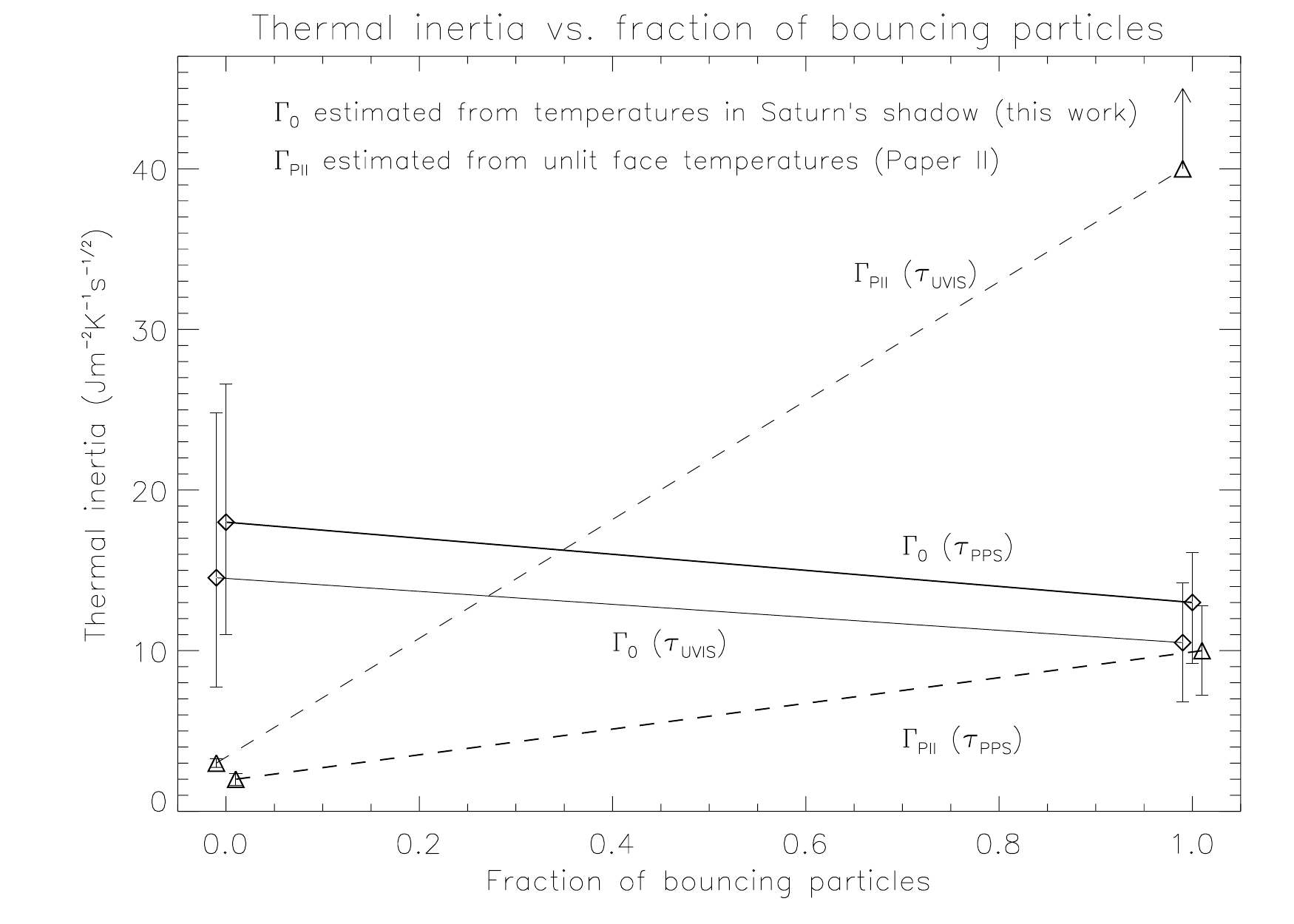}

\end{center}

Fig.~12. Morishima et al.

\end{figure}

\clearpage

\begin{figure}

\begin{center}
\includegraphics[width=.85\textwidth]{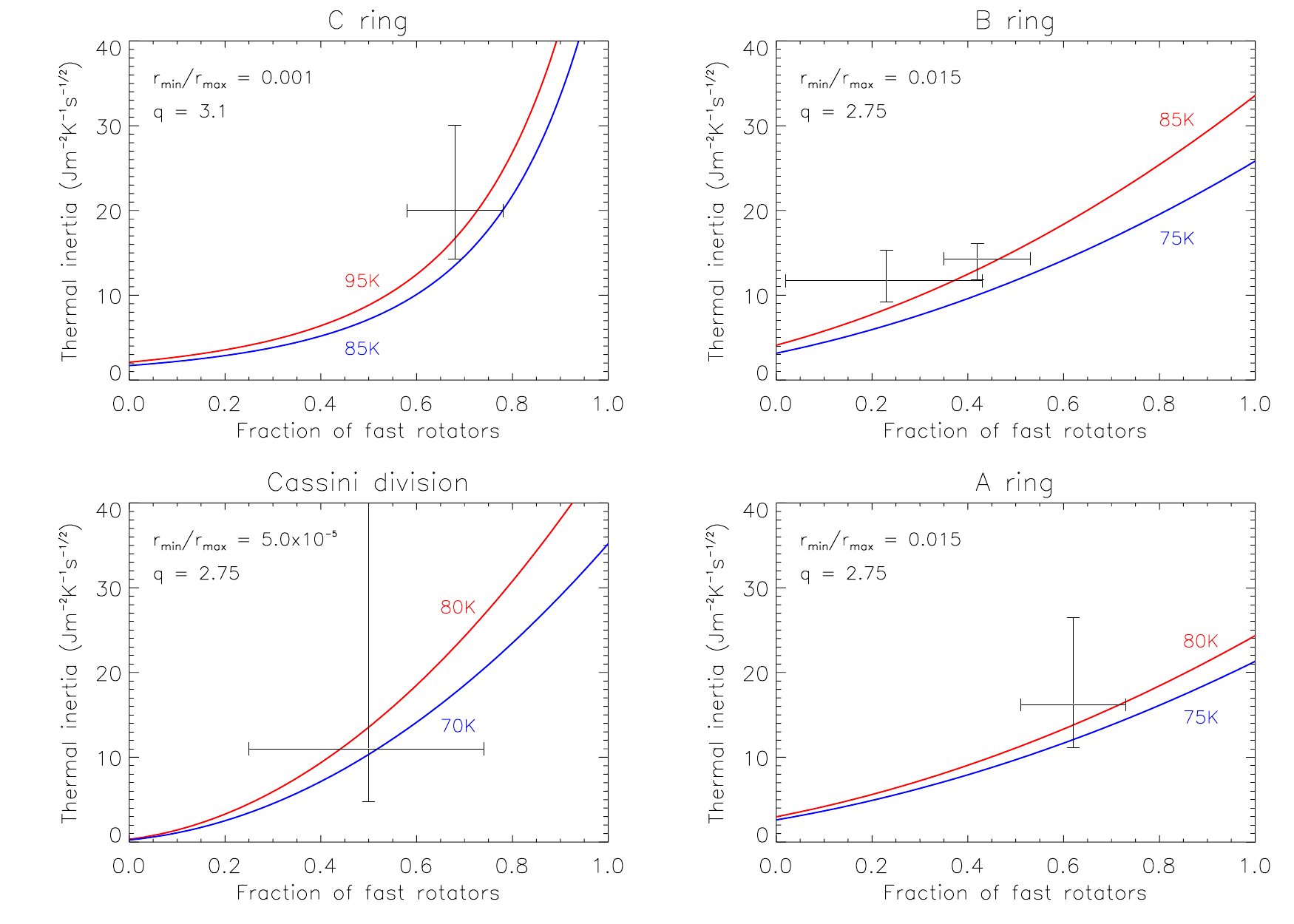}

\end{center}

Fig.~13. Morishima et al.

\end{figure}

\end{document}